\documentclass[aps,prx,twocolumn,amsmath,amssymb,amsfonts]{revtex4-2}
\usepackage{dcolumn}
\usepackage{bm}
\usepackage{amsmath}
\usepackage{graphicx} %eps figures can be used instead
\usepackage{epstopdf}%This line makes .eps figures into .pdf - please comment out if not required.

\usepackage[titletoc,title]{appendix}
\usepackage{hyperref}
\usepackage{xcolor}
\newcommand{\be}{\begin{equation}}
\newcommand{\ee}{\end{equation}}
\newcommand{\bea}{\begin{eqnarray}}
\newcommand{\eea}{\end{eqnarray}}
\newcommand{\rIm}{\mathrm{Im}}
\newcommand{\rRe}{\mathrm{Re}}
\newcommand{\re}{\mathrm{e}}            % Roman e for exponential

\usepackage[normalem]{ulem} %To strikeout text

\DeclareMathOperator\erfc{erfc}
\DeclareMathOperator\erf{erf}

\begin{document}

\title{Lamb-type solution and properties of unsteady Stokes equations}

\author{Itzhak Fouxon$^{1,2,3}$}\email{itzhak8@gmail.com} \author{Alexander Leshansky$^1$}\email{lisha@technion.ac.il} \author{Boris Rubinstein$^4$} \email{bru@stowers.org} \author{Yizhar Or$^{2}$}\email{izi@me.technion.ac.il}
\affiliation{$^1$ Department of Chemical Engineering, Technion, Haifa 32000, Israel}
\affiliation{$^2$ Faculty of Mechanical Engineering, Technion - Israel Institute of Technology, Haifa, 3200003, Israel}
\affiliation{$^3$ Department of Computational Science and Engineering, Yonsei University, Seoul 120-749, South Korea}
\affiliation{$^4$ Stowers Institute for Medical Research, 1000 E 50th st.,Kansas City, MO 64110, USA}

\begin{abstract}

In this paper, we derive the general solution of the unsteady Stokes equations for an unbounded fluid in spherical polar coordinates, in both time and frequency domains. The solution is an expansion in vector spherical harmonics and given as a sum of a particular solution, proportional to pressure gradient exhibiting power-law spatial dependence, and a solution of vector Helmholtz equation decaying exponentially in far field, the decomposition originally introduced by Lamb. The proposed solution representation resembles the classical Lamb's decomposition of the steady Stokes equations solution: the series coefficients are projections of radial component, divergence and curl of the boundary velocity on scalar spherical harmonics. The proposed general representation solution can be applied to construct the transient exterior flow induced by an arbitrary velocity distribution at the spherical boundary, such as arising in the squirmer model of a microswimmer. It can also be used to construct solutions for transient flows driven by initial conditions, unbounded flows driven by volume forces
%\red{I did not understand, we do not have point forces here.}
or disturbance to the unsteady flow due to a stationary spherical particle.
%\red{We might instead of these write the sentence from the text which is This is equivalent to the general solution of unsteady Stokes equations in the exterior of a given sphere, because any solution can be reduced to a sum of a solution for unbounded fluid and a solution driven by a non-trivial boundary condition on the sphere surface. }
The long-time behavior of solution is controlled by the flow component corresponding to average (or collective) motion %\red{This component includes also sphere averages of radial component of the boundary flow which does not correspond to a net motion}
of the boundary. This conclusion is illustrated by the study of decay of transversal wave in the presence of a fixed sphere. We further show that the general representation reduces to the well-known solutions for unsteady flow around a sphere undergoing oscillatory rigid-body (translation and rotation) motion. The proposed solution representation provides an explicit form of the velocity potential far from an oscillating body (``generalized" Darcy's law) and high- and low-frequency expansions. The leading-order high-frequency expansion yields the well-known ideal (inviscid) flow approximation, and the leading-order low-frequency expansion yields the steady Stokes equations. We derive the higher-order corrections to these approximations and discuss d'Alembert paradox. Continuation of the general solution to imaginary frequency yields the general solution of the Brinkman equations describing viscous flow in porous medium.

\end{abstract}

\maketitle

\section{Introduction}

The method of spherical harmonics expansion is a standard method for solving the Laplace equation widely used in the various fields. In this work, we derive a similar expansion for the unsteady Stokes equations describing low-Reynolds-number ($Re<1$) transient or unsteady flows of incompressible viscous fluid \cite{kim,hb,LL}. The smallness of $Re$ allows to drop the \emph{nonlinear} (quadratic) term in the full Navier-Stokes equations yielding the \emph{linear} unsteady Stokes equations, possessing a general solution.  In the present paper we describe the methods of construction of the general solutions of the unsteady Stokes equations and study their properties.

Low-Reynolds-number (viscous) hydrodynamics distinguishes between steady Stokes equations, obtained by dropping all inertia terms (i.e., due to the material derivative of the velocity) in the Navier-Stokes equations, and unsteady Stokes equations obtained by keeping the velocity time-derivative (i.e., the Eulerian acceleration term). The steady Stokes equations read $\eta\nabla^2\bm v =\nabla p$, where $\bm v$ is the incompressible (solenoidal) fluid's velocity field, $p$ is the pressure distribution and $\eta$ is the dynamic viscosity of the fluid; they contain no explicit time dependence and the quasi-static approximation applies. There are several known representations of the general solution of the steady Stokes equations. The seminal Lamb's solution \cite{Lamb} is a sum of three series, whose terms are composed from solid spherical harmonics (see, e.g., \cite{hb}). The first series is a particular solution of the Stokes equations due to the source $\nabla p$, and the remaining two series provide a general solenoidal solution of the vector Laplace equation $\nabla^2\bm v=0$ (see \cite{kim}). Another form of the general solution is given by the so-called adjoint method that uses the expansion of an arbitrary vector field into a complete set of vector functions derived from the spherical harmonics \cite{kim,SF}. This set is rather similar to vector spherical harmonics (VSH) employed here (see \cite{sph}), however has somewhat more cumbersome orthogonality relations. Other forms of the general solution of the Stokes equations are the Papkovich-Neuber \cite{ps,ne,cong}, the Naghdi-Hsu \cite{su,xu,se} and the Boussinesq-Galerkin \cite{bousi,xu} formulations. Finally, \cite{op} provided a general solution derived from poloidal-toroidal decomposition of incompressible flow \cite{chan,cha} (see also \cite{ranger}). Lamb's decomposition is by far the most useful, see, e.g., \cite{Fil} for numerical implementation for many-particle systems.

In striking contrast to the steady Stokes equations, the general solutions of the unsteady Stokes equations, given in the frequency domain by $-i\rho \omega \bm v=\eta\nabla^2\bm v-\nabla p$, attracted far less attention. Here $\rho$ is the fluid's density, $\omega$ is the frequency and $i$ is the imaginary unit. The main reason for less attention is that typically the time and convective derivative terms are of the same order of magnitude. The unsteady Stokes equations apply when the flow is periodic and/or has significant time dependence.  The general solution for axially symmetric case was provided in \cite{rao} in terms of the stream-function. For non-axisymmetric flows, there are two known representations of the general solution. One possible decomposition of the solution is similar to that by Lamb for steady Stokes equations. It was reported by Lamb \cite{Lamb} and is constructed as a sum of the particular solution $-i \omega^{-1}\nabla (p/\rho)$ where $\nabla^2 p=0$, and a general solution of the solenoidal vector Helmholtz equation with an imaginary coefficient, $-i\omega \bm v=\nu\nabla^2\bm v$; here $\nu=\eta/\rho$ is the kinematic viscosity, see also \cite{Yang}. Another decomposition of the general solution uses the poloidal-toroidal decomposition \cite{pad}. The former seems more transparent and will be used below.

A main contribution of this paper is derivation of an explicit form of the solution of unsteady Stokes equations, that allows the detailed study of its properties. That solution provides the flow in terms of prescribed velocity distribution on the surface of a sphere. This is equivalent to the general solution of unsteady Stokes equations in the exterior of a given sphere, because any solution can be reduced to a sum of a solution for unbounded fluid and a solution driven by a non-trivial boundary condition on the sphere surface. We use the standard fundamental set of solutions of the vector Helmholtz equation with real-valued coefficients, $-k^2\bm v=\nu\nabla^2\bm v$, that is provided by the VSH (see, e.g. \cite{abs,St}). The radial dependence of the set's functions is given by spherical Bessel functions. The extension to imaginary coefficient leads to Bessel functions of imaginary argument or modified Bessel functions, which can be reduced to polynomials. The resulting solution of the Helmholtz equation decays exponentially fast away from the origin and is given by the elementary functions. Here we only consider the solution of the \emph{exterior} problem, or the flow external to a sphere, while the corresponding interior problem was solved in \cite{inter}.

The proposed representation provides some important insights. For example, it is well known that at distances beyond the viscous penetration depth $\delta=(2\nu/\omega)^{1/2}$ from an oscillating body the flow is potential \cite{LL}. This property readily follows from the fact that the vorticity $\bm \zeta\equiv \nabla\times \bm v$ obeys the vector Helmholtz equation $-i\omega \bm \zeta=\nu\nabla^2\bm \zeta$, and thus decays exponentially fast away from the object's surface where it is generated (alternatively, this property can be demonstrated by using integral representation of the flow and properties of the fundamental solution \cite{fl18}). In our approach, the emergence of the potential flow is an immediate consequence of the exponential decay of solutions of the Helmholtz equation and a simple, yet fundamental connection between the flow and the pressure, $\bm v\approx -i\omega^{-1} \nabla (p/\rho)$, at distances larger than $\delta$. This result can also be obtained by rewriting the flow equation as $i\omega \bm v-\rho^{-1} \nabla p=\nu\nabla\times \bm \zeta$ and considering the exponential decay of the right-hand-side (RHS) far from the body. It seems that both representations derived here, $\bm v\approx -i\omega^{-1} \nabla (p/\rho)$  and the expression for $p$ via the boundary conditions, are missing in the literature. We provide the solution in the time domain via non-trivial memory kernels, more complex than scalar kernels derived previously in the axially symmetric case \cite{Ishimoto}. We demonstrate that the long-term memory is determined by the lowest order terms in the solution series. Thus, the solution obeys universal long-time asymptotic behavior that is determined by the collective velocity of the boundary %\red{The same as in the Abstract - average radial component is not average motion anywhere}.
To illustrate this notion, we provide below the solution for decay of a transversal wave in the presence of a fixed sphere.

The limit of an ideal (or inviscid) flow is a fundamental topic in fluid mechanics \cite{LL,bat}. In this limit, $\delta$ tends to zero and the above considerations imply that $\bm v\approx -i\omega^{-1} \nabla (p/\rho)$ holds everywhere, in accord with the ideal flow approximation. Within this approximation, the potential $ -i\omega^{-1}p/\rho$ is obtained as the solution of the Laplace equation, whose normal derivative at the surface coincides with the corresponding component of the velocity at the boundary (see, e.g., \cite{simha,fl18}). We are not aware of a rigorous proof of this representation of the potential. We show here that the ideal flow approximation is the leading term of the expansion of the general solution in the viscosity coefficient.  The expansion parameter is $\sqrt{\nu}$, rather than $\nu$, indicating that adding viscosity yields a singular perturbation. The dimensionless expansion parameter is $1/\sqrt{Ro}$ where $Ro\!=\!a^2\omega/\nu$ is the Roshko number defined with the radius $a$ of the sphere at which the boundary conditions are prescribed. This asymptotic series can alternatively be considered as the \emph{high-frequency expansion}.

The Roshko number, $Ro$, provides an estimate for the ratio of the Eulerian time-derivative and viscous terms in the unsteady Stokes equations and can be written as product of the Strouhal number, $Sl=t_s/T$, and Reynolds number, $Re\!=\!v_0 a/\nu$. Here $t_s=a/v_0$ is the Stokes time with $v_0$ being a characteristic velocity and $T$ is the characteristic time $1/\omega$. If $Ro\!\ll\!1$, then the quasi-static (low-frequency) approximation provided by the Stokes equations applies. We demonstrate below that both the Lamb's and the adjoint method's solutions of the Stokes equation can be obtained in the limit
$Ro\! \to\! 0$. However, we demonstrate that care needs to be exercised when applying the approximation. Corrections to the low-frequency quasi-static approximation are given by a series expansion in $\sqrt{Ro}$. Therefore, similarly to the high-frequency expansion, low-frequency expansion also proves to be singular. Moreover, the expansion is as well non-uniform in space.

We notice that the reduced form of the unsteady Stokes equations in the frequency domain, $\bm v\approx -i\omega^{-1} \nabla (p/\rho)$, that holds at large distances from the body is remarkably similar to the Darcy's law governing the flow through porous medium \cite{r}. In fact, it can be called the generalized Darcy's law because it constitutes its analytic continuation, as we demonstrate in Sec.~\ref{unBr} via the general solution of Brinkman equations \cite{r,dru}.

We believe that our work is a significant step towards general understanding of the unsteady Stokes equations and several important closely related topics in viscous hydrodynamics. An application to a model for microswimming is considered in Sec.~\ref{mco}, while other potential applications are discussed in the Conclusions section.

\section{General solution of unsteady Stokes equations} \label{st}

In this work, we derive the general solution of unsteady Stokes equations
\begin{eqnarray}&&\!\!\!\!\!\!\!\!\!\!\!\!\!
\partial_t \bm v(t, \bm x)\!=\!-\rho^{-1}\nabla p\!+\!\nu \nabla^2\bm v,\ \ \nabla\cdot\bm v\!=\!0, \label{unsad}
\end{eqnarray}
in spherical coordinates. We shall study the solution in the exterior of a sphere with radius $a$. The solution for the interior problem was provided in \cite{inter}.  The exterior and interior problems are quite different due to the presence of the far-field region in the former case.

The equations (\ref{unsad}) are dissipative, so we can assume with no loss of generality that the flow vanishes at $|t|=\infty$. We also assume that the fluid velocity is prescribed at the surface of the sphere at $r\!=\!a$, where $r$ is the radial coordinate, and that the fluid is quiescent at infinity, $r \!\rightarrow \!\infty$.

We shall demonstrate below that the most general formulation in Eqs.~(\ref{unsad}) can be reduced to the boundary value problem, so that the solution studied here provides the universal description of the unsteady Stokes flow exterior to a spherical boundary.

\subsection{Reduction of a general unsteady flow to the boundary value problem} \label{flok}

In this subsection, we shall demonstrate how various unsteady Stokes flows can be reduced to the solution to the boundary value problem. We start from considering solutions of Eqs.~(\ref{unsad}) generated by prescribed initial data. We assume that at $t\!=\!0$ a given exterior flow at $r>a$ is prescribed, while we are interested in its evolution at $t\!>\!0$. We may assume, with no loss of generality, that on the surface of the sphere the velocity vanishes due to the no-slip (i.e., homogeneous Dirichlet) boundary condition. If there is a flow prescribed on the sphere, then the solution can be constructed as a superposition of the flow generated by the initial conditions with homogeneous Dirichlet boundary conditions on the sphere's surface, and the flow generated by the nonhomogeneous velocity distribution at the sphere surface.

Thus, we have the problem of the flow decay of the fluid confined between the stationary boundary at $r\!=\!a$ and infinity. This problem corresponds to introducing a source proportional to the Dirac $\delta-$function (that describes instantaneous injection of momentum into the fluid) into Eqs.~(\ref{unsad})
\begin{eqnarray}
&& \partial_t \bm v\!=\!-\rho^{-1}\nabla p\!+\!\nu \nabla^2\bm v+\delta(t)\bm v_0(\bm x),\ \nabla\cdot\bm v=0, \nonumber \\
&&\bm v(r=a)\!=\!0,
\end{eqnarray}
where $\bm v(t<0)=0$ and $\bm v_0(\bm x)$ are the initial conditions. We can construct the solution, as $\bm v(t, \bm x)=\bm v_{\infty}(t, \bm x)+\bm v'(t, \bm x)$. Here $\bm v_{\infty}(t, \bm x)$ is the solution in the unbounded fluid that is well-known from statistical physics near equilibrium, see, e.g., \cite{reichl}. The remaining component $\bm v'(t, \bm x)$ solves Eqs.~(\ref{unsad}) with the boundary conditions $\bm v'(t, r=a)=-\bm v_{\infty}(t, r=a)$. Thus the initial problem is reduced to the boundary value problem.

The above reduction can be extended to the case, where the flow is driven by an arbitrary distributed volume force. In this case
the $\delta(t)\bm v_0(\bm x)$-term in the RHS of the unsteady Stokes equations is replaced by an arbitrary time- and space-dependent force field. The solution is obtained as a sum of the corresponding flow in unbounded fluid and the flow generated by the boundary conditions.

Finally the unsteady Stokes flow can be disturbed by a stationary particle. In this case, one can construct the solution as a superposition of the exterior flow and the disturbance flow. The disturbance flow satisfies a boundary value problem, see, e.g., \cite{Yang}.

We shall consider below the decay of the transversal wave in presence of a fixed sphere. In the absence of the sphere the transverse wave is given by $\bm v_0={\hat y}\cos(kx)\exp(-\nu k^2 t)$ and $p=0$, which solves the unsteady Stokes equations \cite{reichl}. In the presence of the sphere, the solution satisfies the unsteady Stokes equations with the boundary conditions $\bm v(t, r\to\infty)=\bm v_0$ and $\bm v(r=a)=0$.  Looking for the solution as superposition $\bm v=\bm v_0+\bm v'$, where $\bm v'$ vanishes at infinity and satisfies the unsteady Stokes equations with the boundary condition:
\begin{eqnarray}&&\!\!\!\!\!\!\!\!\!\!\!\!\!
\bm v'(r=a)=-{\hat y}\cos(kx)\exp(-\nu k^2 t)H(t), \label{bcv}
\end{eqnarray}
where $H(t)$ is the step function. We will provide the solution to this example in subsection \ref{transversal}.

To conclude, the solution of the boundary value problem with an arbitrary velocity distribution at $r=a$ can be used to construct the general solution of unsteady exterior Stokes Eqs.~(\ref{unsad}) for different settings. Thus, in what follows we shall focus on the boundary value problem.

\subsection{Fourier space formulation and dimensionless variables}

 %Here $\bm v$ is the fluid velocity, \red{$p$} is the pressure, \red{$\nu$ and $\rho$ are the kinematic viscosity and density of the fluid, respectively}.
We use the Fourier-transformed variables, e.g., the velocity
\begin{eqnarray}&&\!\!\!\!\!\!\!\!\!\!\!\!\!
\bm v(\omega, \bm x)\equiv\int \bm v(t, \bm x) \re^{i\omega t} dt,\label{sc}
\end{eqnarray}
where the same symbol $\bm v$ is used for variables in time and frequency domains throughout the paper with no ambiguity. The Fourier-transformed flow equations then read
\begin{eqnarray}&&\!\!\!\!\!\!\!\!\!\!\!\!\!
-i\omega \bm v\!=\!-\rho^{-1} \nabla p\!+\!\nu \nabla^2\bm v,\ \ \nabla\cdot\bm v\!=\!0. \label{freqs}
\end{eqnarray}
Here we used the fact that the flow vanishes at $|t|=\infty$, which allows integration by parts while Fourier-transforming Eqs.~(\ref{unsad}).

Below, the study of the above equation is performed assuming $\omega>0$. The solution for $\omega<0$ can then be obtained by a simple transformation, cf. \cite{LL}, described in Sec. \ref{time}. Once the solution in frequency domain is determined, we can readily obtain the solution in time domain applying the inverse Fourier transform.  %\red{in the form:
\begin{eqnarray}&&\!\!\!\!\!\!\!\!\!\!\!\!\!
\bm v(t, \bm x)=\int_{-\infty}^{\infty} \bm v(\omega, \bm x)  \re^{-i \omega t} \frac{d\omega}{2\pi}. \label{inverse}
\end{eqnarray}
%where $Re$ stands for real part and we used the property $\bm v(-\omega)=\bm v^*(\omega)$}.

For given frequency, $\omega$ the solution can be characterized by the \emph{viscous penetration depth}, $\delta\!=\!\sqrt{2\nu/\omega}$. It provides the typical attenuation length of vorticity generated at the sphere's surface into the fluid bulk, see e.g. \cite{LL}. The relative importance of the frequency term and the viscosity term in Eqs.~(\ref{sc}) is estimated by the dimensionless Roshko number, $Ro\equiv a^2\omega/\nu$, see the Introduction. For $Ro\gtrsim 1$, the unsteady term cannot be neglected.

The limit of Stokes equations holds at $Ro\to 0$. In the small frequency limit, one could attempt to construct the solution of Eqs.~(\ref{freqs}) as a perturbation series in $\omega$. This would lead to divergent integral in the first order correction (although in the presence of infinite boundaries the series does converge \cite{fl18}). As mentioned in the Introduction, the true expansion parameter at small frequency, similarly to the fundamental solution \cite{kim}, is $\sqrt{Ro}\ll 1$, rather than $Ro\ll 1$.

\textit{Dimensionless variables} -- We define dimensionless variables scaling the velocity with $v_0$, pressure with $\eta v_0/a$ and distance with $a$, where $v_0$ is some characteristic velocity. The dimensionless equations velocity $\bm u$ and pressure $p$ (we use the same letters with no ambiguity) read:
\begin{eqnarray}&&\!\!\!\!\!\!\!
\lambda^{2}\bm u\!=\!-\!\nabla p
\!+\! \nabla^2\bm u,\ \ \nabla\!\cdot\!\bm u\!=\!0; \ \ \lambda^2\!=\!-i Ro\!=\!- \frac{i a^2\omega}{\nu}, \label{is}
\end{eqnarray}
where $\lambda=(1-i)\sqrt{Ro/2}$. We seek for the general solution of Eqs.~(\ref{is}) in spherical coordinates. The solution involves arbitrary constants that have to be determined using the prescribed value of $\bm u$ at the sphere surface at $r\!=\!1$.

\textit{General form of the pressure solution} -- By taking divergence of the first of Eqs.~(\ref{is}) we can readily see that the pressure is a harmonic function. Therefore, it can be represented as,
\begin{eqnarray}&&\!\!\!\!\!\!\!\!\!\!\!
p(\omega, \bm x)\!=\!\sum_{lm}\!\frac{c_{lm}(\omega)Y_{lm}(\theta, \phi)}{r^{l+1}}, \ \ \mathrm{where}\ \ \sum_{lm} \!\equiv \!\sum_{l=1}^{\infty}\!\!\sum_{m=-l}^{l}\!,\!\label{pr}
\end{eqnarray}
where we assumed that it vanishes at infinity; the spherical harmonics $Y_{lm}(\theta, \phi)$ are defined by
\begin{eqnarray}&&
Y_{lm}=\sqrt{\frac{(2l+1)}{4\pi}\frac{(l-m)!}{(l+m)!}}P_l^m(\cos\theta)\exp\left(im\phi\right),
\label{fi}
\end{eqnarray}
where $P_l^m$ are the associated Legendre polynomials. We use the multiplicative factor of \cite{sph} in the definition of $Y_{lm}(\theta, \phi)$, so the following orthogonality relations hold:
\begin{eqnarray}&&
\int Y_{lm} Y^*_{l'm'}d\Omega=\int_0^{\pi}\sin\theta d\theta\int_0^{2\pi}d\phi Y_{lm} Y^*_{l'm'}=\delta_{l l'}\delta_{m m'}.\nonumber
\end{eqnarray}
The zero-order term corresponding to $l=0$ is omitted in Eq.~(\ref{pr}), given that there is no net mass flux at infinity (i.e. there is no mass source or sink). The $\lambda$-dependent coefficients $c_{lm}$ should be determined from given boundary conditions for the velocity, $\bm u(r=1)$.

\subsection{Solution decomposition and its far-field behavior}

A particular solution of Eqs.~(\ref{is}) for $\bm u$, where $p$ is considered as a source, is $-\nabla p/\lambda^{2}$. This solution was originally introduced by Lamb, see article $353$ in \cite{Lamb}, however its use was rather limited, cf. \cite{pala,Yang}. One can readily see that this solution diverges at $\omega\to 0$ and does not reduce to the analogous Lamb's particular solution of the Stokes equations (see e.g., \cite{kim}).  Yet, the use of the particular solution in such form in comparison to others is advantageous, since it provides a straightforward decomposition of $\bm u$ into two contributions that possess qualitatively different spatial behavior. The general solution of Eqs.~(\ref{is}) can therefore be constructed as superposition of $-\nabla p/\lambda^{2}$ and a solenoidal field $\bm u_s$ satisfying the Helmholtz equation \cite{Lamb}
\begin{eqnarray}&&\!\!\!\!\!\!\!\!\!\!\!\!\!\!\!\!
\bm u=\bm u_s-\frac{\nabla p}{\lambda^2}, \ \
\lambda^{2}\bm u_s\!=\! \nabla^2\bm u_s,\ \ \nabla\!\cdot\!\bm u_s\!=\!0. \label{helm}
\end{eqnarray}
Solutions of the Helmholtz equation are known to decay exponentially away from the source with exponent given by the real part of $\lambda$, (e.g., $|\bm u_s|\sim \exp(-\lambda r)$ at large distances $r$). Thus, considering the boundary conditions on the sphere as the source, contribution to the far-field solution is negligible at distances $r-1$ from the sphere's surface that are much larger than the dimensionless penetration depth $\delta/a$ (notice that taking the real part of $\lambda$ gives $\rRe\ \lambda=\sqrt{Ro/2}=a/\delta$). The far-field flow is therefore given by $\bm u\approx -\nabla p/\lambda^{2}$ which is potential and decays as a power law, see Eq.~(\ref{pr}). It is quite remarkable that this conclusion can be drawn by inspection, using the above solution decomposition solely.

\subsection{Vector spherical harmonics}

To complete the solution of the unsteady Stokes equations, one has to determine the general solution of the vector Helmholtz equation, see Eqs.~(\ref{pr}) and (\ref{helm}). This is often achieved by using the vector spherical harmonics (VSH) which generalize the standard (scalar) spherical harmonics \cite{abs,St,mors}. We use the definitions of the VSH given in \cite{sph},
\begin{eqnarray}&&\!\!\!\!\!\!
\bm Y_{lm}=\bm{\hat r}Y_{lm},\ \ \bm \Psi_{lm}=r\nabla Y_{lm}=\bm{\hat \theta}
\partial_{\theta}Y_{lm}+\frac{\bm{\hat \phi}\partial_{\phi}Y_{lm}}{\sin\theta},\nonumber\\&&\!\!\!\!\!\!
\bm \Phi_{lm}\!=\!\bm r\!\times \!\nabla Y_{lm}\!=\!-\nabla\!\times\! (\bm r Y_{lm})\!=\!\bm{\hat \phi}\partial_{\theta}Y_{lm}\!-\!\frac{\bm{\hat \theta}\partial_{\phi}Y_{lm}}{\sin\theta}, \label{vsh}
\end{eqnarray}
where $\nabla$ is the three-dimensional gradient operator and $\bm{\hat r}$, $\bm{\hat \theta}$, $\bm{\hat \phi}$ are the unit vectors of the spherical coordinate system. For instance, for $l\!=\!1$, $m\!=\!0$ we have:
\begin{eqnarray}&&\!\!\!\!\!\!\!\!\!\!\!\!
Y_{10}=\sqrt{\frac{3}{4\pi}}\cos\theta,\ \ \bm \Psi_{10}=-{\hat \theta}\sqrt{\frac{3}{4\pi}}\sin\theta. \label{definition}
\end{eqnarray}
We find that
\begin{eqnarray}&&\!\!\!\!\!\!\!\!\!\!\!\!\!\!\!\!\!
-\nabla \left(\frac{Y_{lm}}{r^{l+1}}\right)\!=\!\frac{(l+1)\bm Y_{lm}\!-\!\bm \Psi_{lm}}{r^{l+2}}. \label{slo}
\end{eqnarray}
Thus, the particular solution $-\nabla p/\lambda^{2}$ takes the form:
\begin{eqnarray}&&\!\!\!\!\!\!\!\!\!\!\!\!\!\!\!\!\!
-\frac{\nabla p}{\lambda^2}
\!=\!\sum_{lm} c_{lm}\frac{(l+1)\bm Y_{lm}-\bm \Psi_{lm}}{\lambda^2 r^{l+2}}. \label{ps}
\end{eqnarray}
Any vector field can be represented as a series in VSH \cite{sph} and we can write:
\begin{eqnarray}&&\!\!\!\!\!\!\!\!\!\!
\bm u_s\!=\!\sum_{lm}\! \left(c^r_{lm}(r)\bm Y_{lm}\!+\!c^{(1)}_{lm}(r)\bm \Psi_{lm}\!+\!c^{(2)}_{lm}(r)\bm \Phi_{lm}\right),\label{expansion1}
\end{eqnarray}
where $c^r_{lm}(r)$, $c^{(i)}_{lm}(r)$ are some functions of the radial variable $r$ only.  The $l=0$-term is omitted since $\bm \Psi_{00}$ and $\bm \Phi_{00}$ vanish identically and $c^r_{00}=0$ (see below and Appendix \ref{solutionH}).

\subsection{General solution in VSH} \label{ge}

We determine the solution of the Helmholtz equation for $\bm u_s$ by using its expansion in terms of VSH in Eq.~(\ref{expansion1}), and solving the resulting system of linear ordinary differential equations for $c^r_{lm}(r)$, $c^{(i)}_{lm}(r)$. These $r$-dependent coefficients can be written via elementary functions, yielding the general solution (the details are provided in Appendix \ref{solutionH}):
\begin{eqnarray}&&
\bm u_s\!=\!\frac{\exp(-\lambda r)}{r} \sqrt{\frac{\pi}{2\lambda}}\sum_{lm}\left[\sum_{k=0}^l \frac{(l+k)!}{k!(l-k)!(2\lambda r)^k}
\right.\nonumber\\&&\left.
\times \left(\frac{{\tilde c}^{r}_{lm}\bm Y_{lm}}{r}\!+\!{\tilde c}_{lm}\bm \Phi_{lm}\right)
\!-\!\frac{{\tilde c}^{r}_{lm} \bm \Psi_{lm}}{(l\!+\!1)r}\left(\sum_{k=0}^l \frac{(l\!+\!k)!}{k!(l\!-\!k)!(2\lambda r)^k}
\right.\right.\nonumber\\&&\left.\left.
+ \frac{1}{2l}\sum_{k=0}^{l-1} \frac{(l-1+k)!}{k!(l-1-k)!(2\lambda r)^{k-1}}\right)\right].\label{Helmel}
\end{eqnarray}
The solution involves two sets of constant coefficients ${\tilde c}^{r}_{lm}$ and ${\tilde c}_{lm}$ (one out-of-three set for vector solutions is not shown due to solenoidality). Defining the polynomials:
\begin{eqnarray}&&\!\!\!\!\!
P_l(x)\!\equiv \!
\sum_{k=0}^{l}\! \frac{x^k(2l\!-\!k)!}{k!(l\!-\!k)!2^{l-k}}=x^l\sum_{k'=0}^{l}\! \frac{(l\!+\!k')!}{k'!(l\!-\!k')!(2x)^{k'}},%\ \
%{\cal P}_l(x)\!\equiv \!\sum_{k=0}^l \frac{(l\!+\!k)!x^k}{k!(l\!-\!k)!2^k},
\label{ourp}
\end{eqnarray}
where $k'=l-k$, it can be readily seen that the solution is given by $\exp(-\lambda r)$ times a series which looks quite similar to the solution of the Laplace equation, apart from the fact that each $lm$-term involves a polynomial $P_l(\lambda r)/r^l$ in negative integer powers of $r$ and not just a single power. The polynomial $P_l(x)$ originates from the modified Bessel functions $K_{l+1/2}(x)$ encountered in the solution of the Helmholtz equation with imaginary coefficient,
\begin{eqnarray}&&\!\!\!\!\!\!\!\!
K_{l+1/2}(x)\!=\!\!\sum_{k=0}^l\! \frac{\sqrt{\pi}(l\!+\!k)!\exp(-x)}{k!(l\!-\!k)!(2x)^{k+1/2}}\!=\!\sqrt{\frac{\pi}{2}}e^{-x}\frac{P_l(x)}{x^{l+1/2}}. \label{mod}
\end{eqnarray}
Thus, elementary solutions of the scalar Helmholtz equation which vanish at infinity are
\begin{eqnarray}&&\!\!\!\!\!\!\!\!\!\!\!\!
\chi^{\lambda}_{lm}\equiv -\frac{ K_{l+1/2}(\lambda r)Y_{lm}(\theta, \phi)}{\sqrt{r}};\ \ \nabla^2\chi^{\lambda}_{lm}=\lambda^2\chi^{\lambda}_{lm}.
\label{ek}
\end{eqnarray}
Therefore, we can rewrite Eq.~(\ref{Helmel}) in terms of the modified Bessel functions as
\begin{eqnarray}&&\
\bm u_s\!=\!\sum_{lm} \left(\frac{{\tilde c}^{r}_{lm} K_{l+1/2}(\lambda r)\bm Y_{lm}}{r^{3/2}}+\frac{{\tilde c}_{lm} K_{l+1/2}(\lambda r)\bm \Phi_{lm}}{\sqrt{r}}
\right.\nonumber\\&&\left.
-\frac{{\tilde c}^{r}_{lm} \bm \Psi_{lm}}{l(l+1)r}\left(\frac{l K_{l+1/2}(\lambda r)}{\sqrt{r}}
\!+\!\lambda r^{1/2}K_{l-1/2}(\lambda r)\right)\right). \label{helmh}
\end{eqnarray}
The Bessel functions are also showing in the general solution of the unsteady Stokes equations in the axially symmetric \cite{rao} and general \cite{pad} cases. The use of $P_l(x)$ emphasizes that the solution
is given in terms of the elementary functions and does not involve fractional powers of $r$. (We found the relation between modified Bessel functions of half-integer order and Bell polynomials \cite{na} less useful.) The full solution, given by the superposition in Eq.~(\ref{helm}) reads
\begin{eqnarray}&&\!\!\!\!\!\!\!\!\!\!\!\!
\bm u(\omega, \bm x)\!=\!\sum_{lm}\left[\left(\frac{{\tilde c}^{r}_{lm} K_{l+1/2}(\lambda r)}{r^{3/2}}+\frac{(l+1)c_{lm}}{\lambda^2 r^{l+2}}\right)\bm Y_{lm}
\right.\nonumber\\&&\!\!\!\!\!\!\!\!\!\!\!\!\left.
+\frac{{\tilde c}_{lm} K_{l+1/2}(\lambda r)}{\sqrt{r}}\bm \Phi_{lm}-\!\bm \Psi_{lm}\left\{\frac{c_{lm}}{\lambda^2 r^{l+2}}
\right.\right.\nonumber\\&&\!\!\!\!\!\!\!\!\!\!\!\!\left.\left.
+\frac{{\tilde c}^{r}_{lm} }{l(l+1)r}\left(\frac{l K_{l+1/2}(\lambda r)}{\sqrt{r}}
\!+\!\lambda r^{1/2}K_{l-1/2}(\lambda r)\right)\right\}\right]. \label{ho}
\end{eqnarray}
We shall now proceed to calculation of the coefficients.

\subsection{Coefficients of expansion} \label{sdj}

The constant coefficients $c_{lm}$, ${\tilde c}^{r}_{lm}$ and ${\tilde c}_{lm}$ in Eq.~\ref{ho} have to be determined from the boundary conditions. We consider the case when these are given by the prescribed velocity at the surface of the unit sphere at $r\!=\!1$. We find by projecting the general solution Eq.~(\ref{ho}) onto the VSH and using VSH orthogonality relations \cite{sph} that these coefficients satisfy:
\begin{eqnarray}&&
\int \bm u\cdot \bm Y_{lm}^* d\Omega\!=\!\int u_r Y_{lm}^* d\Omega\!=\! {\tilde c}^{r}_{lm} K_{l+1/2}(\lambda)+\frac{(l+1)c_{lm}}{\lambda^2},\nonumber\\&&
\int \bm u\cdot \bm \Psi_{lm}^* d\Omega\!=\!-{\tilde c}^{r}_{lm} \left(l K_{l+1/2}(\lambda)
\!+\!\lambda K_{l-1/2}(\lambda)\right)
\nonumber\\&&
-\frac{c_{lm}l(l+1)}{\lambda^2}, \nonumber\\&&
\int \bm u\cdot \bm \Phi_{lm}^* d\Omega\!=\!l(l+1){\tilde c}_{lm} K_{l+1/2}(\lambda), \label{cof}
\end{eqnarray}
where we used $\bm u\cdot \bm Y_{lm}^* =u_r Y_{lm}^*$ and $d\Omega$ stands for integration over the solid angle. Multiplying the first of the equations by $l$ and combining it with the second equation gives
\begin{eqnarray}&&
{\tilde c}^{r}_{lm} \!=\!-\frac{l\int u_r Y_{lm}^* d\Omega+\int \bm u\cdot \bm \Psi_{lm}^* d\Omega}{\lambda K_{l-1/2}(\lambda)}. \label{tld}
\end{eqnarray}
Next we find:
\begin{eqnarray}&&
c_{lm}\!=\! \frac{ \lambda\left(l K_{l+1/2}(\lambda)
\!+\!\lambda K_{l-1/2}(\lambda)\right)}{(l\!+\!1) K_{l-1/2}(\lambda)}\int \!u_r Y_{lm}^* d\Omega
\nonumber\\&&
+\frac{\lambda K_{l+1/2}(\lambda)}{(l\!+\!1) K_{l-1/2}(\lambda)}\int\! \bm u\!\cdot\! \bm \Psi_{lm}^* d\Omega.\label{cpr}
\end{eqnarray}
The simple relationship between the coefficients $c_{lm}$ and ${\tilde c}^{r}_{lm}$ can be readily found:
\begin{eqnarray}&&
c_{lm}\!=\! \frac{ \lambda^2}{l\!+\!1}\int\! u_r Y_{lm}^* d\Omega-\frac{\lambda^2K_{l+1/2}(\lambda){\tilde c}^{r}_{lm}}{l\!+\!1}.\label{coper}
\end{eqnarray}
Eqs.~(\ref{cof})-(\ref{cpr}) provide the coefficients of the expansion via projections of the surface velocity onto the VSH. These projections involve vectors, and their direct derivation is cumbersome. The calculations are simplified by using the following identities
\begin{eqnarray}&&\!\!\!\!\!\!\!
\int  Y_{lm}^* \nabla_s\cdot\bm u d\Omega=
2\int Y_{lm}^* u_r  d\Omega - \int \bm u\cdot \bm \Psi_{lm}^* d\Omega
,\nonumber\\&&\!\!\!\!\!\!\!
\int_{r=1}  Y_{lm}^* (\nabla\!\times\! \bm u)_r d\Omega=
-\int \bm u\cdot \bm \Phi_{lm}^* d\Omega,\label{cd}
\end{eqnarray}
derived in Appendix \ref{transformation}. The operator $\nabla_s\cdot\bm u$ taken at the sphere surface at $r=1$ is known as surface divergence and can be written as \cite{kim}:
\begin{eqnarray}&&\!\!\!\!\!\!\!
\nabla_s\!\cdot\!\bm u\!=\!\nabla\cdot \bm u\!-\!\frac{\partial u_r}{\partial r}\!=\!2u_r\!+\!\frac{\partial_{\theta}(\sin\theta u_{\theta})\!+\!\partial_{\phi}u_{\phi}}{\sin\theta}. \label{sp}
\end{eqnarray}
The first of the above representations of the surface divergence involves $\partial_r u_r(r=1)$ which is not readily available from the boundary conditions. This value can be obtained by arbitrary continuation of $u_r$ to $r>1$, as the result is independent of that continuation, as seen from the second representation in (\ref{sp}). In practical problems, such continuation is often straightforward and therefore the first representation in Eq.~(\ref{sp}) is quite useful (see examples below). We notice that the radial component of the curl of the flow $(\nabla\!\times\! \bm u)_r$ is determined uniquely by the given surface velocity $\bm u$ at $r\!=\!1$. The above formulas allow to find the coefficients of the solution by projecting the scalar functions onto the usual spherical harmonics. We have using Eq.~(\ref{mod}) for the coefficients $c_{lm}$ in Eq.~(\ref{pr}) for the pressure
\begin{eqnarray}&&
c_{lm}\!=\! \left(\! \frac{(l\!+\!2)P_l(\lambda)}{(l\!+\!1)P_{l-1}(\lambda)} \!+\!\frac{\lambda^2}{l\!+\!1}\right)\!\int\!Y_{lm}^* u_r d\Omega
\nonumber\\&&
-\frac{P_l(\lambda)}{(l\!+\!1)P_{l-1}(\lambda)}\int \! Y_{lm}^* \nabla_s\!\cdot\!\bm u d\Omega.\label{cope}
%H_l(\lambda)\equiv \frac{\lambda K_{l+1/2}(\lambda)}{(l\!+\!1) K_{l-1/2}(\lambda)}=\frac{P_l(\lambda)}{(l\!+\!1)P_{l-1}(\lambda)}
\end{eqnarray}
We notice that the ratio $P_l(\lambda)/P_{l-1}(\lambda)$ fully determines the non-trivial frequency dependence of the prefactor. Eqs.~(\ref{pr}), (\ref{cope}) constitute remarkable concise result for the pressure, whose form in the time domain is provided later. Similarly we find
\begin{eqnarray}&&\!\!\!\!\!\!\!\!\!\!\!\!
{\tilde c}_{lm}\!=\!-\frac{1}{l(l+1)K_{l+1/2}(\lambda)}\int_{r=1} Y_{lm}^* (\nabla\!\times\! \bm u)_r   d\Omega, \label{coex}
\end{eqnarray}
and
\begin{eqnarray}&&
{\tilde c}^{r}_{lm} \!=\!\frac{\int  Y_{lm}^* \nabla_s\cdot\bm u d\Omega-(l+2)\int u_r Y_{lm}^* d\Omega}{\lambda K_{l-1/2}(\lambda)}
. \label{tild}
\end{eqnarray}
The above equations for the coefficients show the analogy between the unsteady and steady Stokes equations, as the expansion of Lamb's solution is determined by the same projections of the surface flow involving $\hat{\bm r}\cdot\,$, $\nabla_s \cdot\,$ and $\hat{\bm r}\cdot (\nabla\times\,)$ \cite{kim}.

\subsection{Coefficients of the net motion: $l=1$} \label{sdq}

The coefficients with $l\!=\!1$ can be expressed via moments of velocity distribution over the sphere boundary. Thus ${\tilde c}^{r}_{10}$ and ${\tilde c}_{10}$ can be written in terms of the average boundary linear $\bm U$ and angular $\bm \Omega$ velocities, defined by
\begin{eqnarray}&&
\bm U \equiv \int \bm u \frac{d\Omega}{4\pi},\ \ \bm \Omega \equiv \int\bm r\times\bm u \frac{d\Omega}{4\pi}.
\end{eqnarray}
The derivation is somewhat tedious and omitting the technical details %\magenta{[** so why keeping it in the main text? **]} \red{I did not keep it, on the contrary, I only provided the result of the tedious calculation. Please check if this time it is understood}.
the averaged (over the solid angle) velocity components read:
\begin{eqnarray}
\int \!\! u_z d\Omega\!&=&\!\!\int \!\!\left(u_r\cos\theta\!-\!u_{\theta}\sin\theta\right) d\Omega\!=\nonumber \\
&& \!\!\!\!\!\!\!\!\!\!\!\!\!\sqrt{\frac{4\pi}{3}}\!\!\int \!\! \bm u\cdot \left(\bm Y_{10}^*\!+\!\bm \Psi_{10}^*\right)d\Omega, \nonumber \\
\int \!\! u_x d\Omega\!&=&\!\!\int \!\!d\Omega\left(u_r\sin\theta\cos\phi\!+\!u_{\theta}\cos\theta\cos\phi-u_{\phi}\sin\phi\right)\!= \nonumber \\
&& \!\!\!\!\!\!\!\!\!\!\!\!\!-\sqrt{\frac{2\pi}{3}}\!\int\!\! \bm u\!\cdot\!\left(\bm Y_{11}^*\!-\!\bm Y_{1, -1}^*\!+\!\bm \Psi_{11}^*\!-\!\bm \Psi_{1, -1}^*\right) d\Omega; \nonumber \\
\int \!\! u_y d\Omega\!&=&\!\int \!\!\left(u_r\sin\theta\sin\phi\!+\!u_{\theta}\cos\theta\sin\phi\!+\!u_{\phi}\cos\phi\right) d\Omega \nonumber \\
&&\!\!\!\!\!\!\!\!\!\!\!\!\!=\!-i\sqrt{\frac{2\pi}{3}}\int\!\! (\bm u\cdot\left(\bm Y_{11}^*\!+\!\bm Y_{1, -1}^*\!+\!\bm \Psi_{11}^*\!+\!\bm \Psi_{1, -1}^*\right) d\Omega,
\end{eqnarray}
where we used Eqs.~(\ref{fi}), (\ref{vsh}) and
\begin{eqnarray}&&
Y_{11}=-\sqrt{\frac{3}{8\pi}}\sin\theta\exp\left(i\phi\right)=-Y_{1,-1}^*.
\end{eqnarray}
Similarly, we have by using $r=1$ and the formula for Cartesian unit vectors $(\bm{\hat x}, \bm{\hat y}, \bm{\hat z})$ via $(\bm{\hat r}, \bm{\hat \theta}, \bm{\hat \phi})$ that
\begin{eqnarray}
\int \!\!\left(\bm r\!\times\!\bm u\right)_z d\Omega\!&=&\!\int \!\!\bm u\!\cdot\!\left(\bm{\hat z}\!\times\!\bm r\right) d\Omega
\!=\!\int\!\! \sin\theta u_{\phi} d\Omega \nonumber \\
&& \!\!\!\!\!\!=-\sqrt{\frac{4\pi}{3}} \int \!\!\bm u\!\cdot \!\bm \Phi_{10}^*d\Omega, \nonumber \\
\int \!\!\left(\bm u\!\times\!\bm r\right)_x \!d\Omega\!&=&\!\!\!\int\!\! d\Omega\left(u_{\phi}\cos\theta\cos\phi\!+\!u_{\theta}\sin\phi\right)
\nonumber\\
&& \!\!\!\!\!\!\!=-\sqrt{\frac{2\pi}{3}}\int \bm u\cdot\left(\bm \Phi_{11}^*-\bm \Phi_{1, -1}^*\right) d\Omega, \nonumber \\
\int \left(\bm r\times\bm u\right)_y d\Omega &=& \int \bm u\cdot\left(\bm{\hat y}\times\bm r\right) d\Omega \nonumber \\
&& \!\!\!\!\!\!=\int \left(u_{\theta}\cos\phi-u_{\phi}\cos\theta\sin\phi\right) d\Omega
\nonumber\\
&& \!\!\!\!\!\!=i\sqrt{\frac{2\pi}{3}}\int \bm u\cdot\left(\bm \Phi_{11}^*+\bm \Phi_{1, -1}^*\right) d\Omega.
\end{eqnarray}
We conclude by using the equations above, the last of Eqs.~(\ref{cof}) and Eq.~(\ref{tld}) that
\begin{eqnarray}
&& {\tilde c}^{r}_{10}=- 2 U_z e^{\lambda}\sqrt{\frac{6}{\lambda}},\;\;  {\tilde c}^{r}_{1, \pm 1}=2\left(\pm U_x\!-\!iU_y\right)e^{\lambda}\sqrt{\frac{3}{\lambda}},\label{co} \\
&& {\tilde c}_{10}=-\frac{\lambda\Omega_z\sqrt{6\lambda}}{1+\lambda}e^{\lambda},\;\;  {\tilde c}_{1, \pm 1}=\frac{\lambda\left(\pm   \Omega_x\!-\!i\Omega_y\right)\sqrt{3\lambda}}{1+\lambda}e^{\lambda}, \nonumber
\end{eqnarray}
where we used $\lambda K_{1/2}(\lambda)=\sqrt{\lambda \pi/2}\,e^{-\lambda}$  and $K_{3/2}(\lambda)\!=\!\sqrt{\pi/(2\lambda^3)}(1+\lambda)e^{-\lambda}$, see Eqs.~(\ref{ourp})--(\ref{mod}) .

Similarly $c_{1m}$ can be written in terms of $\bm U$ and the average radial flow
\begin{eqnarray}&&
\tilde {\bm U} \equiv \int \tilde {\bm u}\frac{d\Omega}{4\pi},\ \ \tilde {\bm u}\equiv u_r \bm{\hat r}.
\end{eqnarray}
Using Eqs.~(\ref{coper}) and (\ref{co}) we have:
\begin{eqnarray}&&
c_{10}=\sqrt{3\pi}\left[\lambda^2 {\tilde U}_z+(1+\lambda)U_z\right]\,,
\\&&
c_{1, \pm 1}\!=\!-\sqrt{\frac{3\pi}{2}} \left[ \lambda^2(\pm {\tilde U}_x- i{\tilde U}_y)\!+\!(1+\lambda)(\pm U_x- iU_y)\right].\nonumber
\end{eqnarray}
The coefficients with $l\!=\!1$ correspond to average or ``rigid body" motion of the boundary. This notion will be further discussed in Sec. \ref{oscillating}.

\section{Lamb-type form of the solution} \label{Lambs}

In this section, we show that our solution can be written in the form similar to Lamb's solution \cite{Lamb} of the steady Stokes equations $\nabla p=\nabla^2\bm u$. Lamb's solution could be obtained in a way similar to our approach above.  Again, pressure is a harmonic function that
for solutions vanishing at infinity can be written as $p=\sum_{l=1}^{\infty}p_{-l-1}$. Here $p_{-l-1}$ are solid spherical harmonics given by linear combination of $r^{-l-1}Y_{lm}$ with $m$ ranging from $-l$ to $l$, see Eq.~(\ref{pr}) and \cite{kim}. The particular, solenoidal solution of the Stokes equations $\nabla^2 \bm u_n=\nabla p_n$ is $\alpha_n\left( r^2\nabla p_n-2n \bm r p_n/(n+3)\right)$ where $2\alpha_n (2n+3)(n+1)=n+3$, see Exercise $4.1$ in \cite{kim}. The general solution for the steady Stokes flow vanishing at $r\to \infty$ is then written as
\begin{eqnarray}&&\!\!\!\!\!\!\!
\bm u^{Lamb}=\sum_{l=1}^{\infty}\left(-\frac{(l-2)r^2\nabla p_{-l-1}}{2l(2l-1)}+\frac{(l+1)\bm r p_{-l-1}}{l(2l-1)}\right.
\nonumber\\&&\!\!\!\!\!\!\!\left.+\nabla\Phi_{-l-1}+\nabla\times (\bm r\chi_{-l-1})\frac{}{}\right).\label{Lamb}
\end{eqnarray}
The first two terms involving $p_{-l-1}$ give a particular solution of the Stokes equation $\nabla^2 \bm u=\nabla p$, while the last two terms provide the general solenoidal solution of the vector Laplace equation $\nabla^2 \bm u=0$ with $\Phi_{-l-1}$, $\chi_{-l-1}$ being solid spherical harmonics, see \cite{kim}.

\textit{Generalization of Lamb's solution to non-zero frequency}---A main result of the present work is that the general solution of the unsteady Stokes equations \ref{ho} can be written in the form similar to Lamb's general solution of steady Stokes equations:
\begin{eqnarray}&&\!\!\!\!\!\!\!\!\!\!\!\!\!\!\!\!
\bm u=-\frac{\nabla p}{\lambda^2}+e^{\lambda (1-r)} \bm u^H+\nabla\times (\bm r e^{\lambda (1-r)}X), \label{su}
\end{eqnarray}
in terms of some vector, $\bm u^H$, and scalar, $X$, fields. Similarly to Lamb's solution, the first term is a particular solution and the last two terms compose the solution of the homogeneous equation, where one of the terms is toroidal i. e. is given by a curl of radial vector field. Notice that this decomposition differs from the decomposition used by Lamb for unsteady Stokes equations.

\textit{Toroidal component} -- By using the definition $\bm \Phi_{lm}\!=\!-\nabla\!\times\! (\bm r Y_{lm})$ we can write
\begin{eqnarray}&&\!\!\!\!\!\!\!\!
\sum_{lm} \!\frac{{\tilde c}_{lm} K_{l+1/2}(\lambda r)}{\sqrt{r}}\bm \Phi_{lm}\!=\!\sum_{lm}\! {\tilde c}_{lm}\nabla\!\times \!(\bm r\chi^{\lambda}_{lm}),\
\end{eqnarray}
where we used $\chi^{\lambda}_{lm}$ defined in Eq.~(\ref{ek}). The last term provides simple generalization of the last term in Lamb's solution given by Eq.~(\ref{Lamb}) to unsteady Stokes equations. Notice that elementary solutions $\chi^{\lambda}_{lm}$ of the scalar Helmholtz equation generate elementary solenoidal solutions $\nabla\!\times \!(\bm r\chi^{\lambda}_{lm})$ of the vector Helmholtz equation, and therefore, the emergence of the toroidal component in the solution seems natural. Using the identity in Eq.~(\ref{su}), one can rewrite the solution for $\bm u_s$ in Eq.~ (\ref{helmh}) as
\begin{eqnarray}&&
\bm u_s\!=\!\nabla\times (\bm r e^{\lambda (1-r)}X )+e^{\lambda (1-r)} \bm u^H, \label{totr}
\end{eqnarray}
where $X$ in the toroidal component and $\bm u^H$ are given, respectively, by
\begin{eqnarray}
X & \equiv & e^{\lambda  (r-1)}  \sum_{lm}{\tilde c}_{lm} \chi^{\lambda}_{lm}\,, \nonumber\\
\bm u^H & \equiv & e^{\lambda  (r-1)}\sum_{lm}\frac{{\tilde c}^{r}_{lm}}{r^{3/2}}\left(\frac{}{}K_{l+1/2}(\lambda r)\bm Y_{lm}\!\right. \nonumber\\
&& \,\,\left.-\frac{\left(l K_{l+1/2}(\lambda r)+\lambda rK_{l-1/2}(\lambda r)\right) \bm \Psi_{lm}}{l(l+1)}\right).
\end{eqnarray}
The superscript $H$ refers to the fact that $e^{\lambda (1-r)}\bm u^H$ represents a solenoidal solution of the vector Helmholtz equation, $ \nabla^2\left(e^{\lambda (1-r)} \bm u^H\right)=\lambda^2 e^{\lambda (1-r)}\bm u^H$. It follows from Eqs.~(\ref{mod}) and (\ref{coex}) that $X$ can be written as
\begin{eqnarray}&&\!\!\!\!\!\!\!\!\!\!\!\!
X\!=\!
\sum_{lm}\frac{c^t_{lm}P_{l}\left(\lambda r\right)}{P_l(\lambda)}  \frac{Y_{lm}(\theta, \phi)}{r^{l+1}}, \label{X}
\end{eqnarray}
where we singled out the frequency dependence in the coefficients so that $c^t_{lm}$ are frequency-independent:
\begin{eqnarray}&&\!\!\!\!\!\!\!\!\!\!\!\!
c^t_{lm}\!=\!\frac{1}{l(l+1)}\int_{r=1} Y_{lm}^* (\nabla\!\times\! \bm u)_r   d\Omega. \label{cdo}
\end{eqnarray}
From the definition of the polynomials in Eq.~(\ref{ourp}), it follows that at $\lambda\!\to\!0$ we have $P_l(\lambda r)/P_l(\lambda)=1$. Therefore, $X$ is regular in the limit of vanishing $\lambda$, reproducing Lamb's toroidal term in Eq.~(\ref{Lamb}) as
\begin{eqnarray}&&\!\!\!\!\!\!\!\!\!\!\!\!
\sum_{l=1}^{\infty}\chi_{-l-1}=X(\lambda=0)\!=\!%\frac{\exp\left(-\lambda (r\!-\!1)\right)}{r}
\sum_{lm} \frac{c^t_{lm}Y_{lm}(\theta, \phi)}{r^{l+1}}.\label{cdox}
\end{eqnarray}
Thus, the toroidal contribution in Eq.~(\ref{totr}) can be considered as analytic continuation of the corresponding term in Lamb's solution to finite frequency. This term is associated with oscillatory surface velocity, see the solution for oscillatory rotation of a rigid sphere in Section \ref{oscillating}.

\textit{Recovering Lamb's solution from the finite-frequency solution} -- Opposite to the toroidal term, the first two terms in the RHS of Eq.~(\ref{su}) cannot be obtained from analytic continuation to finite frequency of the corresponding terms in Lamb's solution in Eq.~(\ref{Lamb}). For instance, the particular solution given by the first term in the RHS of Eq.~(\ref{su}), diverges at $\lambda\to 0$. It can be shown, however, that the {\it sum} of the first two terms in Eq.~(\ref{su}) does converge to the sum of the first two terms in Eq.~(\ref{Lamb}). Using the above definition of $\bm u^H$ and Eqs.~(\ref{pr}) and (\ref{coper}) for $p$, gives:
\begin{eqnarray}&&
e^{\lambda (1-r)} \bm u^H\!-\!\frac{\nabla p}{\lambda^2}=-\nabla\sum_{lm}\frac{Y_{lm}(\theta, \phi)\int\! u_r Y_{lm}^* d\Omega}{(l+1)r^{l+1}}
\nonumber\\&&
+\sum_{lm}\! C^{r}_{lm}\left(\frac{\delta(r) Y_{lm}(\theta, \phi)\bm{\hat r}}{\lambda K_{l-1/2}(\lambda)r^{l+2} }-\frac{\nabla Y_{lm}(\theta, \phi)}{K_{l-1/2}(\lambda)}
\right.\nonumber\\&&\left.
\times \left(\frac{\delta(r)}{\lambda (l+1) r^{l+1}}-\frac{r^{1/2}K_{l-1/2}(\lambda r)}{l (l+1)}\right)\right).\label{fs}
\end{eqnarray}
Here we introduced $\delta(r)\equiv r^{l+1/2}K_{l+1/2}(\lambda r)\!-\!K_{l+1/2}(\lambda)$ that vanishes at the surface $r\!=\!1$, and rescaled coefficients $C^r_{lm}\!\equiv\! {\tilde c}^{r}_{lm}\lambda K_{l-1/2}(\lambda)$ that by Eq.~(\ref{tild}) obey
\begin{eqnarray}&&\!\!\!\!\!\!\!\!\!
C^{r}_{lm}\!= \!\int\!\!  Y_{lm}^* \nabla_s\!\cdot\!\bm u d\Omega\!-\!(l\!+\!2)\int\!\! u_r Y_{lm}^* d\Omega. \label{Crl}
\end{eqnarray}
We consider the zero frequency limit of Eq.~(\ref{fs}). The leading terms of the Taylor expansion are
\begin{eqnarray}&&
\frac{\lambda K_{l+1/2}(\lambda r) }{ K_{l-1/2}(\lambda)}=\frac{\exp\left(\lambda (1-r)\right)P_l\left(\lambda r\right)}{r^{l+1/2}P_{l-1}\left(\lambda\right)}
= \frac{2l-1}{r^{l+1/2}}
\nonumber\\&& +\frac{\lambda^2}{2}\left(\frac{2l-1}{(2l-3) r^{l+1/2}}- \frac{1}{r^{l-3/2}}\right)+\mathit{o}(\lambda).
%\nonumber\\&&
%\frac{K_{l+1/2}(\lambda r)K_{l-3/2}(\lambda) }{ K^2_{l-1/2}(\lambda)}=\frac{2l-1}{(2l-3) r^{l+1/2}},
\end{eqnarray}
The derivatives of the function in the LHS of the above equation at $\lambda=0$ are obtained by using the identity
\begin{eqnarray}&&\!\!
\left(\!\frac{\lambda K_{l+1/2}(\lambda r) }{ K_{l-1/2}(\lambda)}\!\right)_{\lambda}\!\!\!=\!\frac{\lambda K_{l+1/2}(\lambda r)K_{l-3/2}(\lambda) }{ K^2_{l-1/2}(\lambda)}\!-\!\frac{\lambda r K_{l-1/2}(\lambda r)}{K_{l-1/2}(\lambda)},\nonumber
\end{eqnarray}
where the subscript $\lambda$ stands for the partial derivative. This equation can be derived by using the relations for derivatives of the modified Bessel functions.
We find that
\begin{eqnarray}&&\!\!\!\!\!\!\!\!
e^{\lambda (1-r)} \bm u^H-\frac{\nabla p}{\lambda^2}=\sum_{lm}\left(-\nabla\frac{Y_{lm}(\theta, \phi)\int\! u_r Y_{lm}^* d\Omega}{(l+1)r^{l+1}}
\right.\nonumber\\&&\!\!\!\!\!\!\!\!\left.
+\frac{C^{r}_{lm}}{2r^l}\left(\frac{\bm{\hat r}Y_{lm}}{r^{2}}\! -\!\bm{\hat r}Y_{lm}
\!-\!\frac{\nabla Y_{lm}}{(l+1) r}\!+\!\frac{(l-2)r\nabla Y_{lm}}{l(l+1)}\right)\right).
\end{eqnarray}
The pressure representation $p=\sum_{l=1}^{\infty}p_{-l-1}$ (valid at any frequency), where $p_{-l-1}=r^{-l-1}\sum_{m=-l}^{m=l}c_{lm}Y_{lm}$, see Eq.~(\ref{pr}), gives at $\lambda=0$
\begin{eqnarray}&&\!\!\!\!\!\!\!
\sum_{l=1}^{\infty}\left(-\frac{(l-2)r^2\nabla p_{-l-1}}{2l(2l-1)}+\frac{(l+1)\bm r p_{-l-1}}{l(2l-1)}\right)
\nonumber\\&&\!\!\!\!\!\!\!=\sum_{lm}C^{r}_{lm}\left(\frac{(l-2)\nabla Y_{lm}}{2l(l+1)r^{l-1}}-\frac{\bm{\hat r} Y_{lm} }{2r^{l}}\right)\,, \label{psd}
\end{eqnarray}
where we used that Eq.~(\ref{coper}) gives $c_{lm}\!=\!-(2l-1)C^{r}_{lm}/(l\!+\!1)$ at $\lambda=0$. Thus, we conclude that
\begin{eqnarray}&&
\lim_{\lambda\to 0}\left(e^{\lambda (1-r)} \bm u^H-\frac{\nabla p}{\lambda^2}\right)=\\&&
\sum_{l=1}^{\infty}\left(-\frac{(l-2)r^2\nabla p_{-l-1}}{2l(2l-1)}+\frac{(l+1)\bm r p_{-l-1}}{l(2l-1)}+\nabla\Phi_{-l-1}\right)\,\nonumber ,
\end{eqnarray}
which is the same as Lamb's solution in Eq.~(\ref{Lamb}); here we introduced:
\begin{eqnarray}&&\!\!\!\!\!
\Phi_{-l-1}\!\equiv\! -\sum_{m=-l}^{l}\!\frac{Y_{lm}(\theta, \phi)}{ (l+1)r^{l+1}}
\left(\int\! u_r Y_{lm}^* d\Omega
\!+\!\frac{C^{r}_{lm}}{2}\right)\,.\label{ph}
\end{eqnarray}
Thus we showed that in the limit of $\lambda\to 0$ our solution reproduces the Lamb's solution given by Eq.~(\ref{Lamb}), with components provided by Eqs.~(\ref{cdo})-(\ref{cdox}), (\ref{Crl}), (\ref{psd}) and (\ref{ph}).

\textit{Difficulties in analytic continuation of Lamb's solution} -- Although we demonstrated that Eq.~(\ref{su}) generalizes Lamb's solution for a finite frequency, this continuation is not straightforward. The reason for that is the fact that finite frequency is a singular perturbation of the solution of steady Stokes equations. The solution considered as a function of the frequency $\omega$, is non-analytic at $\omega\!=\!0$, see Sec. \ref{freq}. Other manifestation of the perturbation's singularity is that at distances larger than the viscous penetration depth, $\delta$, the finite-frequency and zero-frequency solutions are very different regardless of how small $\lambda$ is. Notice also that, as mentioned above and demonstrated here explicitly, the first term in Eq.~(\ref{su}) possesses a power-law dependence, while the remaining terms decay exponentially with $r$, as both terms, $X$ and $\bm u^H$ decay algebraically with $r$, see Eq.~(\ref{X}) and the equivalent form of $\bm u^H$ given by
\begin{eqnarray}&&
\bm u^H\!=\!\sum_{l m} C^{r}_{lm}  \left(\frac{\bm {\hat r}  P_{l}\left(\lambda r\right)Y_{lm}(\theta, \phi)}{\lambda^2 P_{l-1}(\lambda) r^{l+2}}
\right.\label{uH}\\&&\left.
-\!\left(\frac{lP_{l}\left(\lambda r\right)}{\lambda^2 P_{l-1}(\lambda)}
\!+\!\frac{r^2P_{l-1}\left(\lambda r\right)}{P_{l-1}(\lambda)}\right)\frac{ \nabla Y_{lm}(\theta, \phi)}{l(l+1)r^{l+1}}\right),\nonumber
\end{eqnarray}
as can be seen from Eqs.~(\ref{ourp})-(\ref{mod}),  (\ref{ho}) and (\ref{tild}). The difference in spatial behavior holds because the pressure at finite frequency yet solves the Laplace equation, while the other two terms in Eq.~(\ref{su}) solve the vector Helmholtz equation.
Finally, we notice that gradients of elementary solutions in Eq.~(\ref{ek}), $\nabla \chi^{\lambda}_{lm}$, solve the vector Helmholtz equation, but exhibit  finite divergence at non-zero frequency. This is the reason why $\bm u^H$-term cannot be constructed as superposition of $\nabla \chi^{\lambda}_{lm}$ terms.

\section{Solutions in the time domain} \label{time}

In this Section, we consider solutions in the time domain. Throughout the previous calculations we implicitly assumed $\omega>0$, cf. \cite{LL}. Here, to perform the inverse Fourier transform of the solution one has to consider negative frequencies also. Thus, in this Section $\lambda$ is defined with the absolute value of frequency, $\lambda\equiv (1-i) a\sqrt{|\omega|/(2\nu)}$. With this definition, the solution for $\omega<0$ or, equivalently, $\omega=-|\omega|$ can be obtained by replacing $\lambda$ with its complex conjugate $\lambda^*$ in all \sout{the} formulas of the previous Sections. Indeed, we have then $(\lambda^*)^2=-\lambda^2$ so that Eqs.~(\ref{is}) holds for $-|\omega|$ replacing $\omega$. We also have that $\rRe\ \lambda^*=\rRe\ \lambda>0$ so that the conditions of decay at infinity holds (one could also use complex conjugation of the solution itself, which however is less useful below).

The solution, presented in the previous Sections, is determined uniquely by flow values on any spherical boundary. The uniqueness holds also if the values are prescribed on a surface of any other shape. This fact is similar to that for solutions of the Laplace equation where harmonic functions are fully determined by their values on the boundaries and it could also be demonstrated via Green's type identities. In this work we focus on (Dirichlet) boundary conditions on the sphere whose radius is set to one. It is useful to describe these conditions by using the projection coefficients in time and frequency domains, respectively, defined by
\begin{eqnarray}&&
I^{t}_{lm}(\bm W) =\int_{r=1} \bm u(t, \bm x)\cdot \bm W_{lm}^* d\Omega,\nonumber\\&&
I^{\omega}_{lm}(\bm W) =\int_{r=1} \bm u(\omega, \bm x)\cdot \bm W_{lm}^* d\Omega,
\label{def00}
\end{eqnarray}
where $\bm W$ is an arbitrary vector spherical harmonics, see the definition in Eq.~(\ref{sc}). We have from Eqs.~(\ref{cd})
\begin{eqnarray}&&
I_{lm}(\bm Y)\!=\!\int Y_{lm}^* u_r  d\Omega;\ \
I_{lm}(\bm \Phi)\!=\!
-\int_{r=1}  Y_{lm}^* (\nabla\!\times\! \bm u)_r d\Omega
,\nonumber\\&&
2I_{lm}(\bm Y) - I_{lm}(\bm \Psi)\!=\!\int  Y_{lm}^* \nabla_s\cdot\bm u d\Omega,\label{kl}
\end{eqnarray}
where we do not write the superscript in Eq.~(\ref{kl}) since the relations hold both in $t$ and $\omega$ domains.

The solution in the frequency domain provided in the previous sections furnishes the general form of periodic solutions of the unsteady Stokes equations, as given by the real part of $\re^{-i\omega t}\bm v(\bm x)$. This is the case of time-periodic $I^{t}_{lm}(\bm W)$. In the case where $I^{t}_{lm}(\bm W)$, defined in Eq.~(\ref{def00}), and thus also the flow itself, are general time-dependent functions, the solution in the time domain is obtained by the inverse Fourier transform, see Eq.~(\ref{inverse}). This leads to quite cumbersome calculations which provide the form of the solution in the time domain described below.

\subsection{The pressure field}

The inverse Fourier transform of the pressure field requires transforming the coefficients $c_{lm}(\omega)$, cf. Eq.~(\ref{pr}). These coefficients are given by Eq.~(\ref{cope}) and can be rewritten via $I_{lm}$ as (here $\omega>0$)
\begin{eqnarray}&&\!\!\!\!\!
c_{lm}(\omega)\!=\! \frac{P_l(\lambda)\left(l I_{lm}^{\omega}(\bm Y)\!+\!I_{lm}^{\omega}(\bm \Psi)\right)}{(l\!+\!1)P_{l-1}(\lambda)} +\!\frac{\lambda^2 I_{lm}^{\omega}(\bm Y)}{l\!+\!1}, \label{clm}
\end{eqnarray}
where we used the first of Eqs.~(\ref{cd}) and stressed the frequency dependence of $c_{lm}$. We introduce into the above expression the factor of one in the form of $(-i\omega)/(\epsilon-i\omega)$ where $\epsilon$ is infinitesimal convergence factor introduced in order to have well-defined inverse Fourier transform, see below and cf. \cite{LL}. Inverse Fourier transform of Eq.~(\ref{clm}) gives by using the definition of $\lambda^2$ and Eq.~(\ref{pr})
\begin{eqnarray}&&
p(t, \bm x)\!=\!\sum_{lm}\!\frac{c_{lm}(t)Y_{lm}(\theta, \phi)}{r^{l+1}};\ \ c_{lm}(t)\!=\!\frac{a^2}{\nu(l\!+\!1)}\frac{dI_{lm}^{t}(\bm Y)}{dt}
\nonumber\\&&
+\int_{-\infty}^{\infty}q_l\left(\frac{\nu(t\!-\!t')}{a^2}\right)\frac{d}{dt'} \left(l I_{lm}^{t'}(\bm Y)\!+\!I_{lm}^{t'}(\bm \Psi)\right)dt',\label{prtime}
\end{eqnarray}
where we defined the memory kernel ($\tau\equiv \nu t/a^2$)
\begin{eqnarray}&&\!\!\!\!\!\!\!\!
q_l(\tau)\equiv - \rIm \int_{0}^{\infty}\!\!  \frac{P_l\left((1-i)\sqrt{\omega/2}\right) e^{-i \omega \tau} d\omega}{\pi(\omega\!+\!i\epsilon)(l\!+\!1)P_{l-1}\left((1\!-\!i)\sqrt{\omega/2}\right)}.\label{ist}
\end{eqnarray}
We used that $c_{lm}(-|\omega|)$ are obtained by changing $\lambda$ with $\lambda^*$ in $c_{lm}(|\omega|)$. It is seen that the convergence factor is necessary since $P_l(0)/P_{l-1}(0)=2l-1$, see Eq.~(\ref{ourp}), and at $\epsilon=0$ the integral diverges. For $\tau<0$ we can close the contour in the upper half plane so that the integral equals that found by introducing $\omega=ix$, giving
\begin{eqnarray}&&\!\!\!
q_l(\tau)\!=\!-\rIm \int_{0}^{\infty}\!\frac{P_l(\sqrt{x}) e^{- x|\tau|}}{(x\!+\!\epsilon)(l+1)P_{l-1}(\sqrt{x})}\frac{dx}{\pi}\!=\!0,
\end{eqnarray}
where we used that $P_l(\sqrt{x})$ is real. The above equation demonstrates that the integral in Eq.~(\ref{prtime}) is over $t'<t$ only, obeying the causality demand.

For $\tau>0$, we cannot close the contour in the lower half plane at the imaginary axis due to the pole at $\omega=-i\epsilon$. Thus we rewrite Eq.~(\ref{ist}) as
\begin{eqnarray}&&
q_l(\tau)\equiv - \rIm \int_{0}^{\infty}  \left[\frac{P_l\left((1-i)\sqrt{\omega/2}\right)}{P_{l-1}\left((1-i)\sqrt{\omega/2}\right)}-\frac{P_l\left(0\right)}{P_{l-1}\left(0\right)}\right]
\nonumber\\&&
 \frac{e^{-i \omega \tau} d\omega}{\pi\omega(l+1)}
-\frac{2l-1}{l+1} \rIm \int_{0}^{\infty}  \frac{e^{-i \omega \tau}}{\omega+i\epsilon} \frac{d\omega}{\pi}, \label{tars}
\end{eqnarray}
where the last integral obeys
\begin{eqnarray}&&\!\!\!\!\!\!
\rIm \int_{0}^{\infty}\!\!  \frac{e^{-i \omega \tau}}{\omega\!+\!i\epsilon} \frac{d\omega}{\pi}\!=\!\int_{-\infty}^{\infty}  \frac{e^{-i \omega \tau}}{\omega\!+\!i\epsilon} \frac{d\omega}{2\pi i}\!=\!-1,
\end{eqnarray}
whereas for the first line in Eq.~(\ref{tars}) we can transform the contour in the lower half-plane and use $\omega=-ix$. We obtain
\begin{eqnarray}&&\!\!\!\!\!\!
q_l(\tau)\!= \!\!-\int_{0}^{\infty}\!\!\rIm \frac{P_l\left(-i\sqrt{x}\right) e^{-x \tau}}{(l\!+\!1)P_{l-1}\left(-i\sqrt{x}\right)} \frac{dx}{x\pi}
\!+\!\frac{2l\!-\!1}{l+1}, \label{chec}
\end{eqnarray}
where we used $\rIm\  P_l\left(i\sqrt{x}\right)/P_{l-1}\left(i\sqrt{x}\right)=0$ at $x=0$. We can also rewrite Eq.~(\ref{tars}) by using an infinitesimal parameter $\delta$ as
\begin{eqnarray}&&
q_l(\tau)\!=\!\!\frac{2l\!-\!1}{l+1}+\int_{0}^{\infty}\!\!\!\!\frac{d\omega}{2\pi i (l+1)\omega} \left[\frac{P_l\left(\sqrt{\delta\!+\!i\omega}\right)e^{i \omega \tau}}{P_{l-1}\left(\sqrt{\delta\!+\!i\omega}\right)}
\right.\nonumber\\&&\left.
-\frac{P_l\left(\sqrt{\delta\!-\!i\omega}\right)e^{-i \omega \tau}} {P_{l-1}\left(\sqrt{\delta\!-\!i\omega}\right)}\right]
\nonumber\\&& =\frac{2l\!-\!1}{l+1} +\int_{\delta-i\infty}^{\delta+i\infty}  \frac{P_l\left(\sqrt{s}\right)e^{s \tau}}{(l+1)sP_{l-1}\left(\sqrt{s}\right)} \frac{ds}{2\pi i}, \label{lap}
\end{eqnarray}
that has the form of inverse Laplace transform. Similar form was reported in the study of axially symmetric solutions of unsteady Stokes equations \cite{Ishimoto}, where the Laplace transform was applied instead of the Fourier transform used here. Although this representation is not used here, it shall be exploited later on.

The analysis of $q_l(\tau)$ below bears much similarity with \cite{Ishimoto}, providing somewhat more details.
We first consider $l=1$ which is different from the case $l>1$. We have, using the definition in Eq.~(\ref{mod}), that $P_1(\lambda)=(1+\lambda)P_0(\lambda)$
and substitution in Eq.~(\ref{chec}) gives
\begin{eqnarray}&&\!\!\!\!\!\!\!\!\!
q_1(\tau)\!=\!\!\int_{0}^{\infty}\!\!\! e^{-y^2\tau}\frac{dy}{\pi}+\frac{1}{2}\!=\!\frac{1}{2\sqrt{\pi\tau}}+\frac{1}{2},
\end{eqnarray}
It can be readily shown that Eq.~(\ref{lap}) produces the same result.
The $l=1$ component determines the leading order far-field behavior of the pressure
\begin{eqnarray}&&
p(t, \bm x)\!\approx \!\frac{c_{1m}(t)Y_{1m}(\theta, \phi)}{r^2};\ \ r\gg 1,\label{f}
\end{eqnarray}
where from Eq.~(\ref{prtime}) we have
\begin{eqnarray}&&
c_{1m}(t)\!=\!\frac{a^2 \dot{I}_{1m}^{t}(\bm Y)}{2\nu}+\frac{I_{1m}^{t}(\bm Y)\!+\!I_{1m}^{t}(\bm \Psi)}{2}
\label{c1m}\\&&
+\int_{-\infty}^{t}\frac{a\left(\dot{I}_{1m}^{t'}(\bm Y)\!+\!\dot{I}_{1m}^{t'}(\bm \Psi)\right)dt'}{2\sqrt{\pi \nu (t-t')}} . \nonumber
\end{eqnarray}
This compact formula allows to determine far-field pressure from the boundary conditions. We see from the above equation that the memory kernel is given by the usual Basset history term \cite{kim,LL}.

We next consider $l>1$. Using the the recurrence relation $\lambda K_{l+1/2}(\lambda)-2(l-1/2)K_{l-1/2}(\lambda)=\lambda K_{l-3/2}(\lambda)$, and the definition in Eq.~(\ref{mod}) we find that
\begin{eqnarray}&&
\frac{\lambda K_{l+1/2}(\lambda)-(2l-1)K_{l-1/2}(\lambda)}{K_{l-1/2}(\lambda)}=\frac{\lambda^2 P_{l-2}(\lambda)}{P_{l-1}(\lambda)}.\label{idsl}
\end{eqnarray}
The last ratio in Eq.~(\ref{idsl}) was studied in \cite{Ishimoto}.
%We introduce polynomials ${\tilde P}_l(\lambda)$ that derive from ${\cal P}_l$ defined in Eq.~(\ref{ourp}) via
%\begin{eqnarray}&&\!\!
%\lambda^l{\cal P}_l\left(\frac{1}{\lambda}\right)\!= \!\sum_{k=0}^l\! \frac{(l\!+\!k)!\lambda^{l-k}}{k!(l\!-\!k)!2^k}
%=\sum_{k=0}^{l}\! \frac{\lambda^k(2l\!-\!k)!}{k!(l\!-\!k)!2^{l-k}}\!\equiv \!{\tilde P}_l(\lambda),\nonumber\\&&\!\!
%K_{l+1/2}(\lambda)\!=\!\sqrt{\frac{\pi}{2\lambda}}e^{-\lambda}\frac{{\tilde P}_l(\lambda)}{\lambda^l}, \label{mod1}
%\end{eqnarray}
%where we used Eq.~(\ref{mod}.
We obtain from Eq.~(\ref{chec}) using Eqs.~(\ref{cope}) and (\ref{idsl})  that
\begin{eqnarray}&&\!\!\!\!\!\!
q_l(\tau)\!=\!\frac{2l\!-\!1}{l+1}%= \int_{0}^{\infty} \rIm\ \frac{P_{l-2}(\lambda)}{ P_{l-1}(\lambda)}|_{\lambda=-i\sqrt{x}}\re^{-x\tau }
\!+\!\int_{0}^{\infty}\!\! \rIm\ \frac{P_{l-2}(-i\sqrt{x})}{P_{l-1}(-i\sqrt{x})}\re^{-x\tau }\frac{dx}{(l\!+\!1)\pi}.
\end{eqnarray}
We find that $l$ roots of $P_l(\lambda)$ are negative, real and simple so that $P_l(\lambda)=\prod_{k=1}^l(\lambda+a_k^l)$ where we designated the roots by $-a_k^l$ with $a_k^l>0$ (these zeros are of course also zeros of $K_{l+1/2}(\lambda)$, see Eq.~(\ref{mod})). This implies that the partial fraction decomposition for the ratio of two polynomials is
\begin{eqnarray}&&\!\!
\frac{P_{l-1}(\lambda)}{P_{l}(\lambda)}\!=\!\sum_{k=1}^{l} \! \frac{c_k^{l}}{\lambda\!+\!a_k^{l}},\label{dec}\\&&\!\! c_k^{l}\!\equiv \!\frac{P_{l-1}(-a_k^{l})}{P'_{l}(-a_k^{l})}\!=\!\frac{\prod_{r=1}^{l-1}(a_r^{l-1}\!-\!a_k^{l})}{\prod_{m=1,\ m\neq k}^{l}(a_m^{l}\!-\!a_k^{l})},\nonumber
\end{eqnarray}
where $c_k^{l}$ are real constants, cf. \cite{Ishimoto}. It therefore follows that
\begin{eqnarray}&&\!\!\!\!
q_l(\tau)\!=\!\frac{2l\!-\!1}{l+1}\!+\!\sum_{k=1}^{l-1}\! c_k^{l-1}\ \rIm \int_{0}^{\infty} \!\!\frac{e^{-x\tau}}{a_k^{l-1}\!-\!i\sqrt{x}}\frac{dx}{(l\!+\!1)\pi}\nonumber\\&&
=\sum_{k=1}^{l-1} c_k^{l-1} \int_{0}^{\infty} \frac{y^2 e^{-y^2\tau}}{(a_k^{l-1})^2+y^2}\frac{2dy}{(l+1)\pi}
\end{eqnarray}
The last integral gives \cite{grad}:
\begin{eqnarray}
q_l(\tau)&=&\frac{2l\!-\!1}{l+1}\!+\! \frac{\sum_{k=1}^{l-1} c_k^{l-1}}{(l+1)\sqrt{\pi \tau}}-\nonumber \\
&& \sum_{k=1}^{l-1} \frac{c_k^{l-1} a_k^{l-1}}{l+1}
\re^{\tau (a_k^{l-1})^2}\erfc\left(\sqrt{\tau}a_k^{l-1}\right), \label{pressurekernel}
\end{eqnarray}
where $\erfc x\!\equiv\! \frac{2}{\sqrt{\pi}}\int_x^{\infty}\! \re^{-t^2} dt$ is the complementary error function, cf. the derivation of this kernel in \cite{rao,Ishimoto}.

We have found that the pressure expansion coefficients $c_{lm}(t)$ with $l>1$ in Eq.~(\ref{prtime}) are given by
\begin{eqnarray}&&
c_{lm}(t)\!=\!\frac{a^2\dot{I}_{lm}^{t}(\bm Y)}{\nu(l\!+\!1)}+\sum_{k=1}^{l-1} \frac{c_k^{l-1}}{l\!+\!1}\int_{-\infty}^{t}\left(l \dot{I}_{lm}^{t'}(\bm Y)\!+\!\dot{I}_{lm}^{t'}(\bm \Psi)\right)
\nonumber\\&&
\cdot \left[\frac{a}{\sqrt{\pi \nu (t-t')}}-a_k^{l-1}\exp\left(\frac{\nu (t-t') (a_k^{l-1})^2}{a^2}\right)\label{fpr}
\right.\\&&\left.
\erfc\left(\!\frac{\sqrt{\nu (t\!-\!t')}a_k^{l-1}}{a}\right)\!\right]dt'\!+\!\frac{(2l\!-\!1)\left(lI_{lm}^{t}(\bm Y)\!+\!I_{lm}^{t}(\bm \Psi)\right)}{l+1}. \nonumber
\end{eqnarray}
In the limit of steady flow, the terms with $\dot I$ drop. The last term then provides the solution for the steady Stokes flow, considered in more detail in the next section.

Notice that $q_l(\tau)$ for $l>1$ differs significantly from $q_1(\tau)$. The asymptotic behavior at small $\tau$,
\begin{eqnarray}&&
q_l(\tau)\sim  \frac{\sum_{k=1}^{l-1} c_k^{l-1}}{(l+1)\sqrt{\pi \tau}}-\sum_{k=1}^{l-1} \frac{c_k^{l-1} a_k^{l-1}}{l+1}\!+\!\frac{2l\!-\!1}{l+1},\ \ \tau\to 0,\nonumber
\end{eqnarray}
is similar to that for $l=1$. However, the memory decays much faster. By using the asymptotic expansion of $\erfc\left(\sqrt{\tau}a_k^{l-1}\right)$ for large $\tau$ we find
\begin{eqnarray}&&
q_l(\tau)\!-\!\frac{2l\!-\!1}{l+1}\sim \frac{1}{2\sqrt{\pi}(l+1)\tau^{3/2}}\sum_{k=1}^{l-1} \frac{c_k^{l-1}}{(a_k^{l-1})^2},\ \ \tau\to \infty.\nonumber
\end{eqnarray}
This implies that at large $t-t'$, the memory kernel in the integral in Eq.~(\ref{fpr}) follows integrable $(t-t')^{-3/2}$ decay law. This is in contrast with the non-integrable $(t-t')^{-1/2}$ decay for $l=1$. Behavior of $l=1$ and $l>1$ components is qualitatively different and more examples of this will be provided below. It appears that $l=1$ components roughly correspond to average (rigid-body) motion of the boundary, see Sec.~\ref{oscillating}. For these components different portions of the spherical surface move coherently generating flow with longer memory.

Eqs.~(\ref{prtime}), (\ref{f}), (\ref{fpr}), providing the general solution for the pressure in terms of the velocity at the boundary, are among the main results of this paper. The expressions are written in terms of the zeros $-a_k^l$ of the modified Bessel function $K_{l+1/2}(\lambda)$. We are not aware of a general formula for the zeros of the modified Bessel function of arbitrary half-integer order, see e.g. \cite{watson}. For any given $l$ the numerical calculation of $a_k^l$ is straightforward. For instance $P_1(\lambda)=1+\lambda$ gives $a_1^1=c_1^1=1$, producing
\begin{eqnarray}&&\!
c_{2m}(t)\!=\!\frac{a^2\dot{I}_{2m}^{t}(\bm Y)}{2\nu}+\sum_{k=1}^{l-1} \int_{-\infty}^{t}\frac{dt'}{2}\left(l \dot{I}_{2m}^{t'}(\bm Y)\!+\!\dot{I}_{2m}^{t'}(\bm \Psi)\right)
\nonumber\\&&\!
\left[\frac{a}{\sqrt{\pi \nu (t-t')}}\!-\!\exp{\left(\frac{\nu (t\!-\!t')}{a^2}\right)}\erfc\left(\frac{\sqrt{\nu (t\!-\!t')}}{a}\right)\right]  \nonumber
\\&&\!+2I_{2m}^{t}(\bm Y)\!+\!I_{2m}^{t}(\bm \Psi).
\end{eqnarray}
We believe that finding the explicit closed-form expressions of $a_k^l$ is possible, however it is beyond the scope of the present paper.

\subsection{The flow field}

We now consider the flow in the time domain. The solution can be written in the frequency domain via $I_{lm}^{\omega}$ by rewriting $\bm u$ in Eq.~(\ref{ho}) as
\begin{eqnarray}&&
\bm u(\omega, \bm x) \!= \!\sum_{lm}\bm U_{lm}(\omega, \bm x),\label{ugn}\\&&
\bm U_{lm}(\omega, \bm x)\equiv \bm u_{lm}(\omega, \bm x)-\frac{I^{\omega}_{lm}(\bm Y)}{l+1}\nabla \left(\frac{Y_{lm}}{r^{l+1}}\right), \nonumber
\end{eqnarray}
where we defined $\bm u_{lm}(\omega, \bm x)$ via
\begin{eqnarray}&&\!\!\!\!\!\!
\bm u_{lm}(\omega, \bm x)\!=\! A_{lm}^{Y}(\omega, r)\bm Y_{lm}(\theta, \phi)\!+\!A_{lm}^{\Psi}(\omega, r)\bm \Psi_{lm}(\theta, \phi)
\nonumber\\&&\!\!\!\!\!\!
+A_{lm}^{\Phi}(\omega, r)\bm \Phi_{lm}(\theta, \phi).\label{ugen}
\end{eqnarray}
Here we have introduced the following auxiliary functions of frequency and radial coordinate
\begin{eqnarray}&&
A_{lm}^{Y}(\omega, r)\! = \!\frac{\left[ I_{lm}^{\omega}(\bm \Psi) \!+\! l I_{lm}^{\omega}(\bm Y)\right]\left[A_{l}(\lambda, 1)\!-\!r^l A_{l}(\lambda, r)\right]}{r^{l+2}};
%+\frac{I_{lm}(\bm Y)}{r^{l+2}},
\nonumber\\&&
A_{lm}^{\Psi}(\omega, r)\!=\!
\frac{\left[ I_{lm}^{\omega}(\bm \Psi)\! +\! l I_{lm}^{\omega}(\bm Y)\right] B_{l-1}(\lambda, r)}{rl(l+1)} %\right.
%\label{APsi}\\&&
%\left.
-\frac{A_{lm}^{Y}(\omega, r)}{l+1};
\nonumber\\&&
A_{lm}^{\Phi}(\omega, r)\! =\!  \frac{I_{lm}^{\omega}(\bm \Phi) B_l(\lambda, r)}{rl(l+1)},
\label{APhis}
\end{eqnarray}
with $A_{l}(\lambda, r)$ and $B_{l}(\lambda, r)$ defined by
\begin{eqnarray}&&\!\!\!\!\!\!\!
A_l(\lambda, r)\!=\!\frac{\sqrt{r}K_{l+1/2}(\lambda r)}{\lambda K_{l-1/2}(\lambda)},\
B_l(\lambda, r)\!=\!\frac{\sqrt{r}K_{l+1/2}(\lambda r)}{K_{l+1/2}(\lambda)}. \label{AB}
\end{eqnarray}
The domain of definition is $l\geq 1$ for $A_l$ and $l\geq 0$ for $B_l$. We remark that $A_{l}(\lambda, 1)=P_l(\lambda)/(\lambda^2 P_{l-1}(\lambda))$. The above formulas hold at $\omega>0$. As explained in the beginning of the Section, for $\omega<0$ we can still use Eqs.~(\ref{ugen})-(\ref{AB}) only instead of $A_l(\lambda, r)$ and  $B_l(\lambda, r)$ we must use $A_l(\lambda^*, r)=A_l^*(\lambda, r)$ and $B_l(\lambda^*, r)=B_l^*(\lambda, r)$, respectively.

By taking inverse Fourier transform, we find that in the time domain the flow reads
\begin{eqnarray}&&\!\!\!\!\!\!
\bm u(t, \bm x) \!= \!\sum_{lm}\left(\bm u_{lm}(t, \bm x)-\frac{I^{t}_{lm}(\bm Y)}{l+1}\nabla \left(\frac{Y_{lm}}{r^{l+1}}\right)\right),\label{ug}
\end{eqnarray}
%where the first and the last terms in the brackets provide the non-local and local in time components of the solution, respectively.
We have the representation
\begin{eqnarray}&&\!\!\!\!\!\!
\bm u_{lm}(t, \bm x) \!=\! A_{lm}^{Y}(t, r)\bm Y_{lm}(\theta, \phi)\!+\!A_{lm}^{\Psi}(t, r)\bm \Psi_{lm}(\theta, \phi) \nonumber\\&&\!\!\!\!\!\!
 +A_{lm}^{\Phi}(t, r)\bm \Phi_{lm}(\theta, \phi),\label{ugent}
\end{eqnarray}
where the functions $A_{lm}^{Y}(t, r)$, $A_{lm}^{\Psi}(t, r)$ and $A_{lm}^{\Phi}(t, r)$ are defined by
\begin{eqnarray}&&
A_{lm}^{Y}(t, r)\! = \!\int_{-\infty}^{\infty}\! \frac{F_l(t\!-\!t', r) [ \dot{I}_{lm}^{t'}(\bm \Psi) \!+\! l \dot{I}_{lm}^{t'}(\bm Y)] dt'}{r^{l+2}},\label{APhit}
\\&&
A_{lm}^{\Psi}(t, r)\!=\! \int_{-\infty}^{\infty} \!\frac{G_{l-1}(t\!-\!t', r) [ \dot{I}_{lm}^{t'}(\bm \Psi) \!+\! l \dot{I}_{lm}^{t'}(\bm Y)] dt'}{r^{l}l(l+1)}
\nonumber\\&&
-\frac{A_{lm}^{Y}(t, r)}{l+1},\ \
A_{lm}^{\Phi}(t, r)\! =\!\int_{-\infty}^{\infty} \!\frac{G_l(t\!-\!t', r) \dot{I}_{lm}^{t'}(\bm \Phi)dt'}{r^{l+1}l(l+1)},\nonumber
\end{eqnarray}
with the memory kernels
\begin{eqnarray}&&
F_l(t, r)\equiv -\rIm \int_{0}^{\infty} [A_{l}(\lambda, 1)\!-\!r^l A_{l}(\lambda, r)] \re^{-i \omega t}\frac{d\omega}{(\omega\!+\!i\epsilon) \pi},\nonumber\\&&
%=\int_{0}^{\infty} (A_{l}(1, \sqrt{a^2(\epsilon-i\omega)/\nu})\!-\!r^l A_{l}(r, (1-i)\sqrt{a^2\omega/(2\nu)})) \re^{-i \omega t}\frac{d\omega}{2\pi}
%\nonumber\\&&
%+\int_{0}^{\infty} (A_{l}(1, \sqrt{a^2(\epsilon+i\omega)/\nu})\!-\!r^l A_{l}(r, (1+i)\sqrt{a^2\omega/(2\nu)})) \re^{i \omega t}\frac{d\omega}{2\pi}
%\nonumber\\&&
%=\int_{\epsilon-i\infty}^{\epsilon+i\infty}A_{l}(1, \sqrt{a^2 s/\nu})\!-\!r^l A_{l}(r, \sqrt{a^2 s/\nu})) \re^{s t}\frac{ds}{2\pi i}
%\nonumber\\&&
G_l(t, r)\equiv  -\rIm  \int_{0}^{\infty}r^l B_l(\lambda, r)\re^{-i \omega t}\frac{d\omega}{(\omega+i\epsilon) \pi}. \label{ut}
\end{eqnarray}
We have made the decomposition into derivatives of $I_{lm}$ and memory kernels as for the pressure above, for similar reasons.

The significance of $F_l(t, r)$ and $G_l(t, r)$ can be seen by studying the flow which is created by a sudden boundary flow acting at time $t_0$. The boundary flow is fully described via the coefficients $I^t_{lm}(\bm W)=\delta(t-t_0)I_{lm}(\bm W)$ where $\bm W$ is an arbitrary VSH. The use of $I^t_{lm}(\bm W)$ in Eqs.~(\ref{APhit}) demonstrates that this impulse boundary velocity creates a flow at time $t$ at point $r$, whose $lm$ component is proportional to $F_l(t-t_0, r)$ and $G_l(t-t_0, r)$. Thus, $F_l$ and $G_l$ can be called (linear) response functions. Moreover, since the induced flow must vanish at $t<t_0$ by causality, then consistency demands that $G_l(t)$ and $F_l(t)$ also vanish for $t<0$, similarly as for the pressure above. \\ \\

\subsection{Inverse Laplace transform form}

We shall now demonstrate that the kernels $F_l(t, r)$ and $G_l(t, r)$ can be written as inverse Laplace transforms which allows to use the tables for the integrals' evaluation, cf. \cite{Ishimoto}.

Before analyzing $F_l(t, r)$ and $G_l(t, r)$, we notice that these kernels are not independent. By using the definition in Eq.~(\ref{AB}) we find that
\begin{eqnarray}&&\!\!\!\!\!
A_{l}(1)\!-\!r^l A_{l}(r)=-\int_1^r \frac{d}{dr'}\frac{r'^{l+1/2}K_{l+1/2}(\lambda r')}{\lambda K_{l-1/2}(\lambda)}dr'
\nonumber\\&&\!\!\!\!\!
=\int_1^r \frac{r'^{l+1/2}K_{l-1/2}(\lambda r')}{K_{l-1/2}(\lambda)}dr'=\!\int_1^r r'^l B_{l-1}(\lambda, r')dr',
\end{eqnarray}
where we used the identity \cite{grad}
\begin{eqnarray}&&
-\frac{d}{dr}\left( r^{\nu}K_{\nu}(\lambda r)\right)=\lambda r^{\nu}K_{\nu-1}(\lambda r). \label{idr}
\end{eqnarray}
Therefore, it follows that
\begin{eqnarray}&&
F_l(t, r)=\int_1^r r' G_{l-1}(t, r')dr', \label{unf}
\end{eqnarray}
using which one kernel could be deduced from the other.

We introduce the dimensionless kernel $g_l$
\begin{eqnarray}&&\!\!\!\!\!\!\!\!\!\!
g_l(\tau, r)\!=\!  -\rIm  \!\!\int_{0}^{\infty}\!\!\!\!\!\re^{-i \omega \tau} \frac{r^{l+1/2} K_{l+1/2}((1\!-\!i)r \sqrt{\omega/2} )d\omega}{(\omega\!+\!i\epsilon) \pi K_{l+1/2}((1\!-\!i) \sqrt{\omega/2})},\label{sta}
\end{eqnarray}
so that $G_l(t, r)= g_l\left(\nu t/a^2, r\right)$. For $\tau<0$ we can close the contour in the upper half plane so that introducing $\omega=ix$ we have (cf. the study of the pressure)
\begin{eqnarray}&&\!\!\!\!\!\!\!
g_l\!=\!-\rIm \int_{0}^{\infty}\!\re^{- x|\tau|} \frac{r^{l+1/2}K_{l+1/2}(r \sqrt{x} )}{K_{l+1/2}(\sqrt{x})}\frac{dx}{(x+\epsilon) \pi}\!=\!0.
\end{eqnarray}
The above equation confirms the causality requirement $G_l(t<0)=0$, and thus also $F_l(t<0)=0$, see Eq.~(\ref{unf}).

For the study of positive values of $\tau$, we can rewrite $g_l(\tau, r)$ in Eq.~(\ref{sta}) as an inverse Laplace transform
\begin{eqnarray}&&
g_l(\tau, r)=-\int_{0}^{\infty}\re^{-i \omega \tau} \frac{r^{l+1/2}K_{l+1/2}(r \sqrt{\delta-i\omega})}{K_{l+1/2}(\sqrt{\delta-i\omega})(\omega\!+\!i\epsilon)}\frac{d\omega}{2\pi i }\nonumber\\&&
+\int_{0}^{\infty}\re^{i \omega \tau} \frac{r^{l+1/2}K_{l+1/2}(r \sqrt{\delta+i\omega})}{K_{l+1/2}(\sqrt{\delta+i\omega})(\omega\!+\!i\epsilon)}\frac{d\omega}{2\pi i}\nonumber\\&&
=\int_{\delta-i\infty}^{\delta+i\infty}\frac{r^{l+1/2}K_{l+1/2}(r \sqrt{s})}{K_{l+1/2}(\sqrt{s})}e^{s\tau}\frac{ds}{2\pi i s},\label{transf}
\end{eqnarray}
where in the last line $\epsilon$ can be set to zero. Similarly we find that for $t>0$
\begin{eqnarray}&&\!\!\!\!\!\!\!
F_l(t, r)=f_l\left(\frac{\nu t}{a^2}, 1\right)-f_l\left(\frac{\nu t}{a^2}, r\right), \label{of}
\end{eqnarray}
where
\begin{eqnarray}&&\!\!\!\!\!\!\!
f_l(\tau, r)\!=\! \int_{\delta-i\infty}^{\delta+i\infty}\frac{r^{l+1/2}K_{l+1/2}(r \sqrt{s})}{\sqrt{s} K_{l-1/2}(\sqrt{s})}e^{s\tau}\frac{ds}{2\pi i s}.\label{fl}
\end{eqnarray}
It is readily seen from Eqs.~(\ref{unf}) and (\ref{of}) that
\begin{eqnarray}&&\!\!\!\!\!\!\!
-\partial_r f_l(\tau, r)\!=\! rg_{l-1}(\tau, r), \label{recur}
\end{eqnarray}
that can be also proved directly from definitions in Eqs.~(\ref{transf}), (\ref{fl}) by using Eq.~(\ref{idr}).

There are two separate cases for $l=1$ and $l>1$.

\subsection{The longest memory: $l=1$}

As was for the pressure, the case of $l=1$ stands out,  We have
\begin{eqnarray}&&\!\!\!\!\!\!\!
g_0\!=\!\int_{\epsilon-i\infty}^{\epsilon+i\infty}\!e^{\sqrt{s}(1\!-\!r)\!+\!s\tau}\frac{ds}{2\pi i s}\!=\!\erfc\left(\frac{r\!-\!1}{2\sqrt{\tau}}\right),\label{g0}
\end{eqnarray}
where we used an integral from \cite{prud}. Thus, the kernel $G_0(t, r)=g_0\left(\nu t/a^2, r\right)$ might seem to have no memory decay, as according to the above equation it tends to $1$ at $t\to\infty$. To demonstrate the decay of memory, we rewrite $A_{1m}^{\Psi}$ in Eq.~(\ref{APhit}) as
\begin{eqnarray}&&
A_{1m}^{\Psi}(t, r)\!=\!-\frac{A_{1m}^{Y}(t, r)}{2}\!+\! \frac{I_{1m}^{t}(\bm \Psi) \!+\!I_{1m}^{t}(\bm Y)}{2r}
\label{asls}\\&&
-\int_{-\infty}^{t}dt'\frac{(\dot{I}_{1m}^{t'}(\bm \Psi) \!+\! \dot{I}_{1m}^{t'}(\bm Y))}{2r}\erf\left(\frac{a(r\!-\!1)}{2\sqrt{\nu(t\!-\!t')}}\right),\nonumber
\end{eqnarray}
where $\erf\left(x\right)\equiv 1-\erfc\left(x\right)$ is the regular error function and where we assumed that $I_{1m}^{t}$ vanishes at $t=-\infty$. Written in this form, it can be seen that the memory decays at long times according to
\begin{eqnarray}&&\!\!\!\!\!\!
\erf\left(\frac{a(r\!-\!1)}{2\sqrt{\nu(t\!-\!t')}}\right)\!\simeq\! \frac{a(r\!-\!1)}{\sqrt{\pi\nu(t\!-\!t')}}%;\ \ \frac{a(r\!-\!1)}{\sqrt{\pi\nu(t\!-\!t')}}\!
\ll\! 1,
\end{eqnarray}
which is the scaling of the standard Basset memory term \cite{kim}. Notice that at short times the above $\erf$ kernel equals to $1$ with exponential accuracy.

We next consider the kernel $F_1$. From Eq.~(\ref{fl}) by using the formulae for $K_{1/2}(x)$ and $K_{3/2}(x)$ we have \cite{prud}:
\begin{eqnarray}&&\!\!\!
f_1\!=\! \int_{\epsilon-i\infty}^{\epsilon+i\infty}\! \left(\frac{1}{\sqrt{s}}+\frac{1}{rs}\right) e^{\sqrt{s}(1\!-\!r)\!+\!s\tau} \frac{rds}{2\pi i  s}\, \label{f1a}\\&&\!\!\!
=\!\frac{2\tau\!-\!r^2\!+\!1}{2}\erfc\left(\frac{r\!-\!1}{2\sqrt{\tau}}\right)\!+\!(r\!+\!1)\sqrt{\frac{\tau}{\pi}}e^{-\frac{(r\!-\!1)^2}{4\tau}}. \nonumber
\end{eqnarray}
%where we used integral tables from \cite{prud}.
It is readily seen that $\partial_r f_1=-rg_0$ is in accord with Eq.~(\ref{recur}).
We can write using Eqs.~(\ref{f1a}) and (\ref{of}) that
\begin{eqnarray}&&\!\!\!\!\!\!\!
F_1(t, r)=\frac{r^2\!-\!1}{2}+{\tilde F}_1\left(\frac{\nu t}{a^2}, r\right), \label{f11}
\end{eqnarray}
where ${\tilde F}_1$ which is defined by
\begin{eqnarray}&&\!\!\!\!\!\!\!
{\tilde F}_1\left(\tau, r\right)\equiv (1\!-\!r)\sqrt{\frac{\tau}{\pi}}\!+\!\frac{2\tau\!-\!r^2\!+\!1}{2}\erf\left(\frac{r\!-\!1}{2\sqrt{\tau}}\right)\nonumber\\&&\!\!\!\!\!\!\!+(r\!+\!1)
\sqrt{\frac{\tau}{\pi}}\left(1-e^{-\frac{(r\!-\!1)^2}{4\tau}}\right), \label{asll}
\end{eqnarray}
vanishes at large times as
\begin{eqnarray}&&\!\!\!\!\!\!\!
{\tilde F}_1\left(\tau, r\right)\sim -\frac{(r\!-\!1)^2(2r\!+\!1)}{6\sqrt{\pi \tau}}+O\left(\frac{1}{\tau^{3/2}}\right).\label{f12}
\end{eqnarray}
Then for $A_{lm}^{Y}(t, r)$ in Eq.~(\ref{APhit}) that gives
\begin{eqnarray}&&
A_{1m}^{Y}(t, r)\! = \!\frac{(r^2\!-\!1)[I_{1m}^{t}(\bm \Psi) \!+\!  I_{1m}^{t}(\bm Y)]}{2r^3}
\label{alma}\\&&\!\!\!\!\!\!\!
+\frac{1}{r^3}\int_{-\infty}^t\! {\tilde F}_1\left(\frac{\nu(t\!-\!t')}{a^2}, r\right) [\dot{I}_{1m}^{t'}(\bm \Psi) \!+\!  \dot{I}_{1m}^{t'}(\bm Y)]dt',\nonumber
\end{eqnarray}
demonstrating that at long times the memory decays as the Basset memory term.

We consider the remaining component $A_{1m}^{\Phi}(t, r)$ that is determined by $g_1(\tau, r)$, see the last of Eqs.~(\ref{APhit}). We have from Eq.~(\ref{transf}) that
\begin{eqnarray}&&
g_1(\tau, r)=\int_{\delta-i\infty}^{\delta+i\infty}\frac{r \sqrt{s}+1}{\sqrt{s}+1}e^{\sqrt{s}(1-r)+s\tau}\frac{ds}{2\pi i s}\\&&
=\int_{\delta-i\infty}^{\delta+i\infty}e^{\sqrt{s}(1-r)+s\tau}\left(\frac{1}{s}+\frac{r-1}{\sqrt{s}}-\frac{r-1}{\sqrt{s}+1}\right)\frac{ds}{2\pi i}.\nonumber
\end{eqnarray}
These transforms can be readily evaluated \cite{prud}:
\begin{eqnarray}&&\!\!\!\!\!\!\!\!
g_1\!=\!\erfc\left(\frac{r\!-\!1}{2\sqrt{\tau}}\right)\!+\!(r\!-\!1)\re^{r-1+\tau}\erfc\left(\frac{r\!-\!1}{2\sqrt{\tau}}\!+\!\sqrt{\tau}\right).\label{g1}
\end{eqnarray}
Similarly to Eqs.~(\ref{f11})-(\ref{f12}) we can write from Eq.~(\ref{g1}) that
\begin{eqnarray}&&\!\!\!\!\!\!\!
G_1(t, r)=1+{\tilde G}_1\left(\frac{\nu t}{a^2}, r\right), \label{f111}
\end{eqnarray}
where ${\tilde G}_1$ which is defined by
\begin{eqnarray}&&\!\!\!\!\!\!\!\!\!\!
{\tilde G}_1\!\equiv\! (r\!-\!1)\re^{r-1+\tau}\erfc\left(\frac{r\!-\!1}{2\sqrt{\tau}}\!+\!\sqrt{\tau}\right)\!-\!\erf\left(\frac{r\!-\!1}{2\sqrt{\tau}}\right),
\end{eqnarray}
vanishes at long times as
\begin{eqnarray}&&\!\!\!\!\!\!\!
{\tilde G}_1\left(\tau, r\right)\sim -\frac{r^3-1}{6\tau\sqrt{\pi\tau}}+O\left(\frac{1}{\tau^{5/2}}\right).\label{longta}
\end{eqnarray}
By using the above in the last of Eqs.~(\ref{APhit}) we find:
\begin{eqnarray}&&\!\!\!\!\!\!\!\!\!\!\!\!
A_{1m}^{\Phi}(t, r)\! =\!\frac{I_{1m}^{t}(\bm \Phi)}{2r^2}
\!-\!\!\int_{-\infty}^{t}\! \!\!\!\frac{\dot{I}_{1m}^{t'}(\bm \Phi)}{2r^2}{\tilde G}_1\!\left(\!\frac{\nu (t\!-\!t')}{a^2}, r\!\right)\!dt'.  \label{asphi}
\end{eqnarray}
It follows from Eq.~(\ref{longta}) that the memory of this component decays faster than of the Basset memory term, according to integrable $\tau^{-3/2}$ law.

Equations (\ref{asls}), (\ref{asll}),  (\ref{alma}) and (\ref{asphi}) fully determine the solution for $l=1$. The limit of steady Stokes flow is obtained by setting $\dot I=0$, see the next section.

We next consider the solution in the limit of large distances. It can be deduced from Eq.~(\ref{asls}) by using the large argument asymptotic form of $\erfc(x)$ that
\begin{eqnarray}
&& A_{1m}^{\Psi}(t, r)+\frac{A_{1m}^{Y}(t, r)}{2} \label{expo} \\
&&=\int_{-\infty}^{t} \!\frac{[\dot{I}_{1m}^{t'}(\bm \Psi) \!+\! \dot{I}_{1m}^{t'}(\bm Y)]dt'}{2r}\erfc\left(\frac{a(r\!-\!1)}{2\sqrt{\nu(t-t')}}\right) \nonumber \\
&&\simeq \int_{-\infty}^{t} \!\!\!\frac{[\dot{I}_{1m}^{t'}(\bm \Psi) \!+\! \dot{I}_{1m}^{t'}(\bm Y)]dt'}{a(r\!-\!1)r}
\sqrt{\frac{\nu(t\!-\!t')}{\pi}}\exp\left(-\frac{a^2(r\!-\!1)^2}{4\nu(t-t')}\right),  \nonumber
\end{eqnarray}
where $a^2(r\!-\!1)^2\!\gg\! \nu t_c$ with $t_c$ being a characteristic time scale given by the largest difference $t-t'$ which significantly contributes to the integral (the integral is assumed to be convergent at long times). The time $t_c$ can be the time since the flow was originated by the boundaries, if the flow at the boundary vanishes after some time, or simply characteristic convergence time of the integral. The first term in the last line in (\ref{expo}) describes diffusive decay of momentum away from the particle, see also below. We conclude that at large distances $A_{1m}^{\Psi}(t, r)=-A_{1m}^{Y}(t, r)/2$ holds with exponential accuracy.

It can be seen from Eq.~(\ref{f1a}) that $f_1(\tau, r)$ vanishes at large distances as a Gaussian. Thus, with exponential accuracy we have from Eqs.~(\ref{of}), (\ref{f1a}) that $F_1(t, r)$ has no spatial dependence at $r\to\infty$
\begin{eqnarray}&&\!\!\!\!\!\!\!
F_1(t, r)\approx f_1\left(\frac{\nu t}{a^2}, 1\right)=\frac{\nu t}{a^2}+\frac{2}{a}\sqrt{\frac{\nu t}{\pi}}. \label{f1}
\end{eqnarray}
By using the above in the first of Eqs.~(\ref{APhit}), we find that the leading-order behavior at large distances $r\!-\!1\gg \sqrt{\nu t_c/a^2}$ is given by
\begin{eqnarray}&&
A_{1m}^{Y}(t, r)\! = \!\frac{a_{1m}(t)}{r^3},\ \ A_{1m}^{\Psi}(t, r)=-\frac{A_{1m}^{Y}(t, r)}{2}, \label{ast}
\end{eqnarray}
where we used Eq.~(\ref{expo}) and $a_{1m}(t)$ given by
\begin{eqnarray}&&\!\!\!\!\!\!\!
a_{1m}\!\equiv\!\!\int_{-\infty}^{t}\!\!\!
\left(\!\frac{\sqrt{\nu}}{a\sqrt{\pi(t\!-\!t')}}\!+\!\frac{\nu}{a^2}\!\right)
 (I_{1m}^{t'}(\bm \Psi) \!+\!I_{1m}^{t'}(\bm Y))dt'. \label{ako}
\end{eqnarray}
The exponential decay of $A_{1m}^{\Phi}(t, r)$ at large distances, that is seen readily from the corresponding decay of $g_1$ in Eq.~(\ref{g1}), implies that the only appreciable contribution to the flow is due to components in Eq.~(\ref{ast}). We find that the solution for $\bm U_{1m}(t, \bm x)$ in Eq.~(\ref{ugn}) at $r\!-\!1\gg \sqrt{\nu t_c/a^2}$ can be written as
\begin{eqnarray}&&\!\!\!\!\!\!
\bm U_{1m}(t, \bm x) \!=\!-\left[a_{1m}(t)+I^{t}_{1m}(\bm Y)\right]\nabla \left(\frac{Y_{1m}}{2r^2}\right),\label{cf}
\end{eqnarray}
where we used Eq.~(\ref{slo}). The significance of this result becomes clearer by noticing that the above equation implies that at large distances,
\begin{eqnarray}&&\!\!\!\!
-\frac{a^2}{\nu}\partial_t\bm U_{1m}\!=\!\nabla \left(\frac{Y_{1m}}{r^2}\right)
\left[\int_{-\infty}^{t}\! \frac{a( \dot{I}_{1m}^{t'}(\bm \Psi) \!+\! \dot{I}_{1m}^{t'}(\bm Y))dt'}{2\sqrt{\pi \nu (t-t')}}
\right.\nonumber\\&&\!\!\!\!\left.
+\frac{I_{1m}^{t}(\bm \Psi) \!+\!I_{1m}^{t}(\bm Y)}{2}
+\frac{a^2\dot{I}_{1m}^{t}(\bm Y)}{2\nu}\right]\!=\!\nabla p_{1m},
\end{eqnarray}
where we used Eq.~(\ref{c1m}) and defined $p_{lm}\equiv c_{lm}(t)Y_{lm}(\theta, \phi)/r^{l+1}$, see Eq.~(\ref{pr}). Therefore, at large distances the $l=1$ component of the flow satisfies the generalized Darcy law given by (in dimensional variables):
\begin{eqnarray}&&\!\!\!\!\!
\partial_t \bm v\!=\!-\frac{\nabla p}{\rho}, \quad \nabla\cdot\bm v=0 \quad \mbox{at}\;\;\; \frac{(r\!-\!a)^2}{\nu t_c}\!\gg\! 1. \label{darcy}
\end{eqnarray}
The main content introduced by the form of the law in the time domain (which of course is the inverse Fourier transform of the form in the frequency domain) is that it gives its validity condition via $t_c$. We shall demonstrate below that Eq.~(\ref{darcy}) holds also for other components of the flow.

\subsection{The case of $l>1$}

We now consider the case of $l>1$. Using Eq.~(\ref{mod}) we can rewrite $g_l$ in Eq.~(\ref{transf}) as
\begin{eqnarray}&&
g_l(\tau, r)=\int_{\delta-i\infty}^{\delta+i\infty}\frac{P_l(r \sqrt{s})}{P_l(\sqrt{s})}e^{\sqrt{s}(1-r)+s\tau}\frac{ds}{2\pi i s}. \label{gl}
\end{eqnarray}
We use partial fraction decomposition
\begin{eqnarray}&&\!\!\!\!\!
\frac{P_l(r x)}{x^2P_l(x)}\!=\!\frac{1}{x^2}\!+\!\frac{r\!-\!1}{x}\!+\!\sum_{k=1}^{l} \! \frac{P_l(-r a_k^{l})}{(a_k^{l})^2P'_l(-a_k^{l}) (x\!+\!a_k^{l})}, \label{pdco}
\end{eqnarray}
which relied on the following relations
\begin{eqnarray}&&\!\!\!\!\!
\lim_{x\to 0}\frac{P_l(r x)\!-\!P_l(x)}{xP_l(x)}\!=\!r\!-\!1,\,\ \ P_l(0)\!=\!P_l'(0)\!=\!(2l-1)!!,\nonumber
\end{eqnarray}
where $(2l-1)!!\equiv (2l)!2^{-l}/l!$ is the double factorial, and where we applied Eq.~(\ref{ourp}) and the L'Hopital's rule. We find using Eq.~(\ref{pdco}) with $x=\sqrt{s}$ in Eq.~(\ref{gl}) that
\begin{eqnarray}&&
g_l(\tau, r)=\erfc\left(\frac{r-1}{2\sqrt{\tau}}\right)+\frac{r-1}{\sqrt{\pi \tau}} \exp{\left(-\frac{(r-1)^2}{4\tau}\right)}
\nonumber\\&&
+\sum_{k=1}^{l}\!C_k^{l}(r)\int_{\delta-i\infty}^{\delta+i\infty}\!\! \frac{e^{s\tau\!-\!\sqrt{s}(r-1)}}{\sqrt{s}\!+\!a_k^{l}}\frac{ds}{2\pi i},\label{glt}
\end{eqnarray}
where we used \cite{prud} and introduced
\begin{eqnarray}&&
C_k^{l}(r)\equiv \frac{P_l(-r a_k^{l})}{(a_k^{l})^2P'_l(-a_k^{l})}. \label{df}
\end{eqnarray}
Eq.~(\ref{glt}) can be further simplified to \cite{prud}:
\begin{eqnarray}&&\!\!
g_l(\tau, r)=\erfc\left(\frac{r-1}{2\sqrt{\tau}}\right)+\frac{r-1}{\sqrt{\pi \tau}}\exp\left(-\frac{(r-1)^2}{4\tau}\right)
\nonumber\\&&\!\!
+\sum_{k=1}^{l}\!\frac{C_k^{l}(r)}{\sqrt{\pi \tau}}\exp\left(-\frac{(r-1)^2}{4\tau}\right)-\sum_{k=1}^{l} a_k^{l} C_k^{l}(r)\nonumber
\\&&\!\!
\exp\left(a_k^{l}(r-1)+\tau (a_k^{l})^2\right)\erfc\left(\frac{r-1+2a_k^{l}\tau}{2\sqrt{\tau}}\right). \label{fid}
\end{eqnarray}
Thus we found the kernel $G_l(t, r)$ that equals $g_l(\nu t/a^2, r)$. By comparing the above with Eq.~(\ref{g1}), one can see that the result is also valid for $l=1$ (where $a_k^{l}=1$ and $C_1^{1}(r)=1-r$, see  Eq.~(\ref{df})).

The calculation of $F_l(t, r)$, or $f_l(\tau, r)$ in Eq.~(\ref{of}), is rather cumbersome and is provided in Appendix \ref{ker}. For the case of $l=2$, which differs from $l>2$, we have
\begin{eqnarray}&&\!\!\!\!\!\!\!
f_2(\tau, r)\!=\!
\left(3\tau+\frac{3-r^2}{2}\right)\erfc\left(\frac{r-1}{2\sqrt{\tau}}\right)
\label{fk}\\&&\!\!\!\!\!\!\!
+\frac{3(r-1)\sqrt{\tau}}{\sqrt{\pi}}\exp\left(-\frac{(r-1)^2}{4\tau}\right)
\nonumber\\&&\!\!\!\!\!\!\!
-(r^2-3r+3) e^{r-1+\tau}\, \erfc\left(\frac{r-1}{2\sqrt{\tau}}+\sqrt{\tau}\right).\nonumber
\end{eqnarray}
Differentiation of $f_2$ reproduces $\partial_r f_2=-r g_1$, cf. Eq.~(\ref{recur}), with $g_1$ given by Eq.~(\ref{g1}).
The kernel $F_2(t, r)$ is found from Eq.~(\ref{fk}) by using Eq.~(\ref{of}). The calculation of $f_l(\tau, r)$ with $l>2$, given in Appendix \ref{ker} gives
\begin{eqnarray}&&
f_l(\tau, r)\!=\!
\left[(2l\!-\!1)\tau\!-\!\frac{r^2}{2}\!+\!\frac{2l\!-\!1}{2(2l\!-\!3)}\right]\erfc\left(\frac{r\!-\!1}{2\sqrt{\tau}}\right)
\label{fkl}\\&&
+\frac{1}{\sqrt{\pi\tau}}\exp\left(-\frac{(r-1)^2}{4\tau}\right)\left[(2l\!-\!1)(r-1)\tau+\frac{r^3(l\!-\!2)}{3}
\right.\nonumber\\&&\left. -r^2(l\!-\!1)+\frac{r(l\!-\!1)(2l\!-\!1)}{2l\!-\!3}
-\frac{l(2l\!-\!1)}{3(2l\!-\!3)}\right]
\nonumber\\&&
+\!\sum_{k=1}^{l-1}\!\!\frac{P_{l}(-r a_k^{l-1})}{\left(a_k^{l-1}\right)^4P'_{l-1}(-a_k^{l-1})}
\left[\frac{1}{\sqrt{\pi\tau}}\exp\left(-\frac{(r-1)^2}{4\tau}\right)
\right.\nonumber\\&&\left.
\!-\!a_k^{l-1}\exp\left(a_k^{l-1}(r\!-\!1)\!+\!(a_k^{l-1})^2\tau\right)
\erfc\left(\frac{r\!-\!1}{2\sqrt{\tau}}\!+\!a_k^{l-1}\sqrt{\tau}\right)
\right].\nonumber
\end{eqnarray}
Substitution of $r=1$ in the above equation yields:
\begin{eqnarray}&&
f_l(\tau, 1)=
(2l-1)\tau+\frac{1}{2l-3}
\label{fl1}\\&&
-\sum_{k=1}^{l-1}\frac{P_{l}(-a_k^{l-1}) e^{(a_k^{l-1})^2\tau}
\erfc\left(a_k^{l-1}\sqrt{\tau}\right)}{\left(a_k^{l-1}\right)^3P'_{l-1}(-a_k^{l-1})}
,\nonumber
\end{eqnarray}
where we used the identity
\begin{eqnarray}&&
\sum_{k=1}^{l-1}\frac{P_{l}(-a_k^{l-1})}{\left(a_k^{l-1}\right)^4P'_{l-1}(-a_k^{l-1})}\!=\!0. \label{sumrule}
\end{eqnarray}
derived in Appendix \ref{ker}. Using Eq.~(\ref{of}) and the last two equations gives $F_l(t, r)$ for $l>2$. The above formulae for $f_l(\tau, r)$ with $l>1$ and Eq.~(\ref{f11}) imply that for all $l\geq 1$ we have the representation
\begin{eqnarray}&&\!\!\!\!\!\!\!
F_l(t, r)=\frac{r^2\!-\!1}{2}+{\tilde F}_l\left(\frac{\nu t}{a^2}, r\right), \label{isl}
\end{eqnarray}
with ${\tilde F}_l$ vanishing at large times.

\subsection{Memory and universal long-time decay}

We are interested to investigate the memory decay of the obtained solution. From Eq.~(\ref{fid}) it follows at large $\tau$ $g_l(\tau, r)$ obeys
\begin{eqnarray}&&\!\!\!\!\!\!\!\!\!\!\!
g_l\!\sim \!1\!-\!\frac{(r\!-\!1)^3}{6\tau\sqrt{\pi\tau}}
\!+\!\sum_{k=1}^{l}\!\frac{C_k^{l}(r)\left(1\!+\!2a_k^{l}(r\!-\!1)\right)}{4(a_k^{l})^2\tau\sqrt{\pi\tau}}\!+\!o\left(\!\frac{1}{\tau}\!\right).
\end{eqnarray}
Thus, corrections to the leading order constant term are of order $\tau^{-3/2}$. Similarly to pressure, all memory kernels involving $G_l(t, r)$ with $l>0$ decay according to the integrable $\tau^{-3/2}$ law rather than the
non-integrable decay $\sim \tau^{-1/2}$ holding for $l=0$. This, for $A_{lm}^{\Phi}(t, r)$ in the last of Eqs.~(\ref{APhit}) we have for any $l$ (including $l=1$, see Eq.~(\ref{asphi})) that
\begin{eqnarray}&&\!\!\!\!\!\!\!\!\!\!\!
A_{lm}^{\Phi}\! =\!\frac{I_{lm}^{t}(\bm \Phi)}{r^{l+1}l(l\!+\!1)}
\!+\!\!\int_{-\infty}^{t}\!\!{\tilde g}_l\left(\!\frac{\nu(t\!-\!t')}{a^2}, r\!\right) \!\frac{\dot{I}_{lm}^{t'}(\bm \Phi)dt'}{r^{l+1}l(l\!+\!1)},\label{almp}%\nonumber
\end{eqnarray}
where the kernel ${\tilde g}_l\equiv g_l-1$ in the last integral decays as $(t-t')^{-3/2}$ at large $t-t'$. This leads to a non-trivial conclusion that the toroidal component of the flow $\bm u_{tor}(t, \bm x)$ has faster decaying memory than the rest of the flow components. The toroidal component in the time domain is given by $\bm u_{tor}(t, \bm x)\equiv \sum_{lm}A_{lm}^{\Phi}(t, r)\bm \Phi_{lm}(\theta, \phi)$, cf. Eqs.~(\ref{ug})-(\ref{ugent}) and Sec. \ref{Lambs}. We have using $\bm \Phi_{lm}\!=\!-\nabla\!\times\! (\bm r Y_{lm})$ and Eq.~(\ref{almp}) that
\begin{eqnarray}&&
\bm u_{tor}(t, \bm x) \!= \!-\sum_{lm}\nabla\!\times\! (\bm r Y_{lm}(\theta, \phi) A_{lm}^{\Phi}(t, r))\\&&
=-\sum_{lm}\nabla\!\times\! \left(\frac{I_{lm}^{t}(\bm \Phi)\bm r Y_{lm}(\theta, \phi)}{r^{l+1}l(l\!+\!1)}\right) \nonumber\\&&
-\sum_{lm}\nabla\!\times\! \left(\bm r Y_{lm}(\theta, \phi)\int_{-\infty}^{t}\!\!{\tilde g}_l\left(\!\frac{\nu(t\!-\!t')}{a^2}, r\!\right)\frac{\dot{I}_{lm}^{t'}(\bm \Phi)dt'}{r^{l+1}l(l\!+\!1)}\right).\nonumber
\end{eqnarray}
Let us now consider other kernels in Eqs.~(\ref{APhit}). The long-time asymptotic form of $f_2(\tau, r)$ is found from Eq.~(\ref{fk}) as
\begin{eqnarray}&&\!\!\!\!\!\!\!
f_2(\tau, r)\!=\!3\tau+\frac{3-r^2}{2}-\frac{1}{\sqrt{\pi\tau}}+O\left(\frac{1}{\tau^{3/2}}\right).
\end{eqnarray}
Similarly, the long time asymptotic form of $f_l(\tau, r)$ is derived by a cumbersome expansion of Eq.~(\ref{fkl}) that yields
\begin{eqnarray}&&\!\!\!\!\!
f_l(\tau, r)\!=\!(2l\!-\!1)\tau-\frac{r^2}{2}+\frac{2l\!-\!1}{2(2l\!-\!3)}+O\left(\frac{1}{\tau^{3/2}}\right).
\end{eqnarray}
The last two equations and properties of ${\tilde F}_1$, see Eq.~(\ref{f12}), imply that ${\tilde F}_l$ defined in Eq.~(\ref{isl}) obeys at long times
\begin{eqnarray}&&\!\!\!\!\!\!\!\!\!\!\!\!\!
{\tilde F}_1\!\left(\frac{\nu t}{a^2}, r\right)\!=\!\!O\left(\frac{1}{\tau^{1/2}}\right);\
{\tilde F}_{l}\!\left(\frac{\nu t}{a^2}, r\right)\!=\!\!O\left(\frac{1}{\tau^{3/2}}\right), \label{ln}
\end{eqnarray}
where $l>1$.
The function $A_{lm}^{Y}(t, r)$ in Eq.~(\ref{APhit}) reads
\begin{eqnarray}&&
A_{lm}^{Y}(t, r)\! = \!\frac{(r^2\!-\!1)[I_{lm}^{t}(\bm \Psi) \!+\! l I_{lm}^{t}(\bm Y)]}{2r^{l+2}}
\\&&\!\!\!\!\!\!\!
+\frac{1}{r^{l+2}}\int_{-\infty}^{t}\! {\tilde F}_l\left(\frac{\nu(t\!-\!t')}{a^2}, r\right) [\dot{I}_{lm}^{t'}(\bm \Psi) \!+\!  l\dot{I}_{lm}^{t'}(\bm Y)]dt',\nonumber
\end{eqnarray}
and thus generalizes Eq.~(\ref{alma}) to any $l$. It follows from Eq.~(\ref{ln}) that the memory for $l>1$ decays as $t^{-3/2}$, i.e., faster than $t^{-1/2}$ corresponding to $l=1$. Finally, the last of $A_{lm}$ coefficients in Eqs.~(\ref{APhit}) is given by
\begin{eqnarray}&&
A_{lm}^{\Psi}(t, r)\!=\! \frac{\left(I_{lm}^{t}(\bm \Psi) \!+\! l I_{lm}^{t}(\bm Y)\right)}{r^{l}l(l+1)}-\frac{A_{lm}^{Y}(t, r)}{l+1}
\nonumber\\
&&\!\! +\int_{-\infty}^{\infty} \!{\tilde g}_{l-1}\left(\!\frac{\nu(t\!-\!t')}{a^2}, r\!\right)\frac{\left(\dot{I}_{lm}^{t'}(\bm \Psi) \!+\! l \dot{I}_{lm}^{t'}(\bm Y)\right) dt'}{r^{l}l(l+1)}.\end{eqnarray}
It is seen that the memory for these coefficients, similarly to $A_{lm}^{Y}(t, r)$ decays at $t^{-3/2}$ for $l>1$ and $t^{-1/2}$ for $l=1$. The derived expressions for the flow in the time domain are one of the main results of this work. They demonstrate explicitly how the flow propagates from the surface where it was generated by the boundary conditions, into the fluid bulk. In dimensional variables, the propagation is governed by the exponential factor $\exp\left(-(r\!-\!a)^2/(4\nu t)\right)$ and similar factors within error functions.

The above analysis provides important implications for free decay problems, where the flow decays starting from some initial conditions prescribed on the sphere (or problems reducible to this setting, cf. subsection~\ref{fl}). Apparently, at long times the toroidal component of the transient flow can be neglected to the leading approximation. The flow has only $l=1$ component that decays inversely proportionally to the square root of time. In the next subsection we provide another example where the difference of memory decay rate for $l=1$ and $l>1$ components has important implications.

\subsection{Long-time decay of transversal wave} \label{transversal}

The problem of decay of the transversal wave in presence of a fixed sphere introduced in subsection \ref{flok} provides specific example showing that $l=1$ component of the solution series controls the long time behavior. The solution component $\bm v'$ is separable at $r=1$ where it is given by the product of $-{\hat y}\cos(kx)$ and $e^{-\nu k^2 t} H(t)$, see Eq.~(\ref{bcv}). This allows to introduce the reduced (constant) coefficients that for $l=1$ read
\begin{eqnarray}
&& I^{t}_{1m}(\bm W) = i_m(\bm W)e^{-\nu k^2 t}\theta(t),\nonumber \\
&& i_m(\bm W)=-\int_{r=1} \cos(kx) W_{1m, y}^* d\Omega, \label{ilm}
\end{eqnarray}
cf. Eq.~(\ref{def00}). Similar coefficients can be introduced for $l>1$.

Let us consider the long-time asymptotic form of the solution that holds at $t\gg (\nu k^2)^{-1}$. The flow $\bm v'$, that obeys the boundary condition given by Eq.~(\ref{bcv}), is dominated by the $l\!=\!1$-term of the solution series. The coefficients $A_{1m}$, defined in Eqs.~(\ref{APhit}), are given by sums of a term that originates from the instantaneous flow at the boundary, and an integral contribution due to memory [see Eqs.~(\ref{alma}) and (\ref{asls})]. The terms corresponding to the instantaneous flow are exponentially small and can be neglected. Then the asymptotic expansion of $A^Y_{1m}$ has the form
\begin{eqnarray}&&
A_{1m}^{Y}(t, r)\! = \!\frac{1}{r^3}\int_{-\infty}^t\! \left[{\tilde F}_1\left(\frac{\nu t}{a^2}, r\right)\!-\!\frac{\nu t'}{a^2}\partial_{\tau}{\tilde F}_1(\tau, r)|_{\tau=\nu t/a^2}
\right.\nonumber\\&&\left.
+\ldots \frac{}{}\right][\dot{I}_{1m}^{t'}(\bm \Psi) \!+\!  \dot{I}_{1m}^{t'}(\bm Y)]dt',
\end{eqnarray}
where the integral is determined by $O[(\nu k^2)^{-1}]$--vicinity of $t'=0$ and dots stand for the higher order terms. The lowest order term in the expansion is negligible and using Eq.~(\ref{f12}) we find
\begin{eqnarray}&&\!\!\!\!\!
A_{1m}^{Y}(t, r)\! \approx \!\frac{\partial_{\tau}{\tilde F}_1(\tau, r)|_{\tau=\nu t/a^2}}{r^3(ka)^2} \left(i_m(\bm \Psi) \!+\! i_m(\bm Y)\right)
\nonumber\\&&\!\!\!\!\!
\approx \sqrt{\frac{a^2}{\pi\nu t}}\frac{(r\!-\!1)^2(2r\!+\!1)\left(i_m(\bm \Psi) \!+\! i_m(\bm Y)\right)}{12\nu t k^2 r^3}. \label{a1m}
\end{eqnarray}
As for Eq.~(\ref{f12}), the approximation assumes that $\nu t\gg (r-1)^2$. Similar calculation demonstrates that $A_{lm}^{Y}(t, r)$ for $l>1$ decays at large times as $t^{-5/2}$ that is factor of $t$ faster than for $l=1$.

The asymptotic analysis of $A_{1m}^{\Psi}(t, r)$ in Eq.~(\ref{asls}) proceeds similarly and yields
\begin{eqnarray}&&
A_{1m}^{\Psi}(t, r)+\frac{A_{1m}^{Y}(t, r)}{2}\!=\!\sqrt{\frac{a^2}{\pi \nu t}}\frac{(r\!-\!1)(i_{m}(\bm \Psi) \!+\! i_{m}(\bm Y))}{4\nu t k^2 r},\nonumber
\end{eqnarray}
provided that $\nu t\!\gg\!(r-1)^2$. Next by using Eq.~(\ref{a1m}) we find that
\begin{eqnarray}&&
A_{1m}^{\Psi}(t, r)\!=\!\sqrt{\frac{a^2}{\pi \nu t}}\frac{(r\!-\!1)(i_{m}(\bm \Psi) \!+\! i_{m}(\bm Y))}{4\nu t k^2 r}
\label{ar}\\&&
\times\left[1-\frac{(r\!-\!1)(2r\!+\!1)}{6r^2}\right]\,, \nonumber
\end{eqnarray}
where $\nu t\gg \max{[(r-1)^2,  k^{-2}]}$. Finally, by collecting the above terms yields the flow at long times. We have from Eqs.~(\ref{ug})-(\ref{ugent}) that
\begin{eqnarray}&&
\bm v'\!= \!\sum_{m=-1}^1\left( A_{1m}^{Y}(t, r)\bm Y_{1m}(\theta, \phi)\!+\!A_{1m}^{\Psi}(t, r)\bm \Psi_{1m}(\theta, \phi) \right)\nonumber\\&&
=\sqrt{\frac{a^2}{\pi \nu t}}\frac{(r\!-\!1)}{4\nu t k^2 r}\sum_{m=-1}^1\left(i_m(\bm \Psi) \!+\! i_m(\bm Y)\right)\left(\frac{(r\!-\!1)(2r\!+\!1)}{3 r^2}
\right.\nonumber\\&&\left.
\times\bm Y_{1m}(\theta, \phi)+\left(1-\frac{(r\!-\!1)(2r\!+\!1)}{6r^2}\right)\bm \Psi_{1m}(\theta, \phi) \right).\label{swa}
\end{eqnarray}
The above result holds under the same conditions as for Eq.~(\ref{ar}).

The coefficients $i_m(\bm \Psi) \!+\! i_m(\bm Y)$ in Eq.~(\ref{swa}) are determined next. From the definition in Eq.~(\ref{vsh}) it follows that $\bm Y_{1m}(\theta, \phi)+\bm \Psi_{1m}(\theta, \phi)=\nabla(rY_{1m})$. Using the last equality in Eq.~(\ref{ilm}) we find
\begin{eqnarray}&&\!\!\!\!\!\!
i_m(\bm \Psi) \!+\! i_m(\bm Y)=-\int_{r=1} \cos(kx) \nabla_y(rY^*_{1m}) d\Omega.
\end{eqnarray}
By using the definition of $Y_{1m}$ in Eq.~(\ref{fi}) it follows that $i_0(\bm \Psi) \!+\! i_0(\bm Y)=0$ and
\begin{eqnarray}&&
i_1(\bm \Psi) \!+\! i_1(\bm Y)=\sqrt{\frac{3}{8\pi}}\int_{r=1} \cos(kx) \nabla_y(x+iy) d\Omega
\\&&
=i\sqrt{\frac{3}{8\pi}}\int_{r=1}\!\! \cos(kz) d\Omega%=i 4\pi \sqrt{\frac{3}{8\pi}}\int_{0}^1 \cos(ky)dy
\!=\!\frac{i \sqrt{6\pi} \sin k}{k}\!=\!i_{-1}(\bm \Psi)\!+\! i_{-1}(\bm Y).\nonumber
\end{eqnarray}
This completes the derivation of the long-time asymptotic solution in Eq.~(\ref{swa}) at the leading order.

For $r\!\gg\! 1$, the Eq.~(\ref{swa}) becomes $\bm Y_{1m}+\bm \Psi_{1m}=\nabla(rY_{1m})$, and the solution simplifies considerably,
\begin{eqnarray}&&\!\!\!\!\!\!
\bm v'=\sqrt{\frac{a^2}{\pi \nu t}}\frac{\sin k}{2\nu t k^3}\bm{\hat y}. \label{vt}
\end{eqnarray}
Notice that this component of the transversal wave is space-independent and decays slower than the undisturbed wave $\bm v_0={\hat y}\cos(kx)e^{-\nu k^2 t}$. At long times Eq.~(\ref{vt}) holds in the domain $r\!\ll\! \sqrt{\nu t}$, whose volume grows at the rate $t^{3/2}$. Inside this volume the flow $\bm v=\bm v_0+\bm v'$ obeys $\bm v\approx \bm v'$ with exponential accuracy. The total momentum carried by the flow, given by this volume times $\bm v'$ in Eq.~(\ref{vt}), is constant, as it should be under the momentum-preserving unsteady Stokes equations. The perturbation $\bm v'$ propagates in the same $\bm{\hat y}$-direction as the undisturbed wave $\bm v_0$.

\subsection{Generalized Darcy's law}

Here we study the asymptotic form of the general solution at large distances from the sphere. It is readily seen from Eq.~(\ref{fid}) that far from the particle $g_l(\tau, r)$ decays as a Gaussian.
Similar observation holds for $f_l(\tau, r)$ in Eq.~(\ref{fkl}). As follows from Eq.~(\ref{of}), and similarly to Eq.~(\ref{f1}), we have
\begin{eqnarray}&&
F_{l}(t, r \rightarrow \infty)=f_{l}\left(\frac{\nu t}{a^2}, 1\right)\!=\!\frac{(2l\!-\!1)\nu t}{a^2}+\frac{1}{2l-3}
\nonumber\\&&
-\sum_{k=1}^{l-1}\frac{P_{l}(-a_k^{l-1})\exp\left((a_k^{l-1})^2\tau\right)
\erfc\left(a_k^{l-1}\sqrt{\tau}\right)}{\left(a_k^{l-1}\right)^3P'_{l-1}(-a_k^{l-1})},
\end{eqnarray}
where we also used Eq.~(\ref{fl1}). Using this equation,
from Eqs.~(\ref{ugn}) and (\ref{ug})--(\ref{APhit}) by using Eq.~(\ref{slo}), similarly to $l\!=\!1$ flow component, at $a^2(r-1)^2\gg \nu t_c$ we obtain
\begin{eqnarray}&&\!\!\!\!\!\!
\bm U_{lm}(t, \bm x) \!\approx \!-\left(a_{lm}(t)+I^{t}_{lm}(\bm Y)\right)\nabla \frac{Y_{lm}}{(l+1)r^{l+1}},
\end{eqnarray}
cf. Eq.~(\ref{cf}). Here the function $a_{lm}(t)$ is given by
\begin{eqnarray}&&
a_{lm}(t)\! = \!\int_{-\infty}^{t}\! F_l(t\!-\!t', r=\infty) ( \dot{I}_{lm}^{t'}(\bm \Psi) \!+\! l \dot{I}_{lm}^{t'}(\bm Y))dt'\nonumber\\&&
=\!\frac{(2l\!-\!1)\nu }{a^2}\!\int_{-\infty}^{t}\! \!\!(I_{lm}^{t'}(\bm \Psi) \!+\! l I_{lm}^{t'}(\bm Y))dt'\!+\!\frac{I_{lm}^{t}(\bm \Psi) \!+\! l I_{lm}^{t}(\bm Y)}{2l\!-\!3}
\nonumber\\&&
-\!\sum_{k=1}^{l-1} \! \!\frac{P_{l}(-a_k^{l-1})}{\left(a_k^{l-1}\right)^3P'_{l-1}(-a_k^{l-1})}\int_{-\infty}^{t}\! dt' ( \dot{I}_{lm}^{t'}(\bm \Psi) \!+\! l \dot{I}_{lm}^{t'}(\bm Y))
\nonumber\\&&\times\exp\left(\frac{(a_k^{l-1})^2\nu (t-t')}{a^2}\right)\erfc\left(\frac{a_k^{l-1}\sqrt{\nu (t-t')}}{a}\right). \label{alm}
\end{eqnarray}
This implies by taking derivative and comparing with Eq.~(\ref{fpr}), that
\begin{eqnarray}&&\!\!\!\!
-\frac{a^2}{\nu}\partial_t\bm U_{lm}\!=\!\nabla \frac{c_{lm}Y_{lm}}{r^{l+1}},%\right)
\end{eqnarray}
where we used the identity
\begin{eqnarray}&&
\sum_{k=1}^{l-1}\frac{P_{l}(-a_k^{l-1})}{\left(a_k^{l-1}\right)^3P'_{l-1}(-a_k^{l-1})}\!=\!\frac{1}{2l-3}, \label{sumrul}
\end{eqnarray}
derived in Appendix \ref{ker}. Together with the analogous result obtained above for $l=1$, we conclude that in dimensional variables, at $(r-a)^2\gg \nu t_c$, the solutions of Eqs.~(\ref{unsad}) reduce to the solutions of the generalized Darcy equations (\ref{darcy}) (with the understanding that $t_c$ can differ for different $lm$). The Laplacian term in Eqs.~(\ref{unsad}) is negligible at large distances and the solution is a potential flow, $\bm v=\nabla \phi$ where $\phi$ is a harmonic function obeying
\begin{eqnarray}&&\!\!\!\!\!\!\!\!\!\!\!\!\!
\phi=-\sum_{lm}\frac{\left(a_{lm}(t)+I^{t}_{lm}(\bm Y)\right)Y_{lm}}{(l+1)r^{l+1}}, \label{phil}
\end{eqnarray}
with $a_{1m}$ given by Eq.~(\ref{ako}) and $a_{l>1, m}$ given by Eq.~(\ref{alm}). The results explicitly relate the boundary velocity with the potential flow outside the ``viscous layer" of width $\nu t_c$, where the pressure is given by $-\rho\partial_t\phi$.

\section{Solution as asymptotic series in frequency and corrections to the steady Stokes limit}\label{freq}

In this section we consider the solution as an asymptotic series in half-integer powers of the frequency and study the low-order terms of this series. These describe the finite-frequency corrections to the Stokes limit of zero frequency. As already mentioned, the small frequency expansion is non-uniform in space. At distances from the sphere that are larger than the viscous penetration depth $\delta$, the cases of finite and zero frequency are very different, no matter how small the frequency is. This depth tends to infinity in the zero frequency limit. The actual parameter of the expansion is $r|\lambda| =r\sqrt{Ro} $ and not just $\sqrt{Ro}$, so that the expansion breaks down at $r\sim Ro^{-1/2}$. Some of the formulas below that are formally written as expansions in $|\lambda|$ must be understood accordingly.

\subsection{Solution as a series in frequency}\label{frequencyindependent}

Eqs.~(\ref{ugn})-(\ref{AB}) imply that coefficients of $I_{lm}^{\omega}(\bm W)$ in $A_{lm}^{W}(\omega, r)$ are given by a Taylor series in $\lambda$ where, as previously, $\bm W$ stands for any of the VSH. This allows to write the solution in the frequency domain as
\begin{eqnarray}&&
A_{lm}^{W}(\omega, r)\! = \!\sum_{i=1}^3\sum_{k=0}^{\infty} b_k^i(r)(-i\omega)^k I_{lm}^{\omega}(\bm W_i)
\nonumber\\&&
+\frac{1}{\sqrt{\delta-i\omega}}\sum_{i=1}^3\sum_{k=1}^{\infty} B_k^i(r)(-i\omega)^k I_{lm}^{\omega}(\bm W_i), \label{frs}
\end{eqnarray}
with some coefficients $b_k^i(r)$ and $B_k^i(r)$. Here the first sum in the RHS contains integer powers of $\omega$ while the second contains half-integer powers. This series provides the solution as a series in half-integer powers of the frequency and not integer powers as could be expected from the form of Eq.~(\ref{freqs}). In dimensionless variables this is the series in $|\lambda|=\sqrt{Ro}$ and not $Ro$. This form of the solution is most useful in the case where the boundary flow contains only the frequencies whose Roshko number $a^2\omega/\nu$ is small. Then, truncating the series at some order provides a good approximation at not too large distances: the full solution involves exponentials of $\lambda r$ so the expansion in $\lambda$ is not accurate at $r\gtrsim |\lambda|^{-1}$, cf. above.

Before we consider the low-order terms of the above series, we notice that the real time form of the above solution is
\begin{eqnarray}&&
A_{lm}^{W}(t, r)\! = \!\sum_{i=1}^3\sum_{k=0}^{\infty} b_k^i(r)\frac{d^k I_{lm}^{t}(\bm W_i)}{dt^k}
\nonumber\\&&
+\sum_{i=1}^3\sum_{k=1}^{\infty} \int_{-\infty}^t \frac{dt'}{\sqrt{\pi(t-t')}} B_k^i(r)\frac{d^k I_{lm}^{t'}(\bm W_i)}{dt^{'k}}.  \label{fdo}
\end{eqnarray}
The low-frequency expansion can then be also considered as neglecting higher order time derivatives of $I_{lm}^{t}(\bm W_i)$.

The series up to {$\mathcal O(\lambda^2)$ is determined by the expansions of $\bm U^{lm}$ in Eq.~(\ref{ugn})
\begin{eqnarray}&&
\bm U_{lm}=\bm U^0_{lm}+\lambda\bm U^1_{lm}+\lambda^2\bm U^2_{lm}+\ldots,
\end{eqnarray}
where $\bm U^i_{lm}$ are frequency-independent and dots stand for higher order terms.
The expansion is derived from the low-frequency expansion of the coefficients in Eq.~(\ref{AB})
\begin{eqnarray}&&
r^lA_l(r)-A_l(1) =
-\frac{r^2-1}{2} +
\frac{(r^2-1)^2}{8(2l-3)}\lambda^2
\nonumber\\&&
+\frac{\delta_{l1}}{6}(r-1)^2(1+2r)\lambda(1+\lambda),
\label{Adef}
\end{eqnarray}
and
\begin{eqnarray}&&\!\!\!\!\!\!\!\!\!
B_l(r) = \frac{1}{r^l} +\lambda^2 \frac{(1-r^2)}{2r^l(2l-1)},\ \ l>0,
\label{Bdef}\\&&
\!\!\!\!\!\!\!\!\!B_{l-1}(r) \!=\! \frac{1}{r^{l-1}} \!+\!\lambda(1\!+\!\lambda)\delta_{l1}(1-r) \!+\!\frac{\lambda^2(1\!-\!r^2)}{2r^{l-1}(2l\!-\!3)}.
%B_0=1+\lambda (1-r)+\frac{\lambda^2 (r-1)^2}{2}.
\nonumber
\end{eqnarray}
%(we provide directly $B_{l-1}(r)$ which is more useful below than $B_l(r)$)
%\begin{eqnarray}&&\!\!\!
%%B_l(r) = \frac{1}{r^l} +\lambda^2 \frac{(1-r^2)}{2r^l(2l-1)},\ \ l>0,
%%\label{Bdef}\\&&
%B_{l-1}(r) \!=\! \frac{1}{r^{l-1}} \!+\!\lambda(1\!+\!\lambda)\delta_{l1}(1-r) \!+\!\frac{\lambda^2(1\!-\!r^2)}{2r^{l-1}(2l\!-\!3)}. \label{Bdef}
%%B_0=1+\lambda (1-r)+\frac{\lambda^2 (r-1)^2}{2}.
%\end{eqnarray}
We consider terms of different order in $\lambda$.

\subsection{Zero-order solution: yet another form of general solution of steady Stokes equations}

In Section \ref{Lambs} we derived Lamb's solution for the steady Stokes flow by taking the limit $\lambda\to 0$ of our solution in Eq.~(\ref{su}). In this subsection we study the same limit using a different representation. This corresponds to considering the term of zero order in frequency in Eq.~(\ref{frs}). We have from the above that $\bm U_{lm}^0$, obtained by setting $\lambda=0$, obeys
\begin{eqnarray}&&
\bm U_{lm}^0 = a_{lm}^{Y}\bm Y_{lm} +  a_{lm}^{\Psi}\bm \Psi_{lm} +  a_{lm}^{\Phi}\bm \Phi_{lm},
\label{uge0}
\end{eqnarray}
where
\begin{eqnarray}&&
a_{lm}^{Y} = \frac{ (I_{lm}(\bm \Psi) + l I_{lm}(\bm Y))(r^2-1)+2I_{lm}(\bm Y)}{2r^{l+2}},
%+\frac{I_{lm}(\bm Y)}{r^{l+2}},
\label{uge1}\\&&
a_{lm}^{\Psi} \!=\!\frac{I_{lm}(\bm \Psi) \!+\! l I_{lm}(\bm Y)}{r^{l}l(l\!+\!1)}
\!-\!\frac{a_{lm}^{Y}}{l\!+\!1}, \ \ a_{lm}^{\Phi}\! =\!  \frac{I_{lm}(\bm \Phi) }{r^{l+1}l(l\!+\!1)}.
\nonumber
\end{eqnarray}
Here we do not use the subscript of $I_{lm}$ indicating whether it belongs to time or frequency domains, since the above equality holds both in time, where $I_{lm}=I_{lm}^t$, and frequency, where $I_{lm}=I_{lm}^{\omega}$, domains. Given a boundary condition for the steady Stokes flow, the coefficients $I_{lm}$, defined in Eq.~(\ref{def00}), can be readily obtained from Eqs.~(\ref{cd})-(\ref{sp}).

Eqs.~(\ref{ugn}) and (\ref{uge0})-(\ref{uge1}) give a general solution for steady Stokes flow, which is seemingly missing in the literature. We believe that it has a simpler form than several representations provided in \cite{kim}. For instance, the adjoint method \cite{kim} gives the solution as expansion in terms of the surface vector fields $\bm A_{lm}, \bm B_{lm}, \bm C_{lm}$ defined as follows
\begin{eqnarray}&&
\bm A_{lm} = l \tilde Y_{lm} \hat r + \partial_{\theta}\tilde Y_{lm} \hat \theta
+ \frac{\partial_{\phi}\tilde Y_{lm} }{\sin\theta}\hat \phi,
\label{sph_harm_def1}
\\&&
\bm B_{lm} =-(l+1) \tilde Y_{lm} \hat r + \partial_{\theta}\tilde Y_{lm} \hat \theta
+ \frac{\partial_{\phi}\tilde Y_{lm} }{\sin\theta}\hat \phi,
\nonumber\\&&
\bm C_{lm}\! =\! \frac{\partial_{\phi}\tilde Y_{lm}}{\sin\theta}\hat \theta
\!- \!\partial_{\theta}\tilde Y_{lm} \hat \phi;\ \  \tilde Y_{lm}\! =\! (-1)^m P_l^m(\cos\theta) e^{im\phi}.
\nonumber
\end{eqnarray}
The above fields are linear combinations of the VSH. The transition between the two types of expansions can be readily demonstrated:
\begin{eqnarray}&&
\tilde Y_{lm} \hat r = \frac{\bm A_{lm} - \bm B_{lm}}{2l+1},
\label{sph_harm_def2}\\&&
 \partial_{\theta}\tilde Y_{lm} \hat \theta
+ \frac{\partial_{\phi}\tilde Y_{lm} }{\sin\theta}\hat \phi =
\frac{(l+1)\bm A_{lm} + l\bm B_{lm}}{2l+1}.
\nonumber
\end{eqnarray}
Using the definition in Eq.~(\ref{fi}) we find:
%$$
%Y_{lm}=\sqrt{\frac{(2l+1)}{4\pi}\frac{(l-m)!}{(l+m)!}}P_l^m(\cos\theta)\exp\left(im\phi\right),
%$$
$$
Y_{lm}= \eta_{lm}\tilde Y_{lm},
\ \eta_{lm}=(-1)^m\sqrt{\frac{(2l+1)}{4\pi}\frac{(l-m)!}{(l+m)!}}.
$$
This gives, by comparing Eqs.~(\ref{sph_harm_def2}) and (\ref{vsh}), that
\begin{eqnarray}&&
\bm Y_{lm} = \frac{\eta_{lm}}{2l+1} (\bm A_{lm} - \bm B_{lm}),\
\bm \Phi_{lm} = \eta_{lm} \bm C_{lm},
\nonumber \\&&
\bm \Psi_{lm} = \frac{\eta_{lm}}{2l+1} ((l+1)\bm A_{lm} + l\bm B_{lm}).
\label{YPsiPhi_via_ABC}
\end{eqnarray}
The inverse transformations read:
\begin{eqnarray}&&\!\!\!\!\!
\bm A_{lm} = \frac{\bm \Psi_{lm}+l \bm Y_{lm}}{\eta_{lm}}, \ \
\bm B_{lm} = \frac{\bm \Psi_{lm}-(l+1) \bm Y_{lm}}{\eta_{lm}},
\label{ABC_via_YPsiPhi}
\end{eqnarray}
and $ \bm C_{lm} =\bm \Phi_{lm}/\eta_{lm}$. The representation provided here is simpler than the one that uses surface vector fields, since
VSH (as opposed to $\bm A_{lm}$, $\bm B_{lm}$ and $\bm C_{lm}$) are orthogonal at each point. Most significantly, the expansion coefficients are calculated as scalar integrals by using simple relationships in  Eqs.~(\ref{cd}). Similar advantages hold in comparison to other solution representations in \cite{kim}.

\subsection{Leading-order frequency correction}

%\begin{eqnarray}&&
%\bm u_{lm} =  \frac{(2-l)( I_{lm}(\bm \Psi) + l I_{lm}(\bm Y))\bm \Psi_{lm}}{2l(l+1)r^l}
%\nonumber\\&&
%+( I_{lm}(\bm \Psi) + l I_{lm}(\bm Y))\bm Y_{lm}
%\frac{r^2-1}{2r^{l+2}}
%\nonumber\\&&
%+( I_{lm}(\bm \Psi) + l I_{lm}(\bm Y))\bm \Psi_{lm}
%\frac{1}{2r^{l+2}(l+1)}
%\nonumber\\&&
%\nonumber\\&&
%+\frac{I_{lm}(\bm Y)}{r^{l+2}(l+1)}((l+1)\bm Y_{lm}-\bm \Psi_{lm})+ \frac{ I_{lm}(\bm \Phi)\bm \Phi_{lm}}{l(l+1)r^{l+1}},
%\label{uge}
%\end{eqnarray}

Substituting Eqs.~(\ref{Adef})-(\ref{Bdef}) into Eqs.~(\ref{ugn})-(\ref{AB}), it can be shown
find that the linear term in $\lambda$ appears only for $l=1$, so that $\bm U^1_{lm}=\delta_{l1}\bm U^1_{1m}$. We have
\begin{eqnarray}&&\!\!\!\!\!
\bm U^1_{1m}=\frac{(1-r)}{12r^3}
( I_{1m}(\bm \Psi) + I_{1m}(\bm Y))
\nonumber\\&& \cdot
((1+r+4r^2)\bm \Psi_{1m}-2(1+r-2r^2)\bm Y_{1m}).
\label{lambda1a}
\end{eqnarray}
Thus in the case where $l=1$, projections $I_{1m}$ are non-zero and the lowest order frequency correction to the solution of the steady Stokes equations is $\sim \sqrt{\omega}$. This is similar to correction for the fundamental solution of the unsteady Stokes equations \cite{kim}. In time-domain, the correction is non-local and has the structure of the Basset memory integral \cite{kim}, see Eq.~(\ref{fdo}).

The derived result is in agreement with the study of the solution in the time domain performed in Sec. \ref{time}. There we showed that at long times the solution is fully determined by the $l=1$ component. Long-time asymptotic expansion in the time domain is determined by the low-frequency expansion in the frequency domain. Thus, the statements that low-frequency asymptotic expansion is dominated by $l=1$ term and so is the long time expansion, are equivalent.

\subsection{Next order correction}

We have using Eqs.~(\ref{Adef})-(\ref{Bdef}) in Eqs.~(\ref{ugn})-(\ref{AB})
\begin{eqnarray}&&
\bm U^2_{lm}=a_{lm}^{Y}\bm Y_{lm}\!+\! a_{lm}^{\Psi}\bm \Psi_{lm} \!+  \!a_{lm}^{\Phi}\bm \Phi_{lm}
\end{eqnarray}
where we the coefficients $a_{lm}^{Y,\Phi,\Psi}$ are functions of radial coordinate $r$ (whose domain of definition for $a^{\Psi}$ and $a^{\Phi}$ is $l>0$ due to $\bm \Psi_{00}=\bm \Phi_{00}=0$) given by:
\begin{eqnarray}&&
a_{lm}^{Y}\! = \!-\left(\frac{(r^2-1)^2}{8(2l-3)}+\frac{\delta_{l1}(r-1)^2(1+2r)}{6}\right)
\\&&
\cdot \frac{( I_{lm}(\bm \Psi) \!+\! l I_{lm}(\bm Y))}{r^{l+2}};\ \ a_{lm}^{\Phi} = \frac{(1-r^2) I_{lm}(\bm \Phi)}{2r^{l+1}l(l+1)(2l-1)},
%+\frac{I_{lm}(\bm Y)}{r^{l+2}},
\nonumber\\&&
a_{lm}^{\Psi} \!=\!  ( I_{lm}(\bm \Psi)\! +\! l I_{lm}(\bm Y))
\left(\frac{\delta_{l1}(1\!-\!r)}{rl(l\!+\!1)} \!+\!\frac{(1\!-\!r^2)}{2r^{l}l(l\!+\!1)(2l\!-\!3)}
%\label{APsi}\\&&
%\left.
\right.\nonumber\\&&\left.
+\frac{(r^2-1)^2}{8(2l-3)r^{l+2}(l+1)}+\frac{\delta_{l1}(r-1)^2(1+2r)}{6r^{l+2}(l+1)}
\right).\nonumber
\end{eqnarray}
In the time-domain this sub-leading (and linear in frequency, $\sim \omega$) correction is local, see Eq.~(\ref{fdo}).

\section{Flow due to rigid motions of sphere} \label{oscillating}

The problem of the flow due to an oscillatory small-amplitude motion (translations or rotations) of a rigid sphere was previously solved, see e.g. \cite{kim}. We rederive here the solution from our general solution with the purpose of both validating the general solution and demonstrating which terms in the general decomposition given by Eq.~(\ref{su}) are induced by the coordinate-independent boundary flow.

\subsection{Periodic translations}

The well-known closed-form solution of the unsteady Stokes equations (only a limited number of such analytical solutions is available) is for the flow due to oscillatory translations of a rigid sphere in an unbounded viscous fluid, which was originally derived by Stokes (see e.g., \cite{kim}). Here, the flow on the sphere is given in the frequency-domain by a constant complex vector $\bm U$. In this problem, both the surface divergence and radial component of vorticity at $r\!=\!1$ vanish. Thus, we find from Eq.~(\ref{cd}) that
\begin{eqnarray}&&\!\!\!\!\!\!\!
\int \bm u\cdot \bm Y_{lm}^* d\Omega=\int  U_r Y_{lm}^* d\Omega,\ \ \int \bm u\cdot \bm \Phi_{lm}^* d\Omega=0,
\nonumber\\&&\!\!\!\!\!\!\!
\int \bm u\cdot \bm \Psi_{lm}^* d\Omega=
2\int  U_r Y_{lm}^* d\Omega.
\end{eqnarray}
Thus, finding the coefficients reduces to the calculation of
\begin{eqnarray}&&
\int  U_r Y_{lm}^* d\Omega=U_x\int \sin\theta\cos\phi Y_{lm}^* d\Omega
\nonumber\\&&
+U_y\int \sin\theta\sin\phi Y_{lm}^* d\Omega+U_z\int \cos\theta Y_{lm}^* d\Omega. \label{yc}
\end{eqnarray}
The details of the straightforward, yet cumbersome, calculations are provided in Appendix \ref{tr}. It is found that the flow is determined by the $l\!=\!1$ term of the series solution, i. e., $\bm u_{lm}$ in Eq.~(\ref{ugen}) is proportional to $\delta_{l1}$ (similarly to the counterpart problem of the steady Stokes flow \cite{kim}). The solution has the general structure provided in Eq.~(\ref{su}) with
\begin{eqnarray}&&
p(\omega, \bm x)\!=\!\left(1\!+\!\lambda\!+\!\frac{\lambda^2}{3}\right)\frac{3\bm U\!\cdot \!\bm r}{2r^3},\ \ X=0, \\&&
\bm u^H\!=\!\frac{3(1\!+\!\lambda r)\left(\bm U\!-\!3(\bm U\!\cdot\! \bm{\hat r})\bm{\hat r}\right)}{2\lambda^2 r^3}+
\frac{3\left(\bm U\!-\!(\bm U\!\cdot\! \bm{\hat r})\bm{\hat r}\right)}{2r}.\nonumber
\end{eqnarray}
Thus, time-periodic translations of the sphere generate the flow with zero toroidal component. The agreement with the known solution is confirmed in Appendix \ref{tr}. The surface traction associated with the oscillations is given by the value of $\sigma_{ik}{\hat r}_k$ at the sphere's surface where $\sigma_{ik}$ the stress tensor \cite{maxeyriley,kim}
\begin{eqnarray}&&\!\!\!\!\!\!\!\!\!\!\!\!\!
\sigma_{ik}\!\equiv\! - p\delta_{ik}\!+\!\frac{\partial v_i}{\partial x_k}\!+\!\frac{\partial v_k}{\partial x_i}.\label{dtr}
\end{eqnarray}
It is found by direct calculation that
\begin{eqnarray}&&\!\!\!\!\!\!\!\!\!\!\!\!\!
\bm\sigma \cdot \bm {\hat r}=-\frac{3(1+\lambda)\bm U+\lambda^2 (\bm U\cdot \bm {\hat r})\bm {\hat r}}{2}. \label{tra}
\end{eqnarray}
In the limit of $\lambda\!\to\! 0$, this reproduces the constant traction of the Stokes flow $-3\bm U/2$ whose surface integral gives the Stokes force $-6\pi \bm U$. Integrating the surface traction in Eq.~(\ref{tra}) over the sphere's surface gives the well-known force $\bm F$ that the fluid exerts on the oscillating sphere \cite{kim}
\begin{eqnarray}&&\!\!\!\!\!\!\!\!\!\!\!\!\!
\bm F=-6\pi(1+\lambda+\lambda^2/9) \bm U. \label{ford}
\end{eqnarray}
Here the first term corresponds to pseudo-Stokes (zero-frequency) force, the second term to the Basset (memory) force and the last term to the added mass (high-frequency) contribution to the force.

%We find collecting the terms that
%\begin{eqnarray}&&
%\bm u_s\!=\!\frac{3\exp\left(-\lambda (r-1)\right)}{2\lambda^2}
%\nonumber\\&&
%\times \left(\frac{(1\!+\!\lambda r)\left(\bm U\!-\!3(\bm U\!\cdot\! \bm{\hat r})\bm{\hat r}\right)}{r^3}+
%\frac{\lambda^2\left(\bm U\!-\!(\bm U\!\cdot\! \bm{\hat r})\bm{\hat r}\right)}{r}\right). \label{fis}
%\end{eqnarray}
%given by
%\begin{eqnarray}&&
%\bm u_s\!=\!\frac{3\exp\left(-\lambda (r-1)\right)}{2\lambda^2}
%\\&&
%\times \left(\frac{(1\!+\!\lambda r)\left(\bm U\!-\!3(\bm U\!\cdot\! \bm{\hat r})\bm{\hat r}\right)}{r^3}+
%\frac{\lambda^2\left(\bm U\!-\!(\bm U\!\cdot\! \bm{\hat r})\bm{\hat r}\right)}{r}\right).\nonumber
%\end{eqnarray}
%
%\subsection{Surface traction}
%
%We

\begin{figure}
\includegraphics[width=0.9\columnwidth]{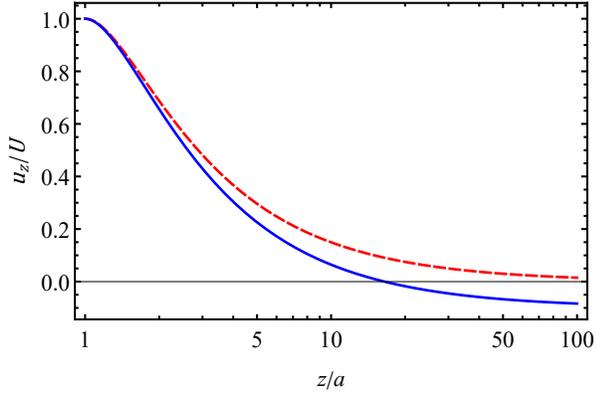}
\caption{Axial velocity component, $u_z/U$, plotted vs. the scaled distance $z/a$ (semi-log plot). The red (dashed) curve corresponds to the Stokes flow around the steadily translating sphere of radius $a$. The blue (solid) line depicts the real part of $u_z/U$ for the oscillatory translation with $\rRe\ \lambda\!=\!0.1$. Notice that already at $z/a\!=\!3$ the two velocities differ by more than $10$~\%.}
\label{fig1}
\end{figure}

\textit{Small-frequency expansion.}
It is of interest to study the small-frequency expansion of the above solution for the flow. After tedious yet straightforward calculation the small $\lambda r$ expansion yields:
\begin{eqnarray}
\bm u\! &=& \!\frac{3(2\!+\!2\lambda\!+\!\lambda^2)\left(\bm U\!+\!(\bm U\!\cdot\! \bm{\hat r})\bm{\hat r}\right)}{8r}
\nonumber \\
&&-\lambda(1\!+\!\lambda)\bm U\!+\! (2\!+\!\lambda)^2\, \frac{(\bm U\!-\!3(\bm U\!\cdot\! \bm {\hat r})\bm {\hat r})}{16r^3} \nonumber \\
&& \!+\!  \frac{3 r\lambda^2(3 \bm U\!-\!(\bm U\!\cdot\! \bm {\hat r})\bm {\hat r})}{16}\!+\!O\left[(\lambda r)^3\right]. \label{sa}
\end{eqnarray}
Notice that the expansion is singular and the expansion parameter is proportional to the square root of the frequency and not the frequency itself, cf. above. It can be readily verified that the flow obeys $\bm u(r\!=\!1)=\bm U$ with $O\left[(\lambda r)^3\right]$ accuracy. At $\lambda\!=\!0$ the flow reduces to the seminal steady Stokes flow $\bm u^\mathrm{St}$, which is induced by the particle moving steadily with a velocity $\bm U$ through the unbounded viscous fluid quiescent at infinity. That is given by \cite{kim,hb,LL,Lamb}
\begin{eqnarray}&&
\bm u^\mathrm{St}\!=
\!\frac{3\left(\bm U\!+\!(\bm U\!\cdot\! \bm{\hat r})\bm{\hat r}\right)}{4r}+\frac{\bm U\!-\!3(\bm U\!\cdot\! \bm {\hat r})\bm {\hat r}}{4r^3}.
\end{eqnarray}
To linear order in $\lambda$ there is a simple relation between the steady and unsteady Stokes flows:
\begin{eqnarray}&&
\bm u\!=(1\!+\!\lambda)\bm u^\mathrm{St}-\lambda\bm U, \label{las}
\end{eqnarray}
see Eq.~(\ref{sa}). Thus the leading-order contribution of weak unsteadiness is two folds: it modifies the steady $\bm u^\mathrm{St}$-flow via the multiplicative factor $(1+\lambda)$ and adds a constant velocity equal to $-\lambda \bm U$. The latter constant velocity has no conflict with the condition at infinity, as Eq.~(\ref{las}) only valid in a finite spatial region with $r\ll |\lambda|^{-1}$. The validity of $\bm u(r=1)=\bm U$ follows instantly from Eq.~(\ref{las}). The corresponding streamlines are given by isolines of streamfunction, $\psi$, defined by
\begin{eqnarray}&&
u_r=\frac{1}{r^2\sin\theta}\frac{\partial \psi}{\partial\theta},\ \ u_{\theta}=-\frac{1}{r\sin\theta}\frac{\partial \psi}{\partial r},
\end{eqnarray}
where we use spherical coordinates with polar angle $\theta$ measured from the direction of $\bm U$. We therefore have
\begin{eqnarray}&&\!\!\!\!\!\!\!\!
\psi=(1\!+\!\lambda) Ur^2\sin^2\theta\left(\frac{3}{4r}-\frac{1}{4r^3}\right)-\frac{\lambda U r^2 \sin^2\theta}{2}.
\end{eqnarray}
It can be readily seen that the condition $|\lambda|\ll 1$ warrants that the correction to $\bm u^\mathrm{St}$ remains small at $r\ll |\lambda|^{-1}$. The low-frequency expansion breaks down at $|\lambda|\sim 1$, where the correction becomes of the same order as the leading order term and the volume of the region where the steady Stokes flow applies shrinks to zero. However, care must be exercised as far as the actual numerical values are concerned. As an example, Fig.~\ref{fig1} depicts the scaled flow component along the $z$-axis, $u_z/U$, as a function of the distance along the $z$-axis, which reads
\begin{eqnarray}&&
\frac{u_z(z)}{U}\!=\frac{3(1\!+\!\lambda)}{2z}-\frac{1\!+\!\lambda}{2z^3}-\lambda.
\end{eqnarray}
It follows from Fig. \ref{fig1}, that already at the distance of $3$ sphere radii, the magnitudes of the steady Stokes flow, that holds at $\lambda\!=\!0$, and the unsteady Stokes flow that corresponds to $\rRe\ \lambda\!=\!0.1$, differ appreciably.

The uniform velocity correction, $-\lambda \bm U$, could lead to a steady drift which may not be negligible in applications. A similar correction holds for the expansion of Green's function of the unsteady Stokes equations \cite{kim}. Remarkably, uniform correction to the Stokes flow is also induced by a small, but finite stratification \cite{fl}. The constant velocity term requires attention since in presence of many particles it could result in a collective flow. For instance, consider a dilute solution of $N$ particles in a fluid volume of a size much smaller than $1/|\lambda|$ (it requires that $|\lambda|^{-1}\gtrsim 100$ to allow for large distances between the particles). Since the inter-particle distance $r\gg 1$, then at the leading order i$^\mathrm{th}$ particle induces uniform flow $-\lambda \bm U_i$ at the locations of other particles. Therefore, if the particles oscillate in-sync in the same direction, for instance, if their oscillatory velocities coincide, $\bm U_i=\bm U$, then there is constructive interference of the flows induced by different particles. The total flow induced by other particles at a position of any single particle may no longer be a small correction to the steady flow and it may result in coherent collective motion (drift) of all particles. The study of such collective motion is beyond the scope of the present study.

\subsection{Oscillatory rotation of a sphere} \label{oscillatory}

Let us now consider another well-known solution of the unsteady Stokes equations for the flow due to oscillatory rotation of a sphere \cite{kim}. The solution is an example of a purely toroidal flow. In the frequency-domain, the flow at the surface of the unit sphere is $\bm \omega\times  \bm {\hat r}$, where $\bm \omega$ is a constant vector of angular velocity. We find readily from Eq.~(\ref{sp}), that the continuation of the surface velocity into the whole space, $\bm u=\bm \omega\times  \bm r$, satisfies $\nabla_s\cdot\bm u=\nabla\cdot \bm u=0$. The radial component of vorticity obeys at the surface $(\nabla\times \bm u)_r=2 \omega_r$. We conclude from Eqs.~(\ref{cd}) that the only non-zero projection is
\begin{eqnarray}&&\!\!\!\!\!\!\!
\int \bm u\cdot \bm \Phi_{lm}^* d\Omega\!=\!-2\int_{r=1}  Y_{lm}^* \omega_r d\Omega\!=\!-2\int_{r=1}  Y_{lm}^* d\Omega
\nonumber\\&&\!\!\!\!\!\!\!\times \left(\omega_x \sin\theta\cos\phi
+\omega_y \sin\theta\sin\phi+\omega_z \cos\theta\right) .
\end{eqnarray}
Calculations in Appendix \ref{ur} demonstrate that the solution has the general form given by Eq.~(\ref{su}). Both $p$ and $\bm u^H$ vanish and the flow is toroidal
\begin{eqnarray}&&
\bm u\!=\!\frac{(1\!+\!\lambda r)\bm \omega\times\bm r}{(1\!+\!\lambda)r^3}e^{-\lambda(r-1)}=\nabla\times (\bm r e^{\lambda (1-r)}X ),\nonumber\\&&
X=\frac{(1\!+\!\lambda r)\bm \omega\cdot\bm r}{(1\!+\!\lambda)r^3}e^{-\lambda(r-1)}. \label{ror}
\end{eqnarray}
This agrees with the solution provided in \cite{LL}. The surface traction is
\begin{eqnarray}&&\!\!\!\!\!\!\!\!\!\!\!\!\!
\bm\sigma\cdot \bm {\hat r}=-\bm \omega\times\bm {\hat r} \left(3+\frac{\lambda^2}{1+\lambda}\right),  \label{trar}
\end{eqnarray}
see e. g. \cite{for}. The torque is readily found by integrating Eq.~(\ref{trar}) over the sphere surface to be:
\begin{eqnarray}&&\!\!\!\!\!\!\!\!\!\!\!\!\!
\bm T=-8\pi\bm \omega-\frac{8\pi\bm \omega \lambda^2}{3(1+\lambda)}\,, \label{toq}
\end{eqnarray}
in a complete agreement with \cite{kim}. The torque dependence on $\lambda$ in this case is more complex than that for the force in Eq.~(\ref{ford}).

\section{Far-field behavior of the general solution}

The decomposition $\bm u=\bm u_s-\lambda^{-2}\nabla p$ introduced in Eq.~(\ref{helm}) is rather useful for understanding the far-field behavior of the solutions. We find that $\bm u_s$ decays exponentially far from the sphere, cf.
Eq.~(\ref{mod}), so that
\begin{eqnarray}&&
\bm u=-\frac{\nabla p}{\lambda^2}+O(\exp(-\lambda r)),\ \ r\gg \delta. \label{exp}
\end{eqnarray}
The formula for $p$ is then provided by Eq.~(\ref{pr}) with coefficients in Eq.~(\ref{cope}). In non-degenerate case, where at least one of $c_{1m}$ is non-zero, we have
\begin{eqnarray}&&\!\!\!\!\!\!\!\!\!\!\!\!\!
p=\sum_{m=-1}^{m=1}\frac{c_{1m}Y_{1m}(\theta, \phi)}{r^{2}}+o\left(\frac{1}{r}\right),
\end{eqnarray}
where the coefficients $c_{1m}$ are given by Eq.~(\ref{cope}). By using the definition of $Y_{1m}$, it can be readily seen that to the leading approximation, the pressure has the form of the solution for the sphere oscillating with an effective velocity $\bm U_\mathrm{eff}$:
\begin{eqnarray}&&
p\!=\!\left(1\!+\!\lambda\!+\!\frac{\lambda^2}{3}\right)\frac{3\bm U_\mathrm{eff}\!\cdot \!\bm r}{2r^3},
\end{eqnarray}
where we introduced
\begin{eqnarray}&&\!\!\!\!\!\!\!\!\!\!\!\!\!
\bm U_\mathrm{eff}\equiv \sqrt{\frac{1}{6\pi}} \left(1\!+\!\lambda\!+\!\frac{\lambda^2}{3}\right)^{-1}\nabla\left(
\left(c_{1, -1}\!-\!c_{11}\right)x
\right.\nonumber\\&&\!\!\!\!\!\!\!\!\!\!\!\!\!\left.
\! -\!i\left(c_{1, -1}\!+\!c_{11}\right)y\!+\!\sqrt{2}c_{10} z\right), \label{ueff}
\end{eqnarray}
cf. Eq.~(\ref{oer}) of Appendix \ref{tr}. Thus, at large distances any velocity distribution at the boundary yields a flow due to an oscillating rigid sphere. This is true provided that $U_\mathrm{eff}$ in Eq.~(\ref{ueff})  does not become trivial for a particular velocity distribution at the boundary, in which case higher order terms must be considered.

\section{Axially symmetric case}
\label{sro}

Next, we consider the reduction of our solution to the case of axially symmetric flows
\begin{eqnarray}&&\!\!\!\!\!\!\!
\bm u\!=\!u(r, \theta)\bm {\hat r}\!+\!v(r, \theta)\bm {\hat \theta}, \label{axisym}
\end{eqnarray}
where $u$ and $v$ are the radial and polar velocity components, respectively. This solution applies if the velocity at $r\!=\!1$ can be written in the form of Eq.~(\ref{axisym}). The coefficients $c^t_{lm}$ in Eq.~(\ref{cdo})
vanish in this case, so the $X$-term in the solution given by Eq.~(\ref{su}) is zero, see Eq.~(\ref{X}). The coefficients $c_{lm}$ and ${\tilde c}^{r}_{lm}$ in Eqs.~(\ref{coper}), (\ref{tild}) vanish for $m\neq 0$. Projecting the solution given by Eq.~(\ref{su}) onto the radial direction, we find that
\begin{eqnarray}&&\!\!\!\!\!\!\!\!\!\!\!\!
u=-\sum_{l=1}^{\infty}\left(\frac{D_l(\lambda)}{r^{l+2}}+\frac{F_l(\lambda)K_{l+1/2}(\lambda r)}{r^{3/2}} \right)
{\cal P}_l(\cos\theta),\label{axs}
\end{eqnarray}
where ${\cal P}_l(x)$ are the Legendre polynomials. We used $Y_{l0}={\cal P}_l(\cos\theta)\sqrt{2l+1/(4\pi)}$ [see Eq.~(\ref{fi})] and
%****
%\begin{eqnarray}&&\!\!\!\!\!\!\!\!\!\!\!\!
%e^{\lambda}\sum_{l m} {\tilde c}^{r}_{lm} \frac{Y_{lm} }{r^{3/2}}K_{l+1/2}(\lambda r)\sqrt{\frac{2\lambda}{\pi}}
%\end{eqnarray}
%\begin{eqnarray}&&\!\!\!\!\!\!\!\!\!\!\!\!
%{\cal P}_l\!\equiv \!\sum_{k=0}^l \frac{(l\!+\!k)!\lambda^k}{k!(l\!-\!k)!2^k},\ \
%K_{l+1/2}(\lambda)\!=\!\sqrt{\frac{\pi}{2\lambda}}e^{-\lambda}{\cal P}_l(\lambda^{-1}).
%\end{eqnarray}
%***
defined the coefficients
\begin{eqnarray}&&\!\!\!\!\!\!\!\!\!\!\!\!\!\!\!
D_l\!\equiv\! -\sqrt{\frac{2l\!+\!1}{4\pi}}\frac{(l\!+\!1)c_{l0}}{\lambda^2},\ \ %\nonumber\\&&\!\!\!\!\!\!\!\!\!\!\!\!
F_l\!\equiv \!-e^{\lambda}\sqrt{\frac{(2l\!+\!1)\lambda}{2}}\frac{ {\tilde c}^{r}_{l0}}{\pi}. \label{Dl}
\end{eqnarray}
The explicit form of the coefficients is found by using Eqs.~(\ref{coper}), (\ref{tild}) for $c_{l0}$ and ${\tilde c}^{r}_{l0}$. Calculations provided in Appendix \ref{as} result in
\begin{eqnarray}&&
D_l\!=\! -\frac{(2l\!+\!1)\left(l P_l(\lambda)
\!+\!\lambda^2 P_{l-1}(\lambda)\right)}{2\lambda^2 P_{l-1}(\lambda)}\int_{-1}^1\!\!  {\cal P}_l(x)  u(x) dx
\nonumber\\&&
-\frac{(2l+1)P_l(\lambda)}{2\lambda^2 P_{l-1}(\lambda)}\int_{-1}^1\!\!  P_l^1(x)  v(x) dx.\label{dl}
\end{eqnarray}
The corresponding formula for $F_l$ can be readily derived from the relation between $c_{lm}$ and ${\tilde c}^{r}_{lm}$ by Eq.~(\ref{coper}). The obtained solution reproduces the known solution of Rao \cite{rao}, see details in Appendix~\ref{as}.

\section{Ideal flow approximation and high-frequency expansion}

In this section we shall demonstrate that the ideal flow approximation, which is often taken for granted, can be in fact rigorously derived from our solution, including the corrections in small, but finite viscosity. The calculations are done in the frequency domain, and then we comment on the form of the results in the time domain.

We demonstrated already that the flow outside the viscous layer with thickness of order of the penetration depth $\delta$ is potential, up to exponentially small corrections in $|\lambda|(r-1)$, cf. Eq.~(\ref{su}). We have the generalized Darcy's law $\bm u=\nabla \psi$ with $\psi=-p/\lambda^2$. The ideal flow limit of small viscosity corresponds to {$|\lambda|\rightarrow \infty$, see the definition in Sec. \ref{st}, where $\delta$ becomes vanishingly small. Thus, in the limit the flow becomes potential everywhere in accord with the ideal flow approximation \cite{bat}. At small but finite $\nu$, the flow is potential outside a thin viscous layer of thickness $\delta$ around the sphere.

We approach the limit of large $|\lambda|$ in two steps. We first neglect the exponentially small corrections and assume $\bm u=\nabla \psi$, i.e. concentrate on the region outside the viscous layer. Then we consider the asymptotic behavior of the potential $-p/\lambda^2$ at large $|\lambda|$. That is given by a power series in $|\lambda|^{-1}$. We have (cf. Eq.~(\ref{exp}))
\begin{eqnarray}&&
\bm u=\nabla \psi,\ \ \psi=\psi_0+\frac{\psi_1}{\lambda}+\frac{\psi_2}{\lambda^2}+\ldots. \label{id}
\end{eqnarray}
The asymptotic expansion is found by making the expansion of the pressure coefficients $c_{lm}$ in Eq.~(\ref{pr}) with respect to $|\lambda|$. We rewrite Eq.~(\ref{cope}) as
\begin{eqnarray}&&
c_{lm}\!=\!\frac{ \lambda^2 b_{lm}}{l\!+\!1}+\frac{P_l(\lambda)\left((l+2)b_{lm}-d_{lm}\right)}{(l\!+\!1) P_{l-1}(\lambda)}, \label{sak}
\end{eqnarray}
where we introduced the coefficients
\begin{eqnarray}&&\!\!\!\!\!\!\!\!\!
b_{lm}\equiv \int_{r=1}Y_{lm}^* u_r d\Omega,\ \  d_{lm}\equiv \int  Y_{lm}^* \nabla_s\!\cdot\!\bm u d\Omega. \label{b}
\end{eqnarray}

\subsection{Ideal flow approximation}

From Eq.~(\ref{mod}) it follows that in the limit $|\lambda|\rightarrow \infty$
\begin{eqnarray}&&
\frac{P_l(\lambda)}{P_{l-1}(\lambda)}=\lambda+l+O\left(\frac{1}{\lambda}\right).
\end{eqnarray}
Using this, we find that the second term in the RHS in Eq.~(\ref{sak}) is $\mathcal{O}(|\lambda|)$.
%\begin{eqnarray}&&\!\!\!\!\!\!\!\!\!
%c_{lm}\!=\! \frac{ \lambda^2 b_{lm}}{l\!+\!1}+O\left(|\lambda|\right). \label{cb}
%\end{eqnarray}
Thus, to the leading approximation we have (we assume that $|d_{lm}|\ll |\lambda b_{lm}|$)
\begin{eqnarray}&&
\bm u=\nabla \psi_0,\ \ \psi_0=-\sum_{l m}\frac{b_{lm}Y_{lm}(\theta, \phi)}{(l+1)r^{l+1}}. \label{idl}
\end{eqnarray}
Therefore, in the limit of vanishing viscosity the solution is fully determined by the normal component of the velocity $u_r$. In fact, we have
\begin{eqnarray}&&\!\!\!\!\!
\partial_r\psi_0(r=1)=\sum_{l m}Y_{lm}(\theta, \phi)\int_{r=1}Y_{lm}^* u_r d\Omega,
\end{eqnarray}
which demonstrates that the normal component of the solution provided by Eq.~(\ref{idl}) coincides with $u_r$ on the sphere's surface. The above formulae hold without change also in the time domain as seen by performing the inverse Fourier transform (it is assumed that $b_{lm}$ and $d_{lm}$ considered as functions of frequency give dominant weight to high frequencies). We have thus recovered the ideal flow representation \cite{LL}.

\subsection{Corrections to ideal flow approximation}

We shall now consider corrections to the ideal flow solution due to small but finite viscosity. Expanding Eq.~(\ref{sak}) for $|\lambda|\rightarrow \infty$ and keeping the first two terms give:
\begin{eqnarray}&&\!\!\!\!\!\!
c_{lm}\!=\! \frac{ \lambda^2 b_{lm}}{l\!+\!1}\!+\!\frac{(\lambda\!+\!l) \left((l\!+\!2)b_{lm}\!-\!d_{lm}\right)}{l\!+\!1}\!+\!O\left(\frac{1}{|\lambda|}\right). \label{cb}
\end{eqnarray}
We find that the corrections to the potential $\psi_1$ and $\psi_2$ in Eq.~(\ref{id}) are given by
\begin{eqnarray}&&\!\!\!\!\!\!\!\!\!
\psi_1=\sum_{l m}\frac{\left(d_{lm}\!-\! (l\!+\!2)b_{lm}\right)Y_{lm}(\theta, \phi)}{(l\!+\!1)r^{l+1}}\,,
\\&&\!\!\!\!\!\!\!\!\!
\psi_2=\sum_{l m}\frac{l\left(d_{lm}\!-\! (l\!+\!2)b_{lm}\right)Y_{lm}(\theta, \phi)}{(l\!+\!1)r^{l+1}}=-\partial_r(r\psi_1)\nonumber.
\end{eqnarray}
Several insights concerning the correction to the ideal flow could be mentioned here. It is proportional to square root of the viscosity rather than the viscosity. It involves tangential components of the surface velocity via the $d_{lm}$ coefficients. Finally, as opposed to the ideal flow approximation, the leading order correction is not local in time. Applying an inverse Fourier transform, the flow in the time-domain at next orders has memory. We have
\begin{eqnarray}&&
\psi_1\!=\!\sum_{l m}\!\frac{Y_{lm}(\theta, \phi)}{(l\!+\!1)r^{l+1}}\int_{-\infty}^t \!\!\frac{ \left(d_{lm}(t')\!-\! (l\!+\!2)b_{lm}(t')\right)dt'}{a\sqrt{\pi(t\!-\!t')/\nu}},\nonumber
\\&&
\psi_2\!=\!\sum_{l m}\frac{\nu lY_{lm}(\theta, \phi)}{(l\!+\!1)a^2 r^{l+1}}\int_{-\infty}^t \left(d_{lm}(t')\!-\! (l\!+\!2)b_{lm}(t')\right)dt'.\nonumber
\end{eqnarray}
The memory kernel at the leading order coincides with that in the Basset force \cite{kim}. Remarkably, at the next order there is no decay of memory: all times have equal weight in the integral defining $\psi_2$. We should recall though, that the derivation assumes that $b_{lm}(t)$ and $d_{lm}(t)$ are fast oscillating functions, so the integral in $\psi_2$ is determined by vicinity of $t$ of order of the characteristic period of oscillations.

\subsection{Force and d'Alembert paradox} \label{focid}

The ideal flow approximation implies that the force exerted on the sphere by the fluid is determined solely by the isotropic component (pressure) of the stress tensor and is given by
\begin{eqnarray}&&\!\!\!\!\!\!
\bm F_\mathrm{ideal}=-\int_{r=1} p\bm{\hat r} d\Omega=- \sum_{lm}c_{lm}\int_{r=1} Y_{lm}(\theta, \phi) \bm{\hat r} d\Omega
\nonumber\\&&\!\!\!\!\!\!
=- \sum_{m}c_{1m}\sqrt{\frac{2\pi}{3}}  \left(\bm{\hat x}(\delta_{m, -1}-\delta_{m, 1})\!-\! i\bm{\hat y}(\delta_{m, -1}+\delta_{m, 1})
\right.\nonumber\\&&\!\!\!\!\!\!\left.
+\bm{\hat z}\sqrt{2}\delta_{m0}\right),
\end{eqnarray}
where we used Eq.~(\ref{fsdp}) from Appendix~\ref{formulae}. It can also be shown (see Appendix~\ref{formulae}) that the full force obeys quadratic dependence on $\lambda$ and can be written as
\begin{eqnarray}&&\!\!\!\!
\bm F\!=\!-\lambda^2\sqrt{\frac{\pi}{6}} \left(\bm{\hat x}(b_{1, -1}\!-\!b_{11})\!-\! i\bm{\hat y}(b_{1, -1}\!+\!b_{11})\!+\!\bm{\hat z}\sqrt{2}b_{10}\right)
\nonumber\\&&\!\!\!\!
+(1\!+\!\lambda)\sqrt{\frac{3\pi}{2}} \left(\bm{\hat x}(d_{1, -1}\!-\!3b_{1, -1}\!-\!d_{11}\!+\!3b_{11})
 \right.\nonumber\\&&\!\!\!\!\left.
- i\bm{\hat y}(d_{1, -1}\!-\!3b_{1, -1}\!+\!d_{11}\!-\!3b_{11})\!+\!\bm{\hat z}\sqrt{2}\left(d_{10}\!-\!3b_{10}\right)\right).\label{fer}
\end{eqnarray}
Comparison of the last two equations, using Eq.~(\ref{cb}), demonstrates that in the leading (quadratic in $\lambda$) order, $\bm F$ and $\bm F_\mathrm{ideal}$ coincide. At higher orders in $\lambda$, $\bm F$ has also the viscous stress contributions, see Appendix \ref{formulae}.

It can be seen that for a steadily moving particle, the coefficients $b$ and $d$ are proportional to $\delta(\omega)$. The force in the ideal flow approximation, which is given by the first term in Eq.~(\ref{fer}), is then proportional to $\lambda^2\delta(\omega)$. It vanishes regardless of how small the viscosity is, which is a manifestation of the d'Alembert paradox. The origin of the paradox is in the degenracy of the support of $b$ and $d$ in the frequency domain that concentrates at zero frequency. In contrast, in cases where the support includes non-zero frequencies, the force in Eq.~(\ref{fer}) is dominated by the first term in the limit of $\nu \to 0$ (that implies $|\lambda|\to \infty$). The force is then given by the ideal flow approximation, provided that the viscosity is small enough. Finally, for steadily moving particle the paradox is resolved, (see, e.g. \cite{LL}) by inclusion of the finite viscosity which is equivalent to inclusion of all terms in Eq.~(\ref{fer}). Then, for $b$ and $d$ proportional to $\delta(\omega)$, we obtain from Eq.~(\ref{fer}) the usual Stokes force.

\section{Application to microswimming} \label{mco}

In this Section, we discuss application of our general solution to the problem of self-propulsion of some microorganisms, such as e.g., Volvox, at low Reynolds numbers. Volvox is a nearly spherical colony of cells that self-propels by effectively moving its surface that at microscopic level is comprised of beating flagella \cite{volvox}. The time-varying deformation of the surface is expressed in body-fixed spherical coordinates as $r=a+\epsilon a f(\theta, \phi, t)$ where $a$ is a mean radius, $\epsilon\ll 1$ and $|f|\sim 1$. In the leading order in $\epsilon$ and dimensionless variables, this results in the velocity boundary condition on the unit sphere
\begin{eqnarray}&&\!\!\!\!\!\!\!
\bm u(\theta, \phi, t)\!=\!\bm v\!+\!\bm \Omega\!\times\!\bm r\!+\!\bm u_s(\theta, \phi, t).
\end{eqnarray}
Here $\bm v$ and $\bm \Omega$ are the translational and angular velocities of the swimmer to linear order in $\epsilon$, to be determined from the solution, and $\bm u_s(\theta, \phi, t)$ is the swimming stroke i.e., the surface velocity in the co-moving frame affixed with the swimmer. The swimmer is assumed to control its shape via the swimming stroke, as its shape and surface velocity determine the interaction of the organism with the fluid via no-slip boundary conditions. This interaction generates hydrodynamic force and torque exerted on the swimmer which determine the resulting translation, $\bm v$, and angular, $\bm \Omega$, velocities. Thus, besides the Navier-Stokes equations with the no-slip boundary conditions, the full problem of modeling the motion also includes equation of motion of the swimmer, i.e., balance of linear and angular momentum. Here we are interested in the flow component of the motion considering $\bm v$ and $\bm \Omega$ as given.

The Reynolds number associated with the motion of Volvox is small, which allows to drop the non-linear term in the Navier-Stokes equations. On the other hand, the unsteady term in the equations, which in most classical studies is dropped \cite{lighthill}, is not necessarily small. It was stressed in \cite{for}, that for colonies whose radius $a$ is larger than $300$~$\mu$m the Roshko number is not small and the unsteady term in the Navier-Stokes equations cannot be neglected. Typically, the swimming stroke is periodic in time, so that $\bm u_s(\theta, \phi, t)$ can be written as the corresponding Fourier series. The resulting solution for $\bm v$ and $\bm \Omega$ in this order is also time-periodic and the flow is also given by a Fourier series, cf. \cite{Ishimoto}. Therefore, the developed solution
can be obtained directly to determine the resulting flow around the swimmer. Notice that it is most common to assume that the stroke and the resulting flow are axisymmetric, resulting in net translation along the symmetry axis without net rotation. However, there are at least some cases where the stroke is not axisymmetric resulting in net rotation of the swimmer. For example, the axially symmetric solution of unsteady Stokes equations presented in \cite{Ishimoto} does not apply to the phenomenon of phototaxis \cite{photo}, and one has to consider the more general situation. So far non axisymmetric problems with finite Roshko number have been only considered in \cite{for}, showing the feasibility of swimming via translation-rotation coupling. However, \cite{for} limited the study to tangential squirmers, whereas the swimming stroke of Volvox involves radial surface waves.

Finally, we remark that in the linear order in $\epsilon$ the swimmer's velocity is a time-periodic function, so the mean swimming velocity vanishes. The finite net swimming velocity arises only when considering the $\epsilon^2$-order problem. The study of that order would require the general solution at the linear order, and its detailed analysis will be published elsewhere.

%%%%%%%%%%%%%%%%%%%%%%

\section{General solution of the Brinkman equations} \label{unBr}

The analysis presented in this paper can be applied directly to the study of the Brinkman equations describing the flow in a porous medium at low volume fraction of solids \cite{r,dru}:
\begin{eqnarray}&&\!\!\!\!
\frac{\nu \bm v }{k}+\rho^{-1} \nabla p=\nu\nabla^2\bm v.
\end{eqnarray}
Here $k$ is the permeability \cite{r}, and it reduces to the Darcy's equation, $\bm v=-(k/a^2)\nabla p$, at length scales much larger than the so-called ``screening length", $\sqrt{k}$, and to the steady Stokes equation at length scales much smaller than the screening length (see \cite{modl} for recent discussion of the coefficients and more references).

We find that Eq.~(\ref{ho}) provides the solution of Eqs.~(\ref{is}) also for the case when $\lambda$ is a complex number whose square is not purely imaginary. In the case of $\omega=i\nu/k$, where $\lambda^2=a^2/k$, Eq.~(\ref{is}) reduces to the Brinkman's equations. Thus, Eq.~(\ref{ho}) with $\lambda=a/\sqrt{k}$ is the general solution of these equations and all of our previous derivations apply by analytic continuation.

In fact, the general solution presented here can be used to study the more general unsteady Brinkman's equations
\begin{eqnarray}&&\!\!\!\!
\left(-i\omega+\frac{\nu}{k}\right) \bm v+\rho^{-1} \nabla p=\nu\nabla^2\bm v\,, \label{ud}
\end{eqnarray}
that apply to the time-dependent flow in porous medium \cite{cos}. It is plausible that they can actually be derived from unsteady Stokes equations in the porous medium similarly to the derivation of the usual Brinkman equations from time-independent Stokes equations, see references in \cite{dru}. Eqs.~(\ref{ud}) can also be studied by analytic continuation of our solution to a complex frequency which is neither purely real nor imaginary.

We remark that a complete solution of the time-independent Brinkman equations was proposed previously in \cite{sek} and its generalization for the unsteady case in \cite{cos}. However, the properties of the solutions are not readily available from the proposed representations. In contrast, the form proposed here makes many properties immediate as explained previously in the context of the unsteady Stokes equations.

\section{Conclusions and future work}

In the present work, we provided a general form of the solution of unsteady Stokes equations in spherical coordinates. The solution is given as a series expansion in vector spherical harmonics (VSH). The coefficients of the expansion are determined uniquely by the values of the velocity on any spherical surface. It allows to determine the flow in the unbounded viscous fluid given the arbitrary velocity distribution at the surface of a sphere. Since any problem in the exterior of the sphere can be reduced to superposition of solution in unbounded fluid and solution driven by the boundary conditions on the sphere, then the series can be used to construct the general solution for any transient Stokes flow in the presence of a sphere.

The solution in the frequency domain is constructed as superposition of (i) a particular solution, which has already been mentioned by Lamb \cite{Lamb}, and (ii) two solutions of the homogeneous Helmholtz equation with an imaginary coefficient, where one of the solutions is a toroidal field. Since the solution bears some similarity with Lamb's solution of steady Stokes equations, it is called ``Lamb-type" solution. The explicit forms of the flow field, both in frequency and time domains, and their simplifications in various limits, are the main contributions of the paper.

Our representation provides important insights into the general behavior of the flow. For instance, it can be readily seen that, in the frequency domain, at distances from the sphere that are much larger than the viscous penetration depth, the generalized Darcy's law holds and the flow is given by $\bm u=-i\nabla p/(\rho\omega)$. The pressure is a series in spherical harmonics with coefficients determined by projections of the surface radial velocity and divergence onto the spherical harmonics. To the leading approximation, the far-field flow is due to a rigid sphere undergoing time-periodic translation with some effective amplitude.

The solution is non-local in the time domain and has memory. The memory in unsteady Stokes flows is well known from the Basset history term \cite{kim}, however, the memory of the flow could also be different. There is a non-trivial dependence of the memory behavior on the order of the VSH. We demonstrate that non-toroidal $l\!=\!1$ harmonics have memory, similar to the Basset history term, cf. \cite{Ishimoto}. That memory decays inversely proportionally to square root of evolution time, $t^{-1/2}$. Thus, time integral of the memory kernel of these $l=1$ terms of the series solution diverges, indicating long memory effects. In contrast to that, memory kernel of terms with $l>1$ decays in integrable fashion, as $t^{-3/2}$. Similar decay holds for the toroidal term with $l=1$. Integrability of the memory kernel is in accord with the approximation in which the kernel is replaced by a $\delta-$function. This approximation by definition is local in time and thus has no memory. Singling out $l\!=\!1$ component with slower decaying memory, and studying it under a separate cover is quite useful, as the flow can be represented as a superposition of  the unsteady, non-local in time flow due to the $l\!=\!1$ component, and a correction which is effectively local in time.

Slower decay of memory for the $l=1$ terms of the solution implies universal long-time asymptotic form of solutions driven by boundary conditions that either vanish or decay fast starting from some moment. In this case, at large times the solution is a superposition of $\bm Y_{1m}$ and $\bm \Psi_{1m}$ terms. The coefficients of these terms are determined by collective components of the boundary velocity, such as translational or rotational velocity averaged over the surface of the sphere. We consider as an example the decay of transversal wave in the presence of a fixed sphere. The solution in unbounded fluid is an eigenmode of linearizied Navier-Stokes equations, well known in statistical physics \cite{reichl}. The sphere creates perturbation that occupies roughly a spherical region whose radius grows as in a diffusion process, proportionally to square root of the evolution time, $\sqrt{t}$. Inside this region the total flow is dominated by the perturbation and is coordinate-independent. The flow amplitude behaves as $t^{-3/2}$ satisfying the net momentum conservation. This flow constitutes the superposition of $\bm Y_{1m}$ and $\bm \Psi_{1m}$ terms in the considered case.

The Darcy's law in a time domain constitutes an interesting result of our paper. The flow outside an appropriately defined viscous layer obeys the reduced version of unsteady Stokes equations, $\rho \partial_t\bm v=-\nabla p$. The Laplacian (viscous) term of the unsteady Stokes equations is negligible and the flow is potential, $\bm v=\nabla\phi$. We provide the formula for $\phi$ in terms of the boundary velocity, solving in this way the boundary layer problem of matching the flow outside the layer with the flow at the surface. The pressure is found via $p=-\rho\partial_t\phi$.

The Darcy's law provides a simple way to understanding the ideal flow approximation and deriving its corrections. That approximation holds in the limit of small viscosity or, equivalently, high frequency. In this limit, the width of the viscous layer (penetration depth) is much smaller than the sphere's radius, so that the flow is potential almost everywhere outside the sphere. Potentiality of the flow is a main characteristics of the ideal flow approximation, where the value of the potential is fixed by the prescribed value of the normal velocity at the sphere's surface. That condition is usually not derived rigorously, but rather postulated \cite{LL,bat}. Here, we actually derive this condition rigorously. We demonstrate that the potential $\phi$ is given by a singular asymptotic expansion in $1/\sqrt{Ro}$, where $Ro=a^2\omega/\nu$ is the dimensionless Roshko number, so that $1/\sqrt{Ro}$ is proportional to the square root of viscosity or, alternatively, the inverse of the square root of frequency.
We demonstrate that the leading-order term of the expansion coincides with the ideal flow approximation. This consistent derivation allows also to derive higher-order corrections, which could be non-local in time. The memory that characterizes the leading order correction is similar to that in the Basset history term \cite{kim}, however in the next order the flow memory is different.

The complementary low-frequency (high-viscosity) limit is given by the singular asymptotic series expansion in $\sqrt{Ro}$. The leading-order term is the well-known Lamb's solution representation of the steady Stokes equations. We also provide an alternative form of the solution of the steady Stokes equations, similar to that derived by the adjoint method \cite{kim}. The sub-leading order correction, proportional to $\sqrt{Ro}$ is noteworthy - it contains only the first-order VSH with $l\!=\!1$. The correction is not local in the time-domain and has memory similar to the Basset history term. The corresponding conclusion regarding the long-time memory being determined by the $l=1$ term was derived above by analyzing solution series in the time domain. The next order correction  proportional to $Ro$ is linear in frequency. This conclusion is local in the time-domain and determined by instantaneous rate-of-change of the surface velocity (the same as the ``added mass" contribution to the force exerted on the particle, see e.g. \cite{LL}).

%Apart from important qualitative insights described above, the explicit form of general solution derived in this paper can be used for solving for unsteady Stokes flows originated by an arbitrary boundary velocity. \red{I believe that the previous sentence can be removed, please check}
We provided two specific examples where the solution is applied to the classical problems of time-periodic motions (translation and rotation) of the rigid sphere. We used the problem of an oscillating rigid sphere to demonstrate that low-frequency corrections might be qualitatively and quantitatively relevant already at a quite low frequency. We also outlined the solution scheme for unsteady flow around a spherical squirmer.
Spherical squirmer is one of the most popular models for low-Reynolds number self-propulsion \cite{lighthill}, which describes, e.g., realistic Volvox colonies \cite{volvox}. For large colonies the usage of unsteady Stokes equations is necessary \cite{for}. Volvox swims by periodically changing its shape that remains nearly spherical. The flow is then determined by the boundary conditions on the moving envelope which, to the leading approximation, reduces to time-periodic velocity at the spherical boundary (recall that in the model of squirmers, whose surface motions are purely tangential, the boundary conditions apply at the spherical boundary and our solution can be used directly without any approximations). The proposed general solution thus can be applied to solve for motion and flow around realistic spherical squirmers (see e.g., \cite{photo}).

We also explained how our formalism can be used to describe the problem of flow decay in presence of a spherical particle. This can be relevant in the problem of initial flow decaying in a dilute solution of particles.

A very common situation concerns a rigid no-slip sphere freely suspended in an external transient Stokes flow, e.g., a shear flow. Representing the solution as the superposition of the external flow and the perturbation due to the particle presence, it can be shown that the solution for the perturbation flow (satisfying non-trivial velocity boundary condition due to rigid-body motion corrected for the external flow) can be readily constructed using the proposed representation. The extra conditions due to the force and torque balance can be readily applied using the formulae for the force and the torque derived in Appendix \ref{formulae}.

It is of interest to study the analytic properties of the solution as a function of the frequency considered as a complex number. This allows to provide immediately the general solution for the Brinkman equations and might also be extended to obtain the solution of the so-called transient Brinkman equations \cite{pala}.

One can also consider the problem of unsteady motion of a slightly deformed sphere. Perturbation theory in small deviations from sphericity leads to the problem of unsteady Stokes equations with non-trivial boundary conditions at the surface of a sphere, similarly to the analogous steady Stokes flow problem \cite{brsl,hb}. Our solution can then be used for term-by-term analysis of the perturbation series.

Other type of problems where our general solution can be useful concern particulate suspensions. Consider a dilute viscous suspension of rigid spheres driven by some external time-periodic flow. For a dilute suspension, the total flow perturbation due to the presence of particles is given by superposition of the individual perturbations that can be described using the proposed solution representation. We demonstrated that constructive interference between the single-particle perturbations is possible, paving way to the study of the collective phenomena. The proposed representation is also a good starting point for the development of the numerical algorithm for simulating time-dependent flows around arbitrary clusters made of spherical beads. The proposed solution can be used in a way analogous to the way Lamb's classical decomposition is exploited to model steady viscous flows around such clusters \cite{Fil}.

Finally, the general solution can be useful in the problem of a flow around a spherical particle suspended in a fluid above an oscillating plane. Such flow is relevant to interpretation of the quartz crystal microbalance measurements \cite{Busca}. Here the use of the method of reflections \cite{hb} demands solving for the flow in the unbounded fluid due to some non-trivial (spatially varying) oscillatory velocity distribution prescribed at the sphere surface. The use of the solution developed in this work allows to obtain the solution at all orders. In the forthcoming work we shall demonstrate that the reflection series converges remarkably fast and that the two-term solution yields a quantitatively accurate approximation even at small proximity of the particle to the oscillating plane.

To conclude, these prospective research directions suggest a wide range of potential applications of the general solution presented in this work.

\section{Acknowledgments}

We thank D.~Palaniappan for pointing out, upon reading the first draft of the paper, that the particular solution of the unsteady Stokes equations employed in this work has been already reported by Lamb. This work was supported, in part, by the Israel Science Foundation (ISF) via the grant No. 1744/17 (A.M.L.), and by Israel Ministry of Science and Technology via grant No. 3-17383 (Y.O.).

\bibliography{finalunsteady_28_4_22.bbl}

%%%%%%%%%%%%%%%%%%%%%%%%%%%%%%%%%%%%%%%%%%%%%

\pagebreak
%\newpage
%\newpage
\begin{appendices}

\section{Solution of vector Helmholtz equation}
\label{solutionH}

In this Appendix we provide details of derivation of solution of vector Helmholtz equation for the subsection \ref{ge}. The curl of Eq.~(\ref{expansion1}) reads \cite{sph},
\begin{eqnarray}&&
\nabla\times \bm u_s=\sum_{l=1}^{\infty}\sum_{m=-l}^l \left(-\frac{l(l+1)c^{(2)}_{lm}\bm Y_{lm}}{r}
\right.\\&&\left.-\left(\frac{dc^{(2)}_{lm}}{dr}+\frac{c^{(2)}_{lm}}{r}\right)\bm \Psi_{lm}+\left(\frac{dc^1_{lm}}{dr}+\frac{c^{(1)}_{lm}}{r}-\frac{c^r_{lm}}{r}\right)
\bm \Phi_{lm}\right).\nonumber
\end{eqnarray}
Taking curl of the curl and using that incompressibility implies $\nabla^2\bm u_s=-\nabla\times (\nabla\times \bm u_s)$ we find,
\begin{eqnarray}&&
\nabla^2\bm u_s=\sum_{l=1}^{\infty}\sum_{m=-l}^l \left(\frac{l(l+1)\bm Y_{lm}}{r}\left(\frac{dc^{(1)}_{lm}}{dr}+\frac{c^{(1)}_{lm}}{r}-\frac{c^r_{lm}}{r}\right)
\right.\nonumber\\&&\left.+\bm \Psi_{lm}\left(\frac{d}{dr}+\frac{1}{r}\right)\left(\frac{dc^{(1)}_{lm}}{dr}+\frac{c^{(1)}_{lm}}{r}-\frac{c^r_{lm}}{r}\right)
\right.\\&&\left.
-\bm \Phi_{lm}\left[\frac{l(l+1)c^{(2)}_{lm}}{r^2}-\left(\frac{d}{dr}+
\frac{1}{r}\right)\left(\frac{dc^{(2)}_{lm}}{dr}+\frac{c^{(2)}_{lm}}{r}\right)\right]
\right).\nonumber
\end{eqnarray}
The coefficients obey by $\lambda^2\bm u_s=\nabla^2\bm u_s$ that
\begin{eqnarray}&&
\lambda^2 c^r_{lm}-\frac{l(l+1)}{r}\left(\frac{dc^{(1)}_{lm}}{dr}+\frac{c^{(1)}_{lm}}{r}-\frac{c^r_{lm}}{r}\right)
=0,\label{daw}\\&&
\lambda^2 c^{(1)}_{lm}-\left(\frac{d}{dr}+\frac{1}{r}\right)\left(\frac{dc^{(1)}_{lm}}{dr}+\frac{c^{(1)}_{lm}}{r}-\frac{c^r_{lm}}{r}\right)=0,\nonumber\\&&
\lambda^2c^{(2)}_{lm}+\frac{l(l+1)c^{(2)}_{lm}}{r^2}-\left(\frac{d}{dr}+\frac{1}{r}\right)\left(\frac{dc^{(2)}_{lm}}{dr}+\frac{c^{(2)}_{lm}}{r}\right)=0.\nonumber
\end{eqnarray}
The last equation can be written as:
\begin{eqnarray}&&\!\!\!\!\!\!\!\!\!\!\!\!
\frac{d^2}{dr^2}\left(rc^{(2)}_{lm}\right)-\left(\lambda^2+\frac{l(l+1)}{r^2}\right)rc^{(2)}_{lm}=0.\label{d1}% c^{(2)}_{lm}={\tilde c}^{(2)}_{lm}r^{-l-1},
\end{eqnarray}
The solution that decays at infinity ($Re \lambda>0$) is,
\begin{eqnarray}&&\!\!\!\!\!\!\!\!\!\!\!\!
c^{(2)}_{lm}=\frac{{\tilde c}_{lm} K_{l+1/2}(\lambda r)}{\sqrt{r}},\label{d2}
\end{eqnarray}
where ${\tilde c}_{lm}$ is a constant and the modified Bessel function $K_{l+1/2}$ is defined in Eqs.~(\ref{ourp})-(\ref{mod}).
%This can be readily confirmed by observing that
%\begin{eqnarray}&&
%r^{3/2}(\sqrt{r} K_{l+1/2}(\lambda r))''\!=\!(\lambda r)^2 K''_{l+1/2}(\lambda r)\!+\!r\lambda K'_{l+1/2}(\lambda r)
%\nonumber\\&&
%-\frac{K_{l+1/2}(\lambda r)}{4}=r^2 \left(\lambda^2+\frac{l(l+1)}{r^2}\right)K_{l+1/2}(\lambda r),
%\end{eqnarray}
%where we used the equation on modified Bessel functions.
The incompressibility condition does not impose any restrictions on ${\tilde c}^{(2)}_{lm}$. We have,
\begin{eqnarray}&&\!\!\!\!\!\!\!\!\!\!\!\!
\nabla\!\cdot\! \bm u\!=\!\sum_{l=1}^{\infty}\sum_{m=-l}^l \left(\frac{dc^r_{lm}}{dr}\!+\!\frac{2c^r_{lm}}{r}\!-\!\frac{l(l\!+\!1)c^{(1)}_{lm}}{r}\right)Y_{lm}.\label{inc}
\end{eqnarray}
The rest of Eqs.~(\ref{daw}) give,
\begin{eqnarray}&&\!\!\!\!\!\!\!\!\!\!\!\!
\frac{d(rc^{(1)}_{lm})}{dr}-c^r_{lm}=\frac{\lambda^2r^2c^{r}_{lm}}{l(l+1)},\nonumber\\&&\!\!\!\!\!\!\!\!\!\!\!\!
\frac{d^2}{dr^2}\left(rc^{(1)}_{lm}\right)-\frac{dc^r_{lm}}{dr}=\lambda^2rc^{(1)}_{lm}.
\end{eqnarray}
Consistency of these equations requires that,
\begin{eqnarray}&&\!\!\!\!\!\!\!\!\!\!\!\!
\frac{d\left(r^2c^r_{lm}\right)}{dr}=l(l+1)rc^{(1)}_{lm}. \label{consist}
\end{eqnarray}
which is equivalent to incompressibility condition, see Eq.~(\ref{inc}). Using this condition we find coupled equations,
\begin{eqnarray}&&\!\!\!\!\!\!\!\!\!\!\!\!
\frac{d(rc^{(1)}_{lm})}{dr}=c^r_{lm}+\frac{\lambda^2r^2c^{r}_{lm}}{l(l+1)},\nonumber\\&&\!\!\!\!\!\!\!\!\!\!\!\!
\frac{d\left(r^2c^r_{lm}\right)}{dr}=l(l+1)rc^{(1)}_{lm}.\label{c1}
\end{eqnarray}
Taking derivative of the last equation and using the first yields,
\begin{eqnarray}&&\!\!\!\!\!\!\!\!\!\!\!\!
\frac{d^2y}{dr^2}-\left[\lambda^2+\frac{l(l+1)}{r^2}\right]y=0, \label{radi}
\end{eqnarray}
where $y=r^2c^r_{lm}$. The solution that vanishes at infinity is,
\begin{eqnarray}&&\!\!\!\!\!\!\!\!\!\!\!\!
c^r_{lm}=\frac{{\tilde c}^{r}_{lm} K_{l+1/2}(\lambda r)}{r^{3/2}},
\end{eqnarray}
where ${\tilde c}^{r}_{lm}$ is a constant. We find using Eq.~(\ref{consist}),
\begin{eqnarray}&&\!\!\!\!\!\!\!\!\!\!\!\!
c^{(1)}_{lm}
=\frac{{\tilde c}^{r}_{lm} }{l(l+1)r}\frac{d\left(r^{1/2}K_{l+1/2}(\lambda r)\right)}{dr}.
\end{eqnarray}
We conclude that the general solution of Eqs.~(\ref{is}) has the form,
\begin{eqnarray}&&\!\!\!\!\!\!\!\!\!\!\!\!
\bm u\!=\!\sum_{l=1}^{\infty}\sum_{m=-l}^l \left(\left(\frac{{\tilde c}^{r}_{lm} K_{l+1/2}(\lambda r)}{r^{3/2}}+\frac{(l+1)c_{lm}}{\lambda^2 r^{l+2}}\right)\bm Y_{lm}
\right.\nonumber\\&&\!\!\!\!\!\!\!\!\!\!\!\!\left.
+\!\left(\frac{{\tilde c}^{r}_{lm} }{l(l+1)r}\frac{d\left(r^{1/2}K_{l+1/2}(\lambda r)\right)}{dr}
-\frac{c_{lm}}{\lambda^2 r^{l+2}}\right)\bm \Psi_{lm}
\right.\nonumber\\&&\!\!\!\!\!\!\!\!\!\!\!\!\left.
+\frac{{\tilde c}_{lm} K_{l+1/2}(\lambda r)}{\sqrt{r}}\bm \Phi_{lm}\right), \label{fomrs}
\end{eqnarray}
where we used Eqs.~(\ref{helm}), (\ref{ps}). We can rewrite the above by using
\begin{eqnarray}&&
\frac{dK_{\nu}(z)}{dz}=-\frac{\nu K_{\nu}(z)}{z}-K_{\nu-1}(z), \nonumber
\end{eqnarray}
which gives
\begin{eqnarray}&&\!\!\!\!\!\!\!\!\!\!\!\!
\left(r^{1/2}K_{l+1/2}(\lambda r)\right)'\!=\!-\frac{l K_{l+1/2}(\lambda r)\!+\!\lambda r K_{l-1/2}(\lambda r)}{\sqrt{r}}
.\nonumber
\end{eqnarray}
The usage of this identity in Eq.~(\ref{fomrs}) gives Eq.~(\ref{ho}).

\section{Transformation of the coefficients} \label{transformation}

In this Appendix, we give details for the identities that underly the transformation of the coefficients of expansion in subsection \ref{sdj}, cf. Appendix of \cite{carra}. We introduce the field $\bm {\tilde u}$ as the field which has zero radial component and has $r-$independent azimuthal
and polar components. It is set that $\bm {\tilde u}$ coincides with $\bm u$ on the sphere $r=1$. We have then
\begin{eqnarray}&&
\int \bm u\cdot \bm \Psi_{lm}^* d\Omega=\int \bm {\tilde u}\cdot \bm \Psi_{lm}^* d\Omega=\int_{r=1}\bm {\tilde u}\cdot\nabla Y_{lm}^* d\Omega
\nonumber\\&&
=\int_{r=1}\nabla\cdot \left(\bm {\tilde u} Y_{lm}^*\right) d\Omega-\int_{r=1}Y_{lm}^* \nabla\cdot \bm {\tilde u} d\Omega.
\end{eqnarray}
We use that by divergence theorem
\begin{eqnarray}&&
0=\int_{x<1}\nabla\cdot \left(\bm {\tilde u} Y_{lm}^*\right) dV=\int_0^1 r dr \left(\nabla\cdot \left(\bm {\tilde u} Y_{lm}^*\right)\right)_{r=1}  d\Omega\nonumber\\&&
=\frac{1}{2}\int_{r=1}\nabla\cdot \left(\bm {\tilde u} Y_{lm}^*\right) d\Omega.
\end{eqnarray}
We conclude from the above that
\begin{eqnarray}&&\!\!\!\!\!\!\!
\int \!\!\bm u\!\cdot\! \bm \Psi_{lm}^* d\Omega\!=\!-\int \! Y_{lm}^* \left(\frac{\partial (u_{\theta}\sin\theta)}{\partial\theta}\!+\!\frac{\partial u_{\phi}}{\partial\phi}\right) d\theta d\phi, \label{projectpsi}
\end{eqnarray}
where we used $\nabla\cdot \bm {\tilde u}$ in spherical coordinates. We develop a similar formula for
\begin{eqnarray}&&\!\!\!\!\!\!\!
\int \!\!\bm u\!\cdot\! \bm \Phi_{lm}^* d\Omega\!=\!-\!\int_{r=1} \bm {\tilde u}\cdot \nabla\!\times\! (\bm r Y^*_{lm}) d\Omega\nonumber\\&&\!\!\!\!\!\!\!
=\int_{r=1} \nabla\cdot \left(\bm {\tilde u}\!\times\! (\bm r Y^*_{lm}) \right)  d\Omega\!-\!
\int_{r=1}  Y^*_{lm} \bm r\cdot\nabla\!\times\! \bm {\tilde u}d\Omega,
\end{eqnarray}
where the radial component of the vector field $\bm {\tilde u}\!\times\! (\bm r Y_{lm})$ is zero and other components depend on $r$ linearly. We have similarly to the above
\begin{eqnarray}&&
0=\int_{x<1} \nabla\cdot \left(\bm {\tilde u}\!\times\! (\bm r Y^*_{lm}) \right)  dV=\int_0^1 r^2 dr   d\Omega\\&&
\cdot\left(\nabla\cdot \left(\bm {\tilde u}\!\times\! (\bm r Y^*_{lm}) \right) \right)_{r=1}=\frac{1}{3}\int_{r=1}\nabla\cdot \left(\bm {\tilde u}\!\times\! (\bm r Y^*_{lm}) \right) d\Omega.\nonumber
\end{eqnarray}
We conclude that
\begin{eqnarray}&&\!\!\!\!\!\!\!
\int \!\!\bm u\!\cdot\! \bm \Phi_{lm}^* d\Omega\!=\!\int_{r=1} \!\! Y_{lm}^* \left(\frac{\partial u_{\theta}}{\partial\phi}\!-\!
\frac{\partial (u_{\phi}\sin\theta)}{\partial\theta}\right) d\theta d\phi,\label{projectphi}
\end{eqnarray}
where we used curl in spherical coordinates. The above gives Eq.~(\ref{cd}).

\section{Calculation of the kernel $F_l(t, r)$}
\label{ker}

Here we provide the calculations for the kernel $F_l(t, r)$ in Eqs.~(\ref{APhit}). Deriving $F_l(t, r)$ from $G_l(t, r)$ according to Eq.~(\ref{unf}) results in cumbersome integrals. Instead, we perform the integration directly by using Eq.~(\ref{fl}) which gives
\begin{eqnarray}&&\!\!\!\!\!\!\!
f_l(\tau, r)\!=\! \int_{\delta-i\infty}^{\delta+i\infty}e^{\sqrt{s}(1-r)}\frac{P_{l}(r \sqrt{s})}{s^{2} P_{l-1}(\sqrt{s})}e^{s\tau}\frac{ds}{2\pi i}. \label{integralrepresentation}
\end{eqnarray}
where we used Eq.~(\ref{mod}). The Eqs.~(\ref{mod}) and (\ref{idr}) imply the identity
\begin{eqnarray}&&\!\!\!\!\!\!\!\!
-\frac{d}{dx}e^{-\lambda x}P_l(\lambda x)=\lambda^2 x e^{-\lambda x}P_{l-1}(\lambda x).
\end{eqnarray}
which allows to prove the identity $-\partial_r f_l\!=\! rg_{l-1}$ given by Eq.~(\ref{recur}) directly from Eqs.~(\ref{gl}) and (\ref{integralrepresentation}).

\subsection{The case $l=2$}

The calculations at $l=2$ and $l>2$ differ as will be seen below. Here we consider $l=2$. We have from the definition in Eq.~(\ref{ourp}) that $P_1(x)=1+x$ and $P_2(x)=3+3x+x^2$. We find
\begin{eqnarray}&&\!\!\!\!\!\!\!
f_2(\tau, r)\!=\! \int_{\delta-i\infty}^{\delta+i\infty}\frac{3+3r \sqrt{s}+r^2s}{s^{2} (1+\sqrt{s})}e^{\sqrt{s}(1-r)+s\tau}\frac{ds}{2\pi i}. \label{fhs}
\end{eqnarray}
We find that
\begin{eqnarray}&&\!\!\!\!\!\!\!\!\!
-\partial_r f_2\!=\!r\!\int_{\delta-i\infty}^{\delta+i\infty}\!\!\frac{1 +r\sqrt{s}}{s(1+\sqrt{s})}e^{\sqrt{s}(1-r)+s\tau}\frac{ds}{2\pi i}=rg_1,
\end{eqnarray}
confirming Eq.~(\ref{recur}) in the considered case. We have
\begin{eqnarray}&&\!\!\!\!\!\!\!
\frac{1}{s^{2} (1+\sqrt{s})}=%\frac{1}{s^2}-\frac{1}{s^{3/2}(1+\sqrt{s})}\\&&=\frac{1}{s^2}-\left(\frac{1}{s^{3/2}}-\frac{1}{s(1+\sqrt{s})}\right)
\frac{1}{s^2}-\frac{1}{s^{3/2}}+\frac{1}{s}-\frac{1}{\sqrt{s}}+\frac{1}{1+\sqrt{s}},
\end{eqnarray}
which gives
\begin{eqnarray}&&\!\!\!\!\!\!\!
\frac{3+3r \sqrt{s}+r^2s}{s^{2} (1+\sqrt{s})}%=\frac{3}{s^2}+\frac{3(r-1)}{s^{3/2}}+\frac{r^2-3r+3}{s}
=\frac{3}{s^2}+\frac{3(r-1)}{s^{3/2}}+ \nonumber \\
&&\!\!\!\!\!\!\!
(r^2-3r+3)\left(\frac{1}{s}-\frac{1}{\sqrt{s}}+\frac{1}{1+\sqrt{s}}\right). \label{lar}
\end{eqnarray}
The use of this decomposition in Eq.~(\ref{fhs}) and of integral tables \cite{prud} gives Eq.~(\ref{fk}) from the main text.

\subsection{Calculation for arbitrary $l$}

The following identities are obtained from Eq.~(\ref{ourp}):
\begin{eqnarray}&&
P_l(0)\!= \!P_l'(0)\!= \!(2l-1)!!,\ \ P_l''(0)=(2l-3)!!(2l\!-\!2),\nonumber \\&&
\left(\frac{P_{l}(r x)}{P_{l-1}(x)}\right)'_{x=0}\!=\!(r\!-\!1)(2l\!-\!1),
\end{eqnarray}
using which we find
\begin{eqnarray}&&
\left(\frac{P_l(r x)}{xP_{l-1}(x)}\!-\!\frac{2l-1}{x}\right)'_{x=0}=\lim_{x\to 0}\left(xrP_{l-1}(x)P'_l(r x)
\right.\nonumber\\&&\left.
-xP_l(r x)P'_{l-1}(x)-P_l(r x)P_{l-1}(x)+(2l-1)P^2_{l-1}(x)\right)
\nonumber\\&&\times\frac{1}{x^2P_{l-1}^2(x)}=\left(xrP_{l-1}(x)P'_l(r x)
-xP_l(r x)P'_{l-1}(x)
\right.\nonumber\\&&\left.
-P_l(r x)P_{l-1}(x)+(2l-1)P^2_{l-1}(x)\right)''_{0}\frac{1}{2P_{l-1}^2(0)}
\nonumber\\&&=2l-1-\frac{rP_l(0)}{P_{l-1}(0)}+\frac{r^2P''_l(0)}{2P_{l-1}(0)}
+\frac{(2l-1)P''_{l-1}(0)}{P_{l-1}(0)}
\nonumber\\&&
-\frac{3P_l(0)P''_{l-1}(0)}{2P_{l-1}^2(0)}=(2l-1)(1-r)+r^2(l-1)
\nonumber\\&&
-\frac{(2l-1)(l-2)}{2l-3}.
\end{eqnarray}
Using the above the partial fraction decomposition for $l>2$ we have
\begin{eqnarray}&&
\frac{P_{l}(r x)}{x^{4} P_{l-1}(x)}\!=\!\frac{2l\!-\!1}{x^4}\!+\!\frac{(r\!-\!1)(2l\!-\!1)}{x^3}\!+\!\frac{1}{x^2}\left(r^2(l\!-\!1)
\right.\nonumber\\&&\left. -r(2l\!-\!1)+\frac{(2l\!-\!1)(l\!-\!1)}{2l\!-\!3}\right)
\!+\!\frac{1}{x}\left(\frac{r^3(l\!-\!2)}{3}
\right.\nonumber\\&&\left. +\frac{r(l\!-\!1)(2l\!-\!1)}{2l\!-\!3}\!-\!r^2(l\!-\!1)
-\frac{l(2l\!-\!1)}{3(2l\!-\!3)}\right)
\nonumber\\&&
+\sum_{k=1}^{l-1}\frac{P_{l}(-r a_k^{l-1})}{\left(a_k^{l-1}\right)^4P'_{l-1}(-a_k^{l-1})\left(x\!+\!a_k^{l-1}\right)}, \label{fk3}
\end{eqnarray}
where from Eq.~(\ref{ourp}) it follows that for $l>2$
\begin{eqnarray}&&
P_l(r x)\!-\!(2l-1)P_{l-1}(x)\!-\!(r\!-\!1)(2l\!-\!1)xP_{l-1}(x)
\nonumber\\&&
\!-\!x^2 P_{l-1}(x)\left(r^2(l\!-\!1)-r(2l\!-\!1)+\frac{(2l\!-\!1)(l\!-\!1)}{2l\!-\!3}\right)
\nonumber\\&&
=x^3P_{l-1}(0)\left(\frac{r^3(l\!-\!2)}{3}\!+\!\frac{r(l\!-\!1)(2l\!-\!1)}{2l\!-\!3}\!-\!r^2(l\!-\!1)
\right.\nonumber\\&&\left.
-\frac{l(2l-1)}{3(2l-3)}\right)+O(x^4).
\end{eqnarray}
We remark that the decomposition given by Eq.~(\ref{fk3}) is valid for $l>2$ only and does not agree with the decomposition at $l=2$ given by Eq.~(\ref{lar}) with $\sqrt{s}=x$. The use of Eq.~(\ref{fk3}) in Eq.~(\ref{integralrepresentation}) and of the integrals in \cite{prud} gives Eq.~(\ref{fkl}) in the main text.

It is useful to provide another observation. Setting $r=1$ in Eq.~(\ref{fk3}) gives
\begin{eqnarray}&&
\frac{P_{l}(x)}{x^{4} P_{l-1}(x)}\!=\!\frac{2l\!-\!1}{x^4}\!+\!\frac{1}{(2l-3)x^2}\nonumber\\&&
+\sum_{k=1}^{l-1}\frac{P_{l}(-a_k^{l-1})}{\left(a_k^{l-1}\right)^4P'_{l-1}(-a_k^{l-1})\left(x\!+\!a_k^{l-1}\right)}. \label{deco}
\end{eqnarray}
Multiplying this equation by $x^4$, taking third derivative and evaluating it at $x=0$ gives
\begin{eqnarray}&&
\frac{\partial^3}{\partial x^3}\left(\frac{P_{l}(x)}{P_{l-1}(x)}\right)|_{x=0}\!=\!0. \label{thi}
\end{eqnarray}
Furthermore we find the partial fraction decomposition
\begin{eqnarray}&&
\frac{P_{l}(x)}{P_{l-1}(x)}\!=\!\sum_{k=1}^{l-1}\frac{P_{l}(-a_k^{l-1})}{P'_{l-1}(-a_k^{l-1})\left(x\!+\!a_k^{l-1}\right)}.\label{dec}
\end{eqnarray}
Taking third derivative of the last identity and evaluating it at $x=0$ gives Eq.~(\ref{sumrule}) from the main text by Eq.~(\ref{thi}). Similarly, multiplying Eq.~(\ref{deco}) by $x^4$, taking second derivative and evaluating at $x=0$ we find
\begin{eqnarray}&&
\frac{\partial^2}{\partial x^2}\left(\frac{P_{l}(x)}{P_{l-1}(x)}\right)|_{x=0}\!=\!\frac{2}{2l-3}.
\end{eqnarray}
Taking second derivative of Eq.~(\ref{dec}) and evaluating it $x=0$ gives Eq.~(\ref{sumrul}) from the main text by using the above equation.

\section{Oscillating sphere problem}\label{tr}

Here we provide the calculations that reproduce the known solution for the flow around an oscillating sphere used in Sec. \ref{oscillating}. We find that the first two terms in the RHS of Eq.~(\ref{yc}) are non-zero only if $m=1$ or $m=-1$. We use
\begin{eqnarray}&&
%\int Y_{lm} Y^*_{l'm'}d\Omega=\int_0^{\pi}\sin\theta d\theta\int_0^{2\pi}d\phi Y_{lm} Y^*_{l'm'}=\delta_{l l'}\delta_{m m'},\nonumber\\&&
Y_{l1}=\sqrt{\frac{(2l+1)}{4\pi l(l+1)}}P_l^1(\cos\theta)\exp\left(i\phi\right)=-Y_{l,-1}^*,\label{sd}%nonumber\\&&
%Y_{l,-1}=-\sqrt{\frac{(2l+1)}{4\pi l(l+1)}}P_l^1(\cos\theta)\exp\left(-i\phi\right)
\end{eqnarray}
and that definition of $P_l^m$ implies
\begin{eqnarray}&&
\int  \sqrt{1-x^2} P_l^1(x) dx=-\int (1-x^2) P_l'(x) dx
\nonumber\\&&
=-2\int x P_l(x) dx=-\frac{4\delta_{l1}}{3}.
\end{eqnarray}
This gives
\begin{eqnarray}&&
\int \sin\theta\cos\phi Y_{lm} d\Omega=\sqrt{\frac{2\pi}{3}} \delta_{l1}(\delta_{m, -1}-\delta_{m, 1}),
\nonumber\\&&
\int \sin\theta\sin\phi Y_{lm} d\Omega=-  i\sqrt{\frac{2\pi}{3}} \delta_{l1}(\delta_{m, -1}+\delta_{m, 1}).
\end{eqnarray}
We also have $\int \cos\theta Y_{lm}^* d\Omega\!=\!2\delta_{l1}\delta_{m0} \sqrt{\pi/3}$ that is readily confirmed by using the first of Eqs.~(\ref{definition}). We conclude that
\begin{eqnarray}&&\!\!\!\!\!\!\!\!
\int\! \bm U\!\cdot\! \bm Y_{lm}^* d\Omega\!=\!\int\! \frac{\bm u\!\cdot \!\bm \Psi_{lm}^* d\Omega}{2}\!=\!\sqrt{\frac{2\pi}{3}} \left(U_x\!+\!iU_y\right)\delta_{l1}\delta_{m, -1}
\nonumber\\&&
- \sqrt{\frac{2\pi}{3}}(U_x-i U_y) \delta_{l1}\delta_{m, 1}+ 2\sqrt{\frac{\pi}{3}}U_z\delta_{l1}\delta_{m0}. \label{cp}
\end{eqnarray}
We now find the coefficients $c_{lm}$. We have from Eq.~(\ref{cope})
\begin{eqnarray}&&
c_{lm}\!=\!\left(\frac{\lambda^2}{2}+\frac{3\lambda K_{3/2}(\lambda)}{2K_{1/2}(\lambda)}\right)\sqrt{\frac{2\pi}{3}}\delta_{l1} \left(\left(U_x+iU_y\right)\delta_{m, -1}
\right.\nonumber\\&&\left.
-(U_x-i U_y) \delta_{m, 1}+ \sqrt{2}U_z\delta_{m0}\right).
\end{eqnarray}
We obtain from Eq.~(\ref{mod}) that
\begin{eqnarray}&&\!\!\!\!\!\!\!\!
K_{1/2}(\lambda)\!=\!\sqrt{\frac{\pi}{2\lambda}}e^{-\lambda},\ \ \lambda K_{3/2}(\lambda)\!=\!\sqrt{\frac{\pi}{2\lambda}}(1\!+\!\lambda)e^{-\lambda}. \label{Mcd}
\end{eqnarray}
We obtain
\begin{eqnarray}&&
c_{lm}\!=\!\left(1+\lambda+\frac{\lambda^2}{3}\right)\sqrt{\frac{3\pi}{2}}\delta_{l1} \left(\left(U_x+iU_y\right)\delta_{m, -1}
\right.\nonumber\\&&\left.
-(U_x-i U_y) \delta_{m, 1}+ \sqrt{2}U_z\delta_{m0}\right).
\end{eqnarray}
We find from Eq.~(\ref{pr}) that the pressure is
\begin{eqnarray}&&
p\!= \!\frac{3}{4r^2} \left(1\!+\!\lambda\!+\!\frac{\lambda^2}{3}\right) \left(\left(U_x\!+\!iU_y\right)\sin\theta e^{-i\phi}+\!(U_x\!-\!i U_y)
\right.\nonumber\\&&\left.
\cdot \sin\theta\exp\left(i\phi\right)\!+\!2U_z \cos\theta\right)\!=\!\left(1\!+\!\lambda\!+\!\frac{\lambda^2}{3}\right)\frac{3\bm U\!\cdot \!\bm r}{2r^3},\label{oer}
\end{eqnarray}
where we used $P_1^1(\cos\theta)=-\sin\theta$. For finding the velocity field, it remains to obtain ${\tilde c}^{r}_{lm}$ since ${\tilde c}_{lm}=0$ by $\int \!\!\bm u\!\cdot\! \bm \Phi_{lm}^* d\Omega\!=\!0$, see the last of Eqs.~(\ref{cof}). We have from Eq.~(\ref{tild}) that
\begin{eqnarray}&&
{\tilde c}^{r}_{lm} \!=\!-\frac{3\int \bm u\cdot \bm Y_{lm}^* d\Omega}{\lambda K_{1/2}(\lambda)}=
\left(-\sqrt{3} \left(U_x+iU_y\right)\delta_{l1}\delta_{m, -1}
\right.\nonumber\\&&\left.
+\sqrt{3}(U_x-i U_y) \delta_{l1}\delta_{m, 1}- \sqrt{6}U_z\delta_{l1}\delta_{m0}\right)\frac{2\exp(\lambda)}{ \sqrt{\lambda}}.
\end{eqnarray}
We find using Eqs.~(\ref{helm}) and (\ref{oer}) that the flow is given by
\begin{eqnarray}&&\!\!\!\!\!\!\!\!\!\!\!\!\!\!\!\!
\bm u=\bm u_s-\nabla \left(1\!+\!\lambda\!+\!\frac{\lambda^2}{3}\right)\frac{3\bm U\!\cdot \!\bm r}{2\lambda^2 r^3},\label{cs}
\end{eqnarray}
where $\bm u_s$ is a solution of the vector Helmholtz equation, which according to Eqs.~(\ref{helmh}) and (\ref{Mcd}) is given by
\begin{eqnarray}&&
\bm u_s\!=\!\sum_{m=-1}^1 \left(\frac{{\tilde c}^{r}_{1m} K_{3/2}(\lambda r){\hat r} Y_{1m}}{r^{3/2}}
\right.\\&&\left.
-\frac{{\tilde c}^{r}_{1m} \lambda r^{1/2}K_{1/2}(\lambda r) }{2}\left(1\!+\!\frac{1}{\lambda r }\!+\!\frac{1}{(\lambda r)^2}\right) \nabla Y_{1m}\right).\nonumber
\end{eqnarray}
The dependence on the coefficients reduces to $\sum_{m=-1}^1  {\tilde c}^{r}_{1m}Y_{1m}$ as seen by rewriting the above as
\begin{eqnarray}&&
\bm u_s\!=\!\frac{\bm{\hat r}}{r^2}\sqrt{\frac{\pi}{2\lambda}}\left(\frac{1}{\lambda r}\!+\!1\right)e^{-\lambda r}\sum_{m=-1}^1  {\tilde c}^{r}_{1m}Y_{1m}
\label{us}\\&&
-\frac{ \lambda^{1/2}}{2}\sqrt{\frac{\pi}{2}}e^{-\lambda r} \left(1\!+\!\frac{1}{\lambda r }\!+\!\frac{1}{(\lambda r)^2}\right) \nabla \sum_{m=-1}^1  {\tilde c}^{r}_{1m}Y_{1m}.\nonumber
\end{eqnarray}
We find from Eqs.~(\ref{cope})-(\ref{tild}) that
\begin{eqnarray}&&
\frac{{\tilde c}^{r}_{1m}}{c_{1m}}=-\frac{6}{\lambda^2\left(3K_{3/2}(\lambda)
\!+\!\lambda K_{1/2}(\lambda)\right)}
\nonumber\\&&
=-\sqrt{\frac{2}{\pi\lambda}}\frac{2\exp(\lambda)}{1+\lambda\!+\!\lambda^2/3}.
\end{eqnarray}
We conclude that
\begin{eqnarray}&&
\sum_{m=-1}^1  {\tilde c}^{r}_{1m}Y_{1m}=-\sqrt{\frac{2}{\pi\lambda}}\frac{2\exp(\lambda)}{1+\lambda\!+\!\lambda^2/3}\sum_{m=-1}^1 c_{1m}Y_{1m}
\nonumber\\&&
=-\sqrt{\frac{2}{\pi\lambda}}\frac{2\exp(\lambda)r^2 p}{1+\lambda\!+\!\lambda^2/3}=-\sqrt{\frac{2}{\pi\lambda}}3\exp(\lambda)\bm U\!\cdot \! \bm{\hat r},
\end{eqnarray}
where we used Eqs.~(\ref{pr}) and (\ref{oer}). We obtain using the above in Eq.~(\ref{us}) that
\begin{eqnarray}&&
\bm u_s\!=\!-\left(\frac{1}{\lambda r}\!+\!1\right)\frac{3\exp(\lambda(1-r))(\bm U\!\cdot\! \bm{\hat r})\bm{\hat r}}{r^2\lambda}
\\&&
+\frac{3\exp(\lambda(1-r))}{2} \left(1\!+\!\frac{1}{\lambda r }\!+\!\frac{1}{(\lambda r)^2}\right) \frac{\bm U\!-\!(\bm U\!\cdot\! \bm{\hat r})\bm{\hat r}}{r},\nonumber
\end{eqnarray}
which can be rewritten as
\begin{eqnarray}&&
\bm u_s\!=\!\frac{3\exp\left(-\lambda (r-1)\right)}{2\lambda^2}
\label{skao}\\&&
\times \left(\frac{(1\!+\!\lambda r)\left(\bm U\!-\!3(\bm U\!\cdot\! \bm{\hat r})\bm{\hat r}\right)}{r^3}+
\frac{\lambda^2\left(\bm U\!-\!(\bm U\!\cdot\! \bm{\hat r})\bm{\hat r}\right)}{r}\right).\nonumber
\end{eqnarray}
We conclude that the flow induced by an oscillating sphere is given by $l=1$ term of the series solution, similarly to the counterpart problem for the steady Stokes flow.

\subsection{Comparison with known solution}

We compare the above with the solution brought in \cite{kim} which can be written as ($\bm {\hat r}\equiv \bm r/r$)
\begin{eqnarray}&&
\bm u\!=\!\frac{3}{2\lambda^2}\left(1\!+\!\lambda\!+\!\frac{\lambda^2}{3}\!+\!\frac{1}{\lambda^2}\left(e^{\lambda}\!-\!1\!-\!\lambda\!-\!\frac{\lambda^2}{3}\right)\nabla^2\right)
\\&&
\cdot \left(\left(1-(1+\lambda r)\exp(-\lambda r)\right)\frac{2(\bm U\cdot \bm {\hat r})\bm {\hat r}}{r^3}
\right.\nonumber\\&&\left.
+ \left((1\!+\!\lambda r\!+\!(\lambda r)^2)\exp(-\lambda r)\!-\!1\right)\frac{\bm U\!-\!(\bm U\!\cdot\! \bm{\hat r})\bm{\hat r}}{r^3}\right).\nonumber
\end{eqnarray}
We remark that \cite{kim} has a typo of multiplicative factor $\lambda$. We find that Laplacian of terms that do not contain the exponential factors is proportional to
\begin{eqnarray}&&
\nabla^2\frac{3{\hat r_i}{\hat r_k}-\delta_{ik}}{r^3}=\nabla^2 \nabla_i\nabla_k \frac{1}{r}=0.
\end{eqnarray}
The Laplacian of the exponential terms is proportional to
\begin{eqnarray}&&
e^{\lambda r}\nabla^2 e^{-\lambda r} \left( (1\!+\!\lambda r)\frac{\bm U\!-\!3(\bm U\!\cdot\! \bm{\hat r})\bm{\hat r}}{r^3}
\!+\!(\lambda r)^2\frac{\bm U\!-\!(\bm U\!\cdot\! \bm{\hat r})\bm{\hat r}}{r^3}\right) \nonumber\\&&
=\left(\lambda^2-\frac{2\lambda}{r}\right) \left( (1\!+\!\lambda r)\frac{\bm U\!-\!3(\bm U\!\cdot\! \bm{\hat r})\bm{\hat r}}{r^3}
\!+\!(\lambda r)^2\frac{\bm U\!-\!(\bm U\!\cdot\! \bm{\hat r})\bm{\hat r}}{r^3}\right)\nonumber\\&&
-2\lambda  \partial_r \left( (1\!+\!\lambda r)\frac{\bm U\!-\!3(\bm U\!\cdot\! \bm{\hat r})\bm{\hat r}}{r^3}
\!+\!(\lambda r)^2\frac{\bm U\!-\!(\bm U\!\cdot\! \bm{\hat r})\bm{\hat r}}{r^3}\right)\nonumber\\&&
+\nabla^2 \left( (1\!+\!\lambda r)\frac{\bm U\!-\!3(\bm U\!\cdot\! \bm{\hat r})\bm{\hat r}}{r^3}
\!+\!(\lambda r)^2\frac{\bm U\!-\!(\bm U\!\cdot\! \bm{\hat r})\bm{\hat r}}{r^3}\right).
\end{eqnarray}
We find taking the derivatives
\begin{eqnarray}&&
e^{\lambda r}\nabla^2 e^{-\lambda r} \left( (1\!+\!\lambda r)\frac{\bm U\!-\!3(\bm U\!\cdot\! \bm{\hat r})\bm{\hat r}}{r^3}
\!+\!(\lambda r)^2\frac{\bm U\!-\!(\bm U\!\cdot\! \bm{\hat r})\bm{\hat r}}{r^3}\right) \nonumber\\&&
=\left(\lambda^2-\frac{2\lambda}{r}\right) \left( (1\!+\!\lambda r)\frac{\bm U\!-\!3(\bm U\!\cdot\! \bm{\hat r})\bm{\hat r}}{r^3}
\!+\!(\lambda r)^2\frac{\bm U\!-\!(\bm U\!\cdot\! \bm{\hat r})\bm{\hat r}}{r^3}\right)\nonumber\\&&
+2\lambda  \left( (3\!+\!2\lambda r)\frac{\bm U\!-\!3(\bm U\!\cdot\! \bm{\hat r})\bm{\hat r}}{r^4}
\!+\!(\lambda r)^2\frac{\bm U\!-\!(\bm U\!\cdot\! \bm{\hat r})\bm{\hat r}}{r^4}\right)\nonumber\\&&
-\frac{4\lambda \left(\bm U\!-\!3(\bm U\!\cdot\! \bm{\hat r})\bm{\hat r}\right)}{r^4}-2\lambda^2\frac{\bm U\!-\!3(\bm U\!\cdot\! \bm{\hat r})\bm{\hat r}}{r^3}\nonumber\\&&
=\frac{\lambda^2 (1\!+\!\lambda r)\left(\bm U\!-\!3(\bm U\!\cdot\! \bm{\hat r})\bm{\hat r}\right)}{r^3}+
\frac{\lambda^4\left(\bm U\!-\!(\bm U\!\cdot\! \bm{\hat r})\bm{\hat r}\right)}{r}.
\end{eqnarray}
We find collecting the terms that
\begin{eqnarray}&&
\bm u\!=\!\frac{3}{2\lambda^2}\left(1\!+\!\lambda\!+\!\frac{\lambda^2}{3}\right) \left(\left(1-(1+\lambda r)\exp(-\lambda r)\right)\frac{2(\bm U\cdot \bm {\hat r})\bm {\hat r}}{r^3}
\right.\nonumber\\&&\left.
+ \left((1\!+\!\lambda r\!+\!(\lambda r)^2)\exp(-\lambda r)\!-\!1\right)\frac{\bm U\!-\!(\bm U\!\cdot\! \bm{\hat r})\bm{\hat r}}{r^3}\right)\nonumber\\&&
+\frac{3}{2\lambda^2}\left(e^{\lambda}\!-\!1\!-\!\lambda\!-\!\frac{\lambda^2}{3}\right)e^{-\lambda r}
\nonumber\\&&
\cdot\left(
\frac{(1\!+\!\lambda r)\left(\bm U\!-\!3(\bm U\!\cdot\! \bm{\hat r})\bm{\hat r}\right)}{r^3}+
\frac{\lambda^2\left(\bm U\!-\!(\bm U\!\cdot\! \bm{\hat r})\bm{\hat r}\right)}{r}\right)\nonumber\\&&
=\frac{3}{2\lambda^2}\left(1\!+\!\lambda\!+\!\frac{\lambda^2}{3}\right) \frac{3(\bm U\cdot \bm {\hat r})\bm {\hat r}-\bm U}{r^3}+\frac{3\exp\left(-\lambda (r-1)\right)}{2\lambda^2}
\nonumber\\&&
\cdot\left(\frac{(1\!+\!\lambda r)\left(\bm U\!-\!3(\bm U\!\cdot\! \bm{\hat r})\bm{\hat r}\right)}{r^3}+
\frac{\lambda^2\left(\bm U\!-\!(\bm U\!\cdot\! \bm{\hat r})\bm{\hat r}\right)}{r}\right). \label{fis}
\end{eqnarray}
It is seen that the final formula is rather simple and $\bm u(r=1)=\bm U$ is readily verified. We can write the solution given by Eq.~(\ref{fis}) in the form given by Eq.~(\ref{helm})
\begin{eqnarray}&&\!\!\!\!\!\!
\bm u=\bm u_s-\frac{\nabla p}{\lambda^2}, \ \ p=-\frac{3}{2}\left(1\!+\!\lambda\!+\!\frac{\lambda^2}{3}\right)  (\bm U\cdot \nabla)\frac{1}{r},
\end{eqnarray}
where $\bm u_s$ is a solution of the Helmholtz equation which can be readily seen to agree with Eq.~(\ref{skao}). We conclude that the solution by the series and the ordinary solutions agree.

We remark that the solution can also be written in concise form which can be inferred from \cite{maxeyriley,brg} as:
\begin{eqnarray}&&\!\!\!\!\!\!
\bm u=(\bm U\cdot\nabla)\nabla\psi-\bm U\nabla^2\psi,%\ \ \psi=\frac{Q_1+Q_2\exp(-\lambda r)}{r}
\nonumber\\&&
\psi= \frac{3}{2\lambda^2 r}\left(1\!+\!\lambda\!+\!\frac{\lambda^2}{3}-\exp(\lambda(1-r))\right). %
%\nonumber\\&&
%Q_1=\frac{3}{2\lambda^2}\left(1\!+\!\lambda\!+\!\frac{\lambda^2}{3}\right),\ \ Q_2=-\frac{3\exp(\lambda)}{2\lambda^2},
%\nonumber\\&&
%\nabla_i\nabla_k\frac{\exp(-\lambda r)}{r}=-\nabla_i\bm{\hat r}_k\frac{(1+\lambda r)\exp(-\lambda r)}{r^2}
%\nonumber\\&&
%=-\delta_{ik}\frac{(1+\lambda r)\exp(-\lambda r)}{r^3}+
%\bm{\hat r}_i\bm{\hat r}_k\frac{(3(1+\lambda r)+\lambda^2r^2)\exp(-\lambda r)}{r^3}
\end{eqnarray}
that can be readily confirmed to agree with other forms. This form expresses the solution via a single scalar function $\psi$ of the radial variable similarly to \cite{LL}. We have
\begin{eqnarray}&&\!\!\!\!\!\!
\bm u_s=\left(\bm U\nabla^2 - (\bm U\cdot\nabla)\nabla\right) \frac{3\exp(\lambda(1-r))}{2\lambda^2 r}.
\end{eqnarray}

\section{Oscillatory rotations of the sphere} \label{ur}

In this Appendix, details of the calculations for the oscillatory rotations of the sphere, considered in subsection \ref{oscillatory}, are provided. It is readily seen that
\begin{eqnarray}&&
\int \sin\theta\cos\phi Y_{lm}^* d\Omega=\pi (\delta_{m, 1}-\delta_{m, -1})\sqrt{\frac{(2l+1)}{4\pi l (l+1)}}
\nonumber\\&&
\cdot\int P_l^1(\cos\theta)\sin^2\theta d\theta,\ \  \int \cos\theta Y_{lm}^* d\Omega\!=\!
2\sqrt{\frac{\pi}{3}}\delta_{l1}\delta_{m0},\nonumber
\end{eqnarray}
where we have used
\begin{eqnarray}&&
%\int Y_{lm} Y^*_{l'm'}d\Omega=\int_0^{\pi}\sin\theta d\theta\int_0^{2\pi}d\phi Y_{lm} Y^*_{l'm'}=\delta_{l l'}\delta_{m m'},\nonumber\\&&
Y_{l1}=\sqrt{\frac{(2l+1)}{4\pi l(l+1)}}P_l^1(\cos\theta)\exp\left(i\phi\right)=-Y_{l,-1}^*.\label{sd}%nonumber\\&&
%Y_{l,-1}=-\sqrt{\frac{(2l+1)}{4\pi l(l+1)}}P_l^1(\cos\theta)\exp\left(-i\phi\right)
\end{eqnarray}
The definition of $P_l^m$ implies
\begin{eqnarray}&&
\int  \sqrt{1-x^2} P_l^1(x) dx=-\int (1-x^2) P_l'(x) dx
\nonumber\\&&
=-2\int x P_l(x) dx=-\frac{4\delta_{l1}}{3},
\end{eqnarray}
which gives
\begin{eqnarray}&&
\int \sin\theta\cos\phi Y_{lm}^* d\Omega=\sqrt{\frac{2\pi}{3}} \delta_{l1}(\delta_{m, -1}-\delta_{m, 1}),
\nonumber\\&&
\int \sin\theta\sin\phi Y_{lm}^* d\Omega=i\sqrt{\frac{2\pi}{3}} \delta_{l1}(\delta_{m, -1}+\delta_{m, 1}).
\end{eqnarray}
We conclude from the above and Eq.~(\ref{cof}) that
\begin{eqnarray}&&
-K_{3/2}(\lambda){\tilde c}_{lm}=\int Y_{lm}^*\omega_r d\Omega\!=\!\sqrt{\frac{2\pi}{3}} \left(\omega_x+i\omega_y\right)\delta_{l1}\delta_{m, -1}
\nonumber\\&&
- \sqrt{\frac{2\pi}{3}}(\omega_x-i \omega_y) \delta_{l1}\delta_{m, 1}+ 2\sqrt{\frac{\pi}{3}}\omega_z\delta_{l1}\delta_{m0}. \label{refe}
\end{eqnarray}
%\begin{eqnarray}&&
% =
%%\sqrt{\frac{2\lambda^3}{\pi}}(1\!+\!\lambda)^{-1}e^{\lambda}
%\sqrt{\frac{2\pi}{3}} \left(\omega_x+i\omega_y\right)\delta_{l1}\delta_{m, -1}
%\nonumber\\&&
%- \sqrt{\frac{2\pi}{3}}(\omega_x-i \omega_y) \delta_{l1}\delta_{m, 1}+ 2\sqrt{\frac{\pi}{3}}\omega_z\delta_{l1}\delta_{m0}.
%\end{eqnarray}
%where we used the formula for $K_{3/2}(\lambda)$ given by Eq.~(\ref{Mcd}).
%\begin{eqnarray}&&\!\!\!\!\!\!\!\!
%K_{1/2}(\lambda)\!=\!\sqrt{\frac{\pi}{2\lambda}}e^{-\lambda},\ \  K_{3/2}(\lambda)\!=\!\sqrt{\frac{\pi}{2\lambda^3}}(1\!+\!\lambda)e^{-\lambda}.
%\end{eqnarray}
This gives on using Eq.~(\ref{helmh}) and the definition of $\bm \Phi$ in Eq.~(\ref{vsh}) that
\begin{eqnarray}&&
\bm u\!=\!-\frac{K_{3/2}(\lambda r)}{K_{3/2}(\lambda)}\sqrt{\frac{2\pi}{3r}}\bm r\!\times \!\nabla
\left( \left(\omega_x+i\omega_y\right)Y_{1, -1}
\right.\\&&\nonumber\left.
-(\omega_x-i \omega_y)Y_{11}+\sqrt{2}\omega_zY_{10}\right).
\end{eqnarray}
We find by using the definitions of $Y_{1m}$ and $Y_{1, -1}=-Y_{11}^*$ that the expression in brackets equals $\bm \omega\cdot\bm {\hat r}$. Finally, using the definition of $K_{3/2}(\lambda)$ in Eq.~(\ref{Mcd}) we obtain Eq.~(\ref{ror}).

\section{Force and torque via the expansion coefficients} \label{formulae}

In this Appendix, we provide the formulae for the force and torque on the sphere, similar to those for the general solution of steady Stokes equations \cite{kim}.

\subsection{Difference from steady Stokes equations}

For steady Stokes equations, that express the condition of zero total force on any volume of inertialess fluid, the force on the sphere of any radius is the same. This is because the total force on the volume enclosed by any two spheres vanishes. This leads to the possibility of finding the force from asymptotic behavior of the flow on an infinitely remote sphere. Thus, the general series solutions of the Stokes equations provide the force as a single coefficient of the series that provides the leading order term at large distances. Similar fact holds for the torque \cite{kim}.

In the case of the unsteady flow, the situation is somewhat different and we have for the force $\bm F$ that the fluid exerts on interior of the unit sphere
\begin{eqnarray}&&
F_i=\int_{r=1} \sigma_{ir}dS=\int_{x=R} \sigma_{ir}dS-\int_{1<r<R} \nabla_k\sigma_{ik} dV
\nonumber\\&&
=\int_{r=R} \sigma_{ir}dS-\lambda^2 \int_{1<r<R}u_i dV. \label{fs0}
\end{eqnarray}
where the stress tensor $\sigma_{ik}$, defined in Eq.~(\ref{dtr}), obeys $\nabla_k\sigma_{ik}=\lambda^2u_i$. The last, volume integral, is proportional to the frequency and absent for steady Stokes flows.

\subsection{Force and torque from reciprocal theorem}

We can circumvent the direct calculation by using reciprocal theorem \cite{kim} similarly to \cite{for}. For any dual flow obeying $\lambda^2{\hat v}_i=\nabla_k{\hat \sigma}_{ik}$ and incompressibility condition, we have the Lorentz identity
\begin{eqnarray}&&\!\!\!\!\!\!\!\!\!\!\!\!\!
{\hat v}_i\frac{\partial \sigma_{ik}}{\partial x_k}\!=\!
u_i\frac{\partial {\hat \sigma}_{ik}}{\partial x_k},\ \ \frac{\partial ({\hat v}_i\sigma_{ik})}{\partial x_k}\!=\!
\frac{\partial (u_i{\hat \sigma}_{ik})}{\partial x_k}. \label{lorentz}
\end{eqnarray}
Here ${\hat \sigma}_{ik}$ is the stress tensor of the dual flow defined similarly to Eq.~(\ref{dtr}). We use as the dual flow the oscillating sphere solution considered in Sec. \ref{oscillating}. We find by integration of the above identity over the sphere exterior that
\begin{eqnarray}&&
\bm U\int_{r=1} \sigma{\hat r}dS=-\int_{r=1} \bm u\cdot \frac{3(1+\lambda)\bm U+\lambda^2 (\bm U\cdot \bm {\hat r})\bm {\hat r}}{2}dS,\nonumber
\end{eqnarray}
where we used Eq.~(\ref{tra}). This equation must hold for any $\bm U$, providing us with a simple formula for the force
\begin{eqnarray}&&\!\!\!\!\!\!
\bm F=-\frac{3(1+\lambda)}{2}\int_{r=1} \bm udS-\frac{\lambda^2}{2}\int_{r=1}  (\bm u\cdot \bm {\hat r})\bm {\hat r}dS. \label{sl}
\end{eqnarray}
We conclude that the most general dependence of the force on $\lambda$ is parabolic. The parabola has only two free coefficients, $\bm F=\lambda^2 \bm F_1+(1\!+\!\lambda)\bm F_2$ where $\bm F_i$ are frequency-independent and given by low moments of velocity field. This is the same dependence on $\lambda$ as for oscillating sphere, see Eq.~(\ref{ford}). The only difference is that there are six free coefficients in the force and not three. The interpretation of the terms and their form in time domain are the same as for the oscillating sphere problem, see e.g. \cite{LL,kim}.

We can similarly obtain the torque. We use as the dual flow the oscillatory rotation of the sphere considered in the previous section. We have
\begin{eqnarray}&&\!\!\!\!\!\!
\int_{r=1}\!\sigma{\hat r}\!\cdot\!  \left(\bm \omega\!\times\! \bm {\hat r} \right) dS\!=\!-\int_{r=1} \!\bm u\!\cdot\!(\bm \omega\!\times\!\bm {\hat r}) \left(3\!+\!\frac{\lambda^2}{1\!+\!\lambda}\right)dS,\nonumber
\end{eqnarray}
where we used Eq.~(\ref{trar}). We find the torque $\bm T$ by requiring that this equation holds for any $\bm \omega$
\begin{eqnarray}&&\!\!\!\!\!\!
\bm T\!=\!-8\pi\bm \omega_{eff}-\frac{8\pi\bm \omega_{eff} \lambda^2}{3(1+\lambda)},  \label{trt}
\end{eqnarray}
where we introduced the effective angular velocity $\bm \omega_{eff}$ as
\begin{eqnarray}&&\!\!\!\!\!\!
\bm  \omega_{eff}\!\equiv \!\frac{3}{8\pi} \int_{r=1} \!\bm {\hat r} \!\times\! \bm u dS, \label{ome}
\end{eqnarray}
so that the torque formula looks as that for oscillating rotations of a rigid sphere, see Eq.~(\ref{toq}). Thus, the torque on the sphere generated by a general surface flow can be described as resulting from rigid sphere rotations.

\subsection{Integrals' calculation}

The above integral formulas for the force and the torque can be written in terms of the coefficients of the expansion in the VSH. This is done by inserting into the integrals the series expansion for $\bm u$ given by Eq.~(\ref{ho}). The calculation requires integrals of the VSH over the sphere. We consider
\begin{eqnarray}&&
\int \bm Y_{lm}d\Omega\!=\!\bm {\hat x}\int \!\!\sin\theta\cos\phi Y_{lm}d\Omega\!+\!\bm {\hat y}\int\!\! \sin\theta\sin\phi Y_{lm}d\Omega
\nonumber\\&&
+\bm {\hat z}\int \cos\theta Y_{lm}d\Omega.
\end{eqnarray}
where we projected $\bm Y_{lm}=\bm {\hat r}Y_{lm}$ on the Cartesian unit vectors. We find using Eqs.~(\ref{refe}) that
\begin{eqnarray}&&
\int \bm Y_{lm}d\Omega\!=\!\delta_{l1}\sqrt{\frac{2\pi}{3}}  \left(\bm {\hat x}(\delta_{m, -1}\!-\!\delta_{m, 1})\!-\! i\bm {\hat y}(\delta_{m, -1}\!+\!\delta_{m, 1})
\right.\nonumber\\&&\left.
+\bm {\hat z}\sqrt{2}\delta_{m0}\right), \label{fsdp}
\end{eqnarray}
This formula allows to find the first term in Eq.~(\ref{fs0}). It is readily seen from the general solution that at large distances, considered in detail later,
\begin{eqnarray}&&\!\!\!\!\!\!\!\!\!\!\!\!\!
p=\sum_{m=-1}^{m=1}\frac{c_{1m}Y_{1m}(\theta, \phi)}{r^{2}}+O(1/r^3),
\end{eqnarray}
and velocity is of order $1/r^3$.  Thus, the integral of the surface traction over the surface of the sphere at infinity is fully determined by the pressure.  The viscous component of the stress does not contribute the integral, in contrast with the steady Stokes equations. We have
\begin{eqnarray}&&
%\sigma_{ir}=-\delta_{ir}\sum_{m=-1}^{m=1}\frac{c_{1m}Y_{1m}(\theta, \phi)}{r^{2}}+O(1/r^3),\ \
\lim_{R\to\infty}\int_{x=R}  \sigma_{ir}dS=-\sum_{m=-1}^{m=1} c_{1m}\int \bm{\hat r}Y_{1m}(\theta, \phi)d\Omega
\nonumber\\&&=-\sqrt{\frac{2\pi}{3}} \sum_{m=-1}^{m=1} \left(\bm{\hat x}(c_{1, -1}-c_{11})\!-\! i\bm{\hat y}(c_{1, -1}+c_{11})
\right.\nonumber\\&&\left.
+\bm{\hat z}\sqrt{2}c_{10}\right),
\end{eqnarray}
where we used Eq.~(\ref{fsdp}) with ${\hat r}Y_{1m}=\bm Y_{1m}$. We consider the remaining integrals
\begin{eqnarray}&&
\int \!\!\bm \Psi_{lm}d\Omega\!=\!\bm{\hat x}\int \!\!\bm{\hat x}\cdot \bm \Psi_{lm}d\Omega\!+\!\bm{\hat y}\int\!\!\bm{\hat y}\cdot \bm \Psi_{lm}d\Omega
\nonumber\\&&
+\bm{\hat z}\int \bm{\hat z}\cdot \bm \Psi_{lm}d\Omega,
\end{eqnarray}
and similar integrals for $\bm \Phi_{lm}$. Performing calculations similar to those that led to Eq.~(\ref{fsdp}), we obtain
\begin{eqnarray}&&
\int \bm \Psi_{lm}d\Omega\!=\!2\delta_{l1}\sqrt{\frac{2\pi}{3}}  \left(\bm{\hat x}(\delta_{m, -1}-\delta_{m, 1})\!-\! i\bm{\hat y}(\delta_{m, -1}+\delta_{m, 1})
\right.\nonumber\\&&\left.
+\bm{\hat z}\sqrt{2}\delta_{m0}\right) =2 \int \bm Y_{lm}d\Omega,
\ \
\int \bm \Phi_{lm}d\Omega\!=0.
\end{eqnarray}
For finding the torque, given by Eq.~(\ref{trt}), we must calculate
\begin{eqnarray}&&\!\!\!\!\!\!
 \int_{r=1} \!\bm {\hat r} \!\times\! \bm \Psi_{lm} dS\!=\!\bm{\hat x}\int \!\!\left(\sin\theta\sin\phi\bm{\hat z} -\cos\theta \bm{\hat y}\right) \cdot \bm \Psi_{lm}d\Omega
\!\!\!\!\!\! \nonumber\\&&
 +\bm{\hat y}\int\!\!\left(\cos\theta \bm{\hat x}-\sin\theta\cos\phi\bm{\hat z}\right) \cdot \bm \Psi_{lm}d\Omega
\!\!\!\!\!\!\nonumber\\&&
+\bm{\hat z}\int \left(\sin\theta\cos\phi\bm{\hat y}-\sin\theta\sin\phi\bm{\hat x}\right) \cdot \bm \Psi_{lm}d\Omega,
\end{eqnarray}
and similar integral for $\bm \Phi_{lm}$. We find calculating the integrals
\begin{eqnarray}&&\!\!\!\!\!\!
 \int_{r=1}\!\!\!\!\!\! \!\bm {\hat r} \!\times\! \bm \Phi_{lm} dS\!=
\!-2 \delta_{l1}\sqrt{\frac{2\pi}{3}}  \left({\hat x}(\delta_{m, -1}-\delta_{m, 1})\!
\right.\nonumber\\&&\!\!\!\!\!\!\left.
- i\bm{\hat y}(\delta_{m, -1}+\delta_{m, 1})
+\bm{\hat z}\sqrt{2}\delta_{m0}\right) = -2 \int_{r=1}\!\!\!\!\!\! \bm Y_{lm} dS,
\nonumber\\&&\!\!\!\!\!\!
 \int_{r=1}\!\!\!\!\!\! \!\bm {\hat r} \!\times\! \bm Y_{lm} dS\!=
 \int_{r=1}\!\!\!\!\!\! \!\bm {\hat r} \!\times\! \bm \Psi_{lm} dS\!=\! 0.
\end{eqnarray}
Finally, for finding the force and the torque via the expansion coefficients we introduce the series representation for $\bm u$ into Eqs.~(\ref{sl}) and (\ref{trt}) and perform term-by-term integration using the formulas above.

\subsection{Force via expansion coefficients}

We obtain by setting $r=1$ in Eq.~(\ref{ho}) and using Eq.~(\ref{Mcd}) that inserting $\bm u$ into Eqs.~(\ref{sl}) gives
\begin{eqnarray}&&
\bm F\!=\!-\sqrt{\frac{2\pi}{3}} \left(\bm{\hat x}(c_{1, -1}\!-\!c_{11})\!-\! i\bm{\hat y}(c_{1, -1}\!+\!c_{11})\!+\!\bm{\hat z}\sqrt{2}c_{10}\right)
\nonumber\\&&
+\pi (1\!+\!\lambda) \sqrt{\frac{\lambda}{3}}e^{-\lambda}
 \left(\bm{\hat x}({\tilde c}^{r}_{1, -1}\!-\!{\tilde c}^{r}_{11})\!-\! i\bm{\hat y}({\tilde c}^{r}_{1, -1}\!+\!{\tilde c}^{r}_{11})
 \right.\nonumber\\&&\left.
 +\bm{\hat z}\sqrt{2}{\tilde c}^{r}_{10}\right). \label{fso}
\end{eqnarray}
It is seen from Eq.~(\ref{fsdp}) that the first line of the above equation equals to the force due to the pressure component of the stress, $-\int_{r=1} p\bm{\hat r} d\Omega$, cf. subsection \ref{focid}. Thus, the last two lines describe the viscous component's contribution to the force.

The correspondence to the results holding in the limit of the steady Stokes flow, $\lambda\to 0$, is seen by observing that Eq.~(\ref{coper}) gives
\begin{eqnarray}&&
{\tilde c}^{r}_{lm} \!\simeq\!-\frac{l+1}{\lambda^2K_{l+1/2}(\lambda)}c_{lm}+o\left(\lambda\right).
\end{eqnarray}
We find from Eq.~(\ref{fso}) and (\ref{Mcd}) that
\begin{eqnarray}&&
\bm F\!\simeq \!-\sqrt{6\pi} \left(\bm{\hat x}(c_{1, -1}\!-\!c_{11})\!-\! i\bm{\hat y}(c_{1, -1}\!+\!c_{11})\!+\!\bm{\hat z}\sqrt{2}c_{10}\right).\nonumber%\label{fsol}
\end{eqnarray}
This equation agrees with the formula implied by Lamb's solution of the steady Stokes equations, see e.g. \cite{kim,hb}, which is
\begin{eqnarray}&&\!\!\!\!\!\!\!\!\!\!\!\!
\bm F=-4\pi \nabla \left(r\sum_{m=-1}^{m=1}c_{1m}Y_{lm}(\theta, \phi)\right)=-\sqrt{6\pi} \nonumber\\&&\!\!\!\!\!\!\!\!\!\!\!\!\!
\cdot \left(\bm{\hat x}(c_{1, -1}-c_{11})\!-\! i\bm{\hat y}(c_{1, -1}+c_{11})+\bm{\hat z}\sqrt{2}c_{10}\right),
\end{eqnarray}
where we used the identity.
\begin{eqnarray}&&\!\!\!\!\!\!\!\!\!\!\!\!\!
\sum_{m=-1}^{m=1}c_{1m}Y_{lm}(\theta, \phi)=\sqrt{\frac{3}{4\pi}}c_{10}\cos\theta-\sqrt{\frac{3}{8\pi}}\sin\theta
\nonumber\\&&\!\!\!\!\!\!\!\!\!\!\!\!\!
\cdot \left((c_{11}-c_{1, -1})\cos\phi+i(c_{11}+c_{1, -1})\sin\phi\right).
\end{eqnarray}
Returning to the general case, Eq.~(\ref{fso}) does not make it obvious that the force dependence on $\lambda$ (see Eq~(\ref{sl})) is parabolic, since it involves $\lambda-$dependent coefficients. We rewrite the force via the frequency-independent coefficients $b_{1m}$ and $d_{1m}$ defined in Eq.~(\ref{b}). By using the definition of $K_{1/2}(\lambda)$ in Eq.~(\ref{tild}) it follows that
\begin{eqnarray}&&\!\!\!\!\!\!
{\tilde c}^{r}_{1m} \!= \!e^{\lambda}\sqrt{\frac{2}{\pi \lambda}}\left(d_{1m}\!-\!3b_{1m}\right). \label{rsu}
\end{eqnarray}
Similarly, using Eq.~(\ref{sak})
and the definition of $ {\cal P}_1(x)$ in Eq.(\ref{ourp}) we have
\begin{eqnarray}&&
c_{1m}\!=\!\frac{ \lambda^2 b_{1m}+(1\!+\!\lambda) \left(3b_{1m}-d_{1m}\right)}{2}.
\end{eqnarray}
We find Eq.~(\ref{fer}) by using the above equations in Eq.~(\ref{fso}).

\subsection{Torque via expansion coefficients}

The torque is found by inserting the expansion of $\bm u$ in the VSH into Eq.~(\ref{trt}) and making term-by-term integration. This gives
\begin{eqnarray}&&\!\!\!\!\!\!
\bm T\!=\!6 \left(1\!+\!\lambda\!+\!\frac{\lambda^2}{3}\right)\frac{\pi}{\sqrt{3\lambda^3}}e^{-\lambda}\left(\bm{\hat x}({\tilde c}_{1, -1}-{\tilde c}_{11})\!
\right.\nonumber\\&&\!\!\!\!\!\!\left.
- i\bm{\hat y}({\tilde c}_{1, -1}+{\tilde c}_{11})
+\bm{\hat z}\sqrt{2}{\tilde c}_{10}\right). \label{rt}
\end{eqnarray}
It is readily seen that in the zero-frequency limit, the above reproduces the result for the steady Stokes flow \cite{kim}. We remark that it is useful to employ in the demonstration that
\begin{eqnarray}&&\!\!\!\!\!\!
\int \bm u\cdot \bm \Phi_{1m}^* d\Omega\!=\!2{\tilde c}_{1m} K_{3/2}(\lambda)\sim 2{\tilde c}_{1m}\sqrt{\frac{\pi}{2\lambda^3}}. \nonumber
\end{eqnarray}
The general $\lambda-$dependence of the torque is obtained by observing that the usage of $K_{3/2}(x)=(1+1/x)\exp(-x)\sqrt{\pi/(2x)}$ in Eq.~(\ref{coex}) gives
\begin{eqnarray}&&\!\!\!\!\!\!\!\!\!\!\!\!
{\tilde c}_{1m}\!=\!-\frac{\exp(\lambda)}{2(1+\lambda)}\sqrt{\frac{2\lambda^3}{\pi}}\int_{r=1} Y_{1m}^* (\nabla\!\times\! \bm u)_r   d\Omega.
\end{eqnarray}
The usage of this equation in Eq.~(\ref{rt}) reproduces Eq.~(\ref{trt}) with
\begin{eqnarray}&&
\bm  \omega_{eff}\!=\!\frac{1}{4}\sqrt{\frac{3}{2\pi}}\left(\bm{\hat x}(e_{-1}-e_{1})\!- i\bm{\hat y}(e_{-1}+e_{1})
+\bm{\hat z}\sqrt{2}e_{0}\right),
\nonumber\\&&
e_{m}\equiv \int_{r=1} Y_{1m}^* (\nabla\!\times\! \bm u)_r   d\Omega.
\end{eqnarray}
which provides explicit form of $\bm  \omega_{eff}$ in Eq.~(\ref{ome}) via projections on $Y_{lm}$.

\section{Axially symmetric case} \label{as}

In this Appendix, we provide the details for the consideration of the axially symmetric case in Sec. \ref{sro}. Using Eq.~(\ref{sp}) and integration over $\phi$ that in the axially symmetric case we find that
\begin{eqnarray}&&\!\!\!\!\!\!
\frac{1}{\sqrt{\pi(2l+1)}}\int  Y_{l0}^* \nabla_s\!\cdot\!\bm u d\Omega=2\int  {\cal P}_l(\cos\theta)  u \sin\theta d\theta
\nonumber\\&&\!\!\!\!\!\!
+\int  {\cal P}_l(\cos\theta) \partial_{\theta}(\sin\theta v) d\theta.
\end{eqnarray}
Performing integration by parts in the last term and using $ \partial_{\theta}\left[{\cal P}_l(\cos\theta) \right]=P_l^1(\cos\theta)$ we find
\begin{eqnarray}&&\!\!\!\!\!\!
\frac{1}{\sqrt{\pi(2l+1)}}\int  Y_{l0}^* \nabla_s\!\cdot\!\bm u d\Omega=2\int  {\cal P}_l(\cos\theta)  u \sin\theta d\theta
\nonumber\\&&\!\!\!\!\!\!
-\int  P_l^1(\cos\theta) v \sin\theta d\theta.
\end{eqnarray}
The use of this formula in Eq.~(\ref{cope}) gives
\begin{eqnarray}&&
c_{l0}\!=\! \frac{\sqrt{\pi(2l+1)} P_l(\lambda)}{(l\!+\!1) P_{l-1}(\lambda)}\int_{-1}^1\!\!  P_l^1(x)  v(x) dx
\\&&
+\frac{\left(l P_l(\lambda)
\!+\!\lambda^2 P_{l-1}(\lambda)\right)\sqrt{\pi(2l\!+\!1)}}{(l\!+\!1) P_{l-1}(\lambda)} \int_{-1}^1\!\!  {\cal P}_l(x)  u(x) dx,\nonumber
\end{eqnarray}
where $u(x)$, $v(x)$ are defined as $u(\theta=\arccos x)$ and $v(\theta=\arccos x)$, respectively. We obtain Eq.~(\ref{dl}) for $D_l$ from the definition in Eqs.~(\ref{Dl}).
%\begin{eqnarray}&&
%c_{lm}\!=\! \frac{ \lambda\left((l+2) {\cal P}_l(\lambda^{-1})
%\!+\!\lambda {\cal P}_{l-1}(\lambda^{-1})\right)}{(l\!+\!1) {\cal P}_{l-1}(\lambda^{-1})}\int_{r=1}Y_{lm}^* u_r d\Omega
%\nonumber\\&&
%-\frac{\lambda {\cal P}_l(\lambda^{-1})}{(l\!+\!1) {\cal P}_{l-1}(\lambda^{-1})} \int  Y_{lm}^* \nabla_s\!\cdot\!\bm u d\Omega.\label{cope}
%\end{eqnarray}
For finding $F_l$ we use
\begin{eqnarray}&&
{\tilde c}^{r}_{l0}\!=\!\sqrt{\frac{2\lambda}{\pi}}
\frac{\lambda^l\exp\left(\lambda\right)}{P_l(\lambda)}\left(\int\! u Y_{l0}^* d\Omega-\frac{(l\!+\!1)c_{l0}}{\lambda^2}\right),
\end{eqnarray}
see Eqs.~(\ref{mod}) and (\ref{coper}). We find
\begin{eqnarray}&&
{\tilde c}^{r}_{l0}\!=\!-
\frac{\lambda^{l-1}\exp\left(\lambda\right)}{P_{l-1}(\lambda)} \sqrt{\frac{2(2l+1)}{\lambda}}\left(\int_{-1}^1\!\!  P_l^1(x)  v(x) dx
\right.\nonumber\\&&\left.
+l \int_{-1}^1\!\!  {\cal P}_l(x)  u(x) dx\right).
\end{eqnarray}
This gives $F_l$ via the last of Eqs.~(\ref{Dl}), completing the solution for the radial velocity $u$ given by Eq.~(\ref{axs}).
%We find using the above relations in Eq.~(\ref{tild}) that
%\begin{eqnarray}&&
%{\tilde c}^{r}_{lm} \!=\!\lambda^{l-1}\sqrt{\pi(2l+1)} \frac{\int  Y_{lm}^* \nabla_s\!\cdot\!\bm u d\Omega-l\int_{-1}^1\!\!  {\cal P}_l(x)  u(x) dx}{P_{l-1}(\lambda)}.\nonumber
%\end{eqnarray}
%thus
%\begin{eqnarray}&&
%c_{lm}\!=\! \frac{ \lambda\left(l K_{l+1/2}(\lambda)
%\!+\!\lambda K_{l-1/2}(\lambda)\right)}{(l\!+\!1) K_{l-1/2}(\lambda)}\int \!\bm u\!\cdot\! \bm Y_{lm}^* d\Omega
%\nonumber\\&&
%+\frac{\lambda K_{l+1/2}(\lambda)}{(l\!+\!1) K_{l-1/2}(\lambda)}\int\! \bm u\!\cdot\! \bm \Psi_{lm}^* d\Omega\!=\!\frac{ \lambda^2}{l\!+\!1}\int\! \bm u\!\cdot\! \bm Y_{lm}^* d\Omega\nonumber\\&&-\frac{\lambda^2K_{l+1/2}(\lambda){\tilde c}^{r}_{lm}}{l\!+\!1}.
%\end{eqnarray}

We compare the solution with the solution of Rao \cite{rao} for axially symmetric flows. That is given via the streamfunction $\psi$ that defines the flow via
\begin{eqnarray}&&\!\!\!\!\!\!\!\!\!\!\!\!
%\bm u=-\nabla \times \frac{\psi}{r\sin\theta}{\hat \phi},\ \
u=\frac{1}{r^2\sin\theta}\frac{\partial \psi}{\partial \theta},\ \ v=-\frac{1}{r\sin\theta}\frac{\partial \psi}{\partial r}.
\end{eqnarray}
It is found that \cite{rao}
\begin{eqnarray}&&
\psi\!=\!\sum_{l=1}^{\infty}\left(\frac{D_l(\lambda)}{r^l}\!+\!F_l(\lambda)r^{1/2}K_{l+1/2}(\lambda r) \right)\int_{-1}^{\cos\theta}{\cal P}_l(x)dx,\nonumber
\end{eqnarray}
which reproduces our formula for $u$. For completeness,
%The radial component of Eq.~(\ref{fomrs}) is,
%\begin{eqnarray}&&\!\!\!\!\!\!\!\!\!\!\!\!
%u^0_r\!=\!\sum_{l=1}^{\infty}\sum_{m=-l}^l \left(\frac{{\tilde c}^{r}_{lm} K_{l+1/2}(\lambda r)}{r^{3/2}}+\frac{l+1}{\lambda^2 r^{l+2}}\right)c_{lm}Y_{lm},
%\end{eqnarray}
%where we used Eq.~(\ref{vsh}). In the axially symmetric case only the terms with $m=0$ are non-vanishing giving
%\begin{eqnarray}&&\!\!\!\!\!\!\!\!\!\!\!\!
%u^0_r\!=\!\\&&\!\!\!\!\!\!\!\!\!\!\!\!
%\sum_{l=1}^{\infty} \left(\frac{{\tilde c}^{r}_{l0} K_{l+1/2}(\lambda r)}{r^{3/2}}+\frac{l+1}{\lambda^2 r^{l+2}}\right)c_{l0}\sqrt{\frac{2l+1}{4\pi}}P_l(\cos\theta).\nonumber
%\end{eqnarray}
%This agrees with the solution given by Eq.~(\ref{axs}) in the axially symmetric case if we identify
%\begin{eqnarray}&&\!\!\!\!\!\!\!\!\!\!\!\!
%D_l(\lambda)=-\frac{(l+1)c_{l0}}{\lambda^2}\sqrt{\frac{2l+1}{4\pi}},\nonumber\\&&\!\!\!\!\!\!\!\!\!\!\!\!
%F_l(\lambda)=-{\tilde c}^{r}_{l0} c_{l0}\sqrt{\frac{2l+1}{4\pi}}.\label{idenrti}
%\end{eqnarray}
we consider the remaining component of the velocity,
\begin{eqnarray}&&\!\!\!\!\!
v\!=\!\sum_{l=1}^{\infty}\!\left(\frac{l D_l}{r^{l+2}}\!-\!\frac{F_l}{r}\frac{d\left(r^{1/2}K_{l+1/2}(\lambda r)\right)}{dr}\!\! \right)\!
\int_{-1}^{\cos\theta}\!\!\frac{{\cal P}_l(x)dx}{\sin\theta}.\nonumber
\end{eqnarray}
We use the identity \cite{lighthill},
\begin{eqnarray}&&\!\!\!\!\!\!\!\!\!\!\!\!
\int_{-1}^{\cos\theta}\!{\cal P}_l(x)\frac{dx}{\sin\theta} \!=\!-\int_{\cos\theta}^{1}\!{\cal P}_l(x)\frac{dx}{\sin\theta}
\!=\!\frac{\partial_{\theta} {\cal P}_l(\cos\theta)}{l(l+1)},
\end{eqnarray}
where we used $\int_{-1}^{1}{\cal P}_l(x)dx=0$ for $l>0$. Thus we can write,
\begin{eqnarray}&&\!\!\!\!\!\!\!\!\!\!\!\!
v\!=\!\sum_{l=1}^{\infty}\!\left(\frac{l D_l}{r^{l+2}}\!-\!\frac{F_l}{r}\frac{d\left(r^{1/2}K_{l+1/2}(\lambda r)\right)}{dr} \right)
%\nonumber\\&&\!\!\!\!\!\!\!\!\!\!\!\!
\frac{\partial_{\theta} {\cal P}_l(\cos\theta)}{l(l\!+\!1)}.\nonumber
\end{eqnarray}
In this form, it can be verified readily that $v$ agrees with the polar component of Eq.~(\ref{fomrs}), finishing demonstration of agreement with \cite{rao}.
%in the axially symmetric case is,
%\begin{eqnarray}&&\!\!\!\!\!\!\!\!\!\!\!\!
%u^0_{\theta}\!=\!\sum_{l=1}^{\infty}\left(\frac{{\tilde c}^{r}_{l0} }{l(l+1)r}\frac{d\left(r^{1/2}K_{l+1/2}(\lambda r)\right)}{dr}
%-\frac{1}{\lambda^2 r^{l+2}}\right)c_{l0}
%\nonumber\\&&\!\!\!\!\!\!\!\!\!\!\!\!
%\sqrt{\frac{2l+1}{4\pi}} \partial_{\theta}{\cal P}_l(\cos\theta),
%\end{eqnarray}
%where we used Eq.~(\ref{vsh}). This agrees with Rao's solution using Eq.~(\ref{idenrti}) completing the proof.

\end{appendices}

\end{document}